%% file: main.tex
\documentclass{aa}  
\usepackage{graphicx}
\usepackage{txfonts}

\usepackage{bm}
\usepackage{natbib,twoopt}
\usepackage{multirow} 
\usepackage{booktabs}
\usepackage{placeins}
\usepackage[breaklinks=true]{hyperref} 
\bibpunct{(}{)}{;}{a}{}{,} 
\makeatletter
\newcommandtwoopt{\citeads}[3][][]{\href{http://adsabs.harvard.edu/abs/#3}%
{\def\hyper@linkstart##1##2{}%
\let\hyper@linkend\@empty\citealp[#1][#2]{#3}}}
\newcommandtwoopt{\citepads}[3][][]{\href{http://adsabs.harvard.edu/abs/#3}%
{\def\hyper@linkstart##1##2{}%
\let\hyper@linkend\@empty\citep[#1][#2]{#3}}}
\newcommandtwoopt{\citetads}[3][][]{\href{http://adsabs.harvard.edu/abs/#3}%
{\def\hyper@linkstart##1##2{}%
\let\hyper@linkend\@empty\citet[#1][#2]{#3}}}
\newcommandtwoopt{\citeyearads}[3][][]%
{\href{http://adsabs.harvard.edu/abs/#3}
{\def\hyper@linkstart##1##2{}%
\let\hyper@linkend\@empty\citeyear[#1][#2]{#3}}}
\makeatother

\usepackage{lscape}

\hypersetup{
  colorlinks=true,   
  urlcolor=blue,     
  linkcolor=blue,     
  citecolor=blue
}

\begin{document} 
  \title{Systematics of planetary ephemeris reference frames inferred from pulsar timing astrometry}
  \author{N. Liu\inst{1,2}
     \and
     Z. Zhu\inst{1}
     \and 
     J. Antoniadis\inst{3,4}
     \and 
     J.-C. Liu\inst{1}
     \and
     H. Zhang\inst{1}
          }

  \institute{School of Astronomy and Space Science,
  		     Key Laboratory of Modern Astronomy and Astrophysics (Ministry of Education), 
  		     Nanjing University, Nanjing 210023, P. R. China\\
              \email{zhuzi@nju.edu.cn}
         \and
             School of Earth Sciences and Engineering, 
             Nanjing University, Nanjing 210023, P. R. China
        \and
        Institute of Astrophysics, Foundation for Research and Technology-Hellas, Voutes, 71110 Heraklion, Greece \\
            \email{john@ia.forth.gr}
        \and
        Max-Planck-Institut für Radioastronomie, Auf dem Hügel 69, D-53121 Bonn, Germany
             }

  \date{Received; accepted}

  \abstract
   {}
   {This study aims to investigate the systematics in planetary ephemeris reference frames through pulsar timing observations.}
   {We used the published data sets from several pulsar timing arrays and performed timing analyses for each pulsar using different planetary ephemerides retrieved from the Jet Propulsion Laboratory's Development Ephemeris (DE), Ephemeris of Planets and the Moon (EPM), and INPOP (Int\'egration Num\'erique Plan\'etaire de l'Observatoire de Paris).
    Then, we compared the timing solutions and modeled the differences in position and proper motion by vector spherical harmonics of the first degree.
    The timing solutions were also compared with those determined by very long baseline interferometry (VLBI) astrometry.
    }
   {
    The orientation offsets between the latest editions of the DE, EPM, and INPOP series do not exceed 0.4 milliarcseconds (mas), while the relative spins between these ephemerides are less than 5 microarcseconds per year ($\mathrm{\mu as\,yr^{-1}}$).
    We do not detect significant glides in either position or proper motion between these ephemerides.
    The orientation of the pulsar timing frames deviates from that of the VLBI frame from zero by approximately $\mathrm{0.4\,mas}$ when considering the formal uncertainty and possible systematics.
    }
   {The orientation of current planetary ephemeris frames is as accurate as at least 0.4\,mas, and the nonrotating is better than $\mathrm{5\,\mu as\,yr^{-1}}$.}


  \titlerunning{Systematics of the planetary ephemerides frames}
  
\maketitle


\section{Introduction}

    Modern numerical planetary ephemerides have been used for various purposes in deep space missions \citepads[e.g.,][]{2022AdSpR..69.1060Y}, fundamental physics (e.g., \citeads{2021A&A...647A.141P}; \citeads{2019PhRvL.123p1103B}; \citeads{2020PhRvD.102b1501B}, \citeyearads{2020PhRvD.102b1501B}, \citeyearads{2022PhRvD.105d4057B}),
    dark matter \citepads[e.g.,][]{2013MNRAS.432.3431P},
    and planetary science (e.g., \citeads{2017A&A...597A..83L}; \citeads{2020MNRAS.492..589F}, \citeyearads{2020A&A...640A...6F}; \citeads{2021A&A...654A..66Y}).
    There are three main ephemeris series, namely, 
    Jet Propulsion Laboratory's (JPL) planetary and lunar ephemerides Development Ephemeris (DE) \citepads{2021AJ....161..105P}, Ephemeris of Planets and the Moon (EPM) from the Institute of Applied Astronomy of the Russian Academy of Sciences \citepads{2022IAUS..364..220P}, and INPOP (Int\'egration Num\'erique Plan\'etaire de l'Observatoire de Paris) ephemeris from the IMCCE (Institut de m\'ecanique c\'eleste et de calcul des \'eph\'em\'erides) at Paris Observatory \citepads{2022IAUS..364...31F}.
    These planetary ephemerides realize the International Celestial Reference System \citepads[ICRS;][]{1998A&A...331L..33F} in a dynamical sense and are aligned onto the International Celestial Reference Frame \citepads[ICRF;][]{2020A&A...644A.159C} with an accuracy of sub-milliarcsecond (mas) by very long baseline interferometry (VLBI) and very long baseline array (VLBA) observations of the planetary spacecraft.
   These observations mainly include tracking measurements of 
   MESSENGER (MErcury Surface, Space ENvironment, GEochemistry, and Ranging) spacecraft orbiting Mercury \citepads{2014A&A...561A.115V};
   Venus Express (VEX) and Magellan spacecraft orbiting Venus \citep{folkner1993results,folkner1994results,folkner1994results2};
   Mars Global Surveyor, Mars Odyssey, and Mars Reconnaissance Orbiter spacecraft orbiting Mars (\citeads{2015AJ....150..121P});
   Galileo and Juno spacecraft orbiting Jupiter \citepads{2021AAS...23714205J};
   and Cassini spacecraft orbiting Saturn (\citeads{2011AJ....141...29J}, \citeyearads{2015AJ....149...28J}, \citeyearads{2020AJ....159...72J}).
   The orbits of inner planets are tied to ICRF by VLBI measurements of Mars-orbiting spacecraft \citepads{2021AJ....161..105P}.
    
    Understanding the systematic errors of the planetary ephemerides is essential for their users.
    The largest uncertainty in planetary positions comes from the lacks in the dynamical modelings \citepads[][]{2002A&A...384..322S}.
    There are at least two methods to determine the uncertainty of the planetary ephemerides \citepads{2009A&A...507.1675F}.
    The first is to compare the extrapolated positions of planets with the observations outside the fit interval of the ephemerides \citepads[e.g.,][]{2013arXiv1301.1510F}.
    The second is to assess the internal agreements between ephemerides from the same series that are constructed by identical or similar dynamical models \citepads[e.g.,][]{1990A&A...233..272S} or external agreements between the state-of-the-art ephemerides from different series \citepads[e.g.,][]{2005tvnv.conf..230P}.
    These agreements are usually evaluated by comparing the heliocentric or geocentric coordinates of planets such as Earth or Mars (e.g., \citeads{2013AASP....3..141T}; \citeads{2014ChA&A..38..330D}).
    
    The pulsar timing technique precisely records the pulse times-of-arrival (TOAs) of pulsars, from which information about the astrometry, rotation, and orbital motion can be inferred.
    One of the main objectives of pulsar timing astrometry is to search for nanohertz (nHz) gravitational wave signals in the timing observations of an ensemble of pulsars forming a pulsar timing array \citepads[PTA;][]{2019A&ARv..27....5B}.
    The planetary ephemerides are used to compute the theoretical TOAs.
    It has been noted for a long time that the pulsar timing positions obtained by using different planetary ephemerides in the pulsar timing analysis vary, for instance, as found in \citetads{1994ApJ...428..713K}.
    Several authors reported that using updated versions of planetary ephemerides yielded an improved solution, leading to a more accurate measurement of astrometric parameters such as the parallax 
    (\citeads{2005ApJ...620..405S}; \citeads{2006MNRAS.369.1502H}; \citeads{2008ApJ...679..675V}).
    As the precision of the timing astrometry is continuously improved to the level of a few tens of microarcseconds ($\mathrm{\mu as}$),
    the systematic error in the planetary ephemerides is recognized as one of the main error sources that may degrade the sensitivity of the detection of gravitational waves \citepads[e.g.,][]{2021MNRAS.508.4970C}.
    
    Pulsar timing astrometry can also contribute to the investigation of systematics in planetary ephemerides with the advantage that pulsar timing observations are fully independent of the creation process of the planetary ephemerides.
    The signal in the residual of TOAs due to the systematics in the planetary ephemerides, such as the uncertainties in the orbits of outer planets, can be modeled in the timing analysis (\citeads{2018MNRAS.481.5501C}; \citeads{2020ApJ...893..112V}).
    The comparison of pulsar positions determined from VLBI and pulsar timing observations provides a direct tie between the planetary frames and the extragalactic frames (\citeads{1996AJ....112.1690B}; \citeads{2017MNRAS.469..425W}; \citeads{2023A&A...670A.173L}).
    It is also possible to compare different ephemerides via pulsar timing astrometry to provide an independent check on the external agreements between planetary ephemerides, as done in \citetads{2011CeMDA.111..363F}.
    
    In this paper, we extend the related work of \citetads{2011CeMDA.111..363F} by using recent data releases from several PTA data sets with increased sample sizes and improved astrometric precision and accuracy (Sect.~\ref{subsect:timing-data}), updated VLBI solutions of pulsars (Sect.~\ref{subsect:vlbi-data}), as well as modeling using the vector spherical harmonics \citepads[VSH;][]{2012A&A...547A..59M} of the ephemeris systematics (Sect.~\ref{subsect:model}).
    Three motivations drive us to perform this study:
    (i) to investigate the realistic accuracy of current planetary ephemeris frames,
    (ii) to provide a transformation relation between planetary ephemerides, especially those produced by different groups,
    and (iii) to evaluate the implication of using different planetary ephemerides in the timing solutions on the derived astrometric parameters for pulsars.
    For these purposes, we compare the timing solutions based on the same sets of timing observations using different ephemerides (Sect.~\ref{subsect:timing-vs-timing}) and compare the timing solutions with the latest VLBI astrometric solutions (Sect.~\ref{subsect:timing-vs-vlbi}).
    Throughout the paper, the planetary ephemeris reference frame is referred to as the reference frame based on Earth's orbit provided by a given planetary ephemeris.
    The formal uncertainties (including the error bars) are always provided at a confidence level of 68 percent.
    The name of pulsars is given by their J2000 coordinates.


\section{Materials and methods}


\subsection{Numerical planetary ephemerides}   

\begin{figure}[htbp]
    \centering
    \includegraphics[width=\columnwidth]{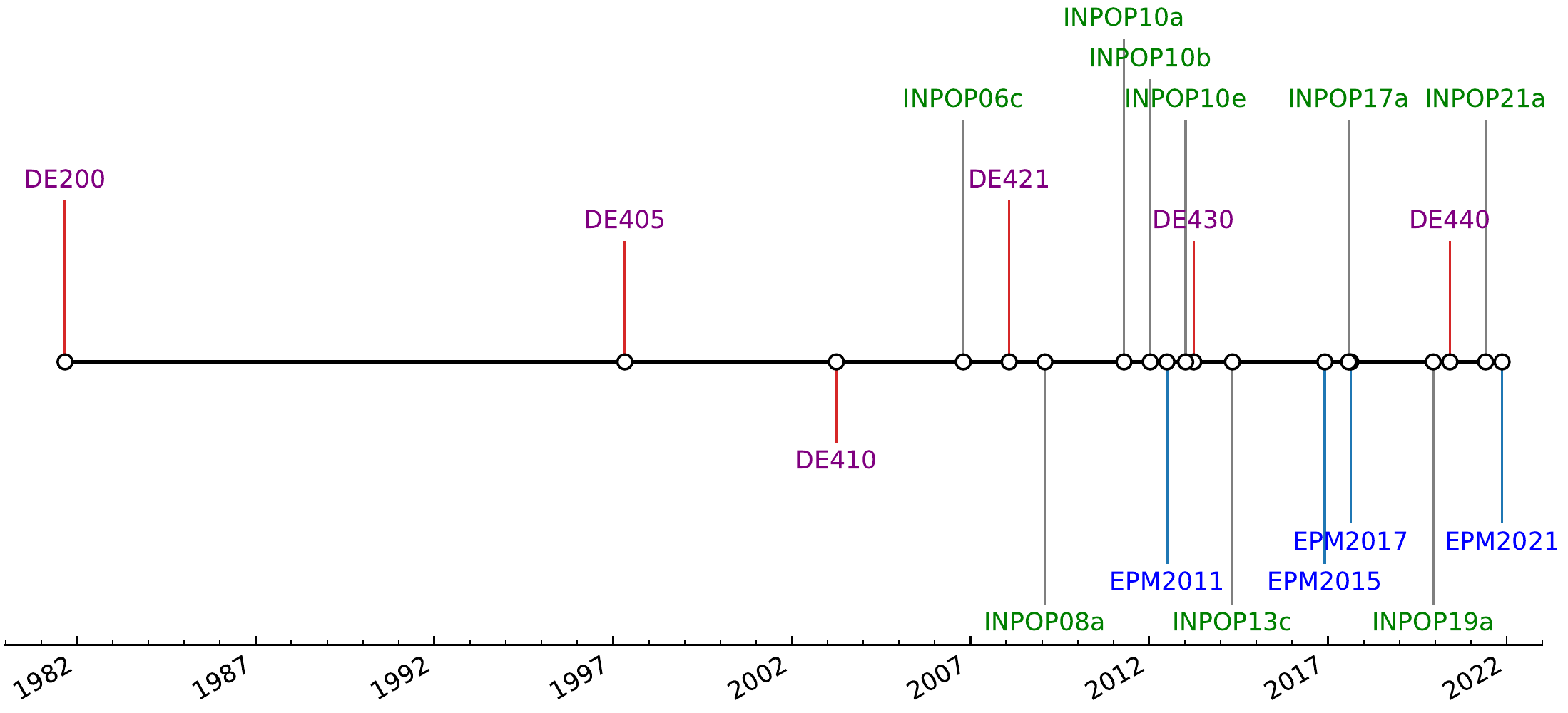}
    \caption[]{\label{fig:ephem-timeline} %
    Release timeline of the planetary ephemerides of the DE, EPM, and INPOP series.
    The release time is chosen to be the relevant date found in the online documents or publications accompanied by each ephemeris and may not be rigorously correct.
}
\end{figure}

%
\begin{table*}
    \caption{\label{table:eph-info}   
        VLBI measurements of Mars spacecraft and asteroid modeling used for each planetary ephemeris considered in this work.} 
    \centering          
    \begin{tabular}{c c c c c c c}     
    \hline
    Ephemeris &\multicolumn{3}{c}{VLBI observations of Mars} & \multicolumn{2}{c}{Dynamical Modeling} & Ref \\ 
    \cmidrule(r){2-4} \cmidrule(r){5-6}
    & Period & Number & Accuracy & Main belt & Kuiper belt  &\\
    & & &($\mathrm{mas}$) &  &   & \\
    \hline                    
    DE200     &\dots         &\dots &\dots  &3               &\dots &1 \\
    DE405     &1989          &2     &10--100&300             &\dots &2 \\
    DE410     &1989--2003    &33    &0.5    &300             &\dots &3 \\
    DE421     &2001--2007    &94    &<0.5   &343             &\dots &4--5 \\
    DE430     &2001--2013    &151   &<0.2   &343             &\dots &6 \\
    DE440     &2001--2020    &316   &0.18--0.25    &343             &30 + 1 ring &7 \\
    \hline
    EPM2011  &1989--2010    &144   &0.8         &301 + 1 ring     &21 + 1 ring  &8   \\
    EPM2015  &1989--2013    &204   &0.18--6.2   &301 + 1 ring     &30 + 1 ring  &9   \\
    EPM2017  &1989--2014    &204   &0.18--6.2   &301 + 3 rings    &30 + 3 rings  &10 \\
    EPM2021  &1989--2014    &204   &0.18--6.2   &277 + 3 rings    &30 + 3 rings  &11 \\
    \hline
    INPOP06c  &1989--2003    &44    &0.5    &300 + 1 ring    &\dots &12 \\
    INPOP08a  &1989--2007    &96    &0.4    &303 + 1 ring    &\dots &13 \\
    INPOP10a  &1989--2007    &96    &0.4    &161 + 1 ring    &\dots &14 \\
    INPOP10b  &1989--2007    &96    &0.4    &161 + 1 ring    &\dots &15 \\
    INPOP10e  &1989--2007    &96    &0.4    &161 + 1 ring    &\dots &16 \\
    INPOP13c  &1989--2007    &96    &0.4    &139             &\dots &17 \\
    INPOP17a  &1989--2013    &194   &0.3    &168             &\dots &18 \\
    INPOP19a  &1989--2013    &194   &0.3    &343             & 3 rings  &19  \\
    INPOP21a  &1989--2013    &194   &0.3    &343             & 500  &20  \\
    \hline                 
    \end{tabular}
    \tablefoot{
    The accuracy of the VLBI observations is given by a priori accuracy or the scatter (root-mean-squared or standard deviation) of postfit residuals in the literature.}
    \tablebib{
(1)~\citetads{1990A&A...233..252S};
(2)~\citet{standish1998de405};
(3)~\citet{standish2003de410};
(4)~\citet{standish2007de418};
(5)~\citetads{2009IPNPR.178C...1F};
(6)~\citetads{2014IPNPR.196C...1F};
(7)~\citetads{2021AJ....161..105P};
(8)~\citetads{2013SoSyR..47..386P};
(9)~\citet{pitjeva2017};
(10)~\citetads{2018CeMDA.130...57P};
(11)~\citetads{2022IAUS..364..220P};
(12)~\citetads{2008A&A...477..315F};
(13)~\citetads{2009A&A...507.1675F};
(14)~\citetads{2011CeMDA.111..363F};
(15)~\citet{fienga2012complementary};
(16)~\citetads{2013arXiv1301.1510F};
(17)~\citetads{2014arXiv1405.0484F};
(18)~\citetads{2017NSTIM.108.....V};
(19)~\citetads{2019NSTIM.109.....F};
(20)~\citetads{2021NSTIM.110.....F}.
}
\end{table*}
%

    We downloaded the successive editions of the planetary ephemerides from the DE series\footnote{\url{https://ssd.jpl.nasa.gov/planets/eph\_export.html}},
    the EPM series\footnote{\url{https://iaaras.ru/en/dept/ephemeris/epm/}},
    and
    the INPOP series\footnote{\url{https://www.imcce.fr/recherche/equipes/asd/inpop/}}.
    Only the planetary ephemerides that worked properly with the timing analysis software were included in this work.
    We only considered the JPL DE ephemerides  since DE403 because they were aligned onto the extragalactic reference frame, except for DE200, which was constructed on its own dynamical reference frame of J2000 \citepads{1982A&A...114..297S}.
    This ephemeris was considered because it was intensively used in the previous timing analyses.
    For many pulsars (especially young, nonrecycled pulsars), the most recent timing solutions were given under the DE200 frame. 
    Including DE200 in the comparison can thus provide a convenient transformation relation from these timing astrometric solutions referred to the DE200 frame to those referred to more recent planetary ephemerides.
    Only these DE ephemerides for a general purpose were considered, that is, DE200, DE405, DE421, DE430, and DE440, except for DE410, which was used for comparison with previous results (Sect.~\ref{subsubsect:disscusion-ephemeris-comparison}).
    The EPM series contained EPM2011, EPM2015, EPM2017, and EPM2021.
    For the INPOP series, we considered the latest version of each generation, which are INPOP06c, INPOP08a, INPOP10e, INPOP13c, INPOP17a, INPOP19a, and INPOP21a.
    In addition, INPOP10a and INPOP10b were included to illustrate the influence of different factors on the planetary ephemeris frames (Sect.~\ref{subsect:eph-accuracy}).
    In total, we considered six DE ephemerides, four EPM ephemerides, and nine INPOP ephemerides.
    
    Figure~\ref{fig:ephem-timeline} shows a rough timeline of the release dates of the ephemerides.
    One can expect that the ephemerides created at similar times may have comparable precision and accuracy, which has been validated in previous studies \citepads[e.g., ][]{2008A&A...477..315F}.
    Table~\ref{table:eph-info} lists the major differences in the VLBI observations of Mars spacecraft and asteroid perturbation modeling of the Main belt objects and Kuiper belt objects between planetary ephemerides. 
    These differences are supposed to be the most relevant to the differences between planetary ephemerides and the alignment accuracy of their frame axes onto ICRF, which are of interest in this work.
    

\subsection{Pulsar timing data sets} \label{subsect:timing-data}  

\begin{figure}[htbp]
    \centering
    \includegraphics[width=\columnwidth]{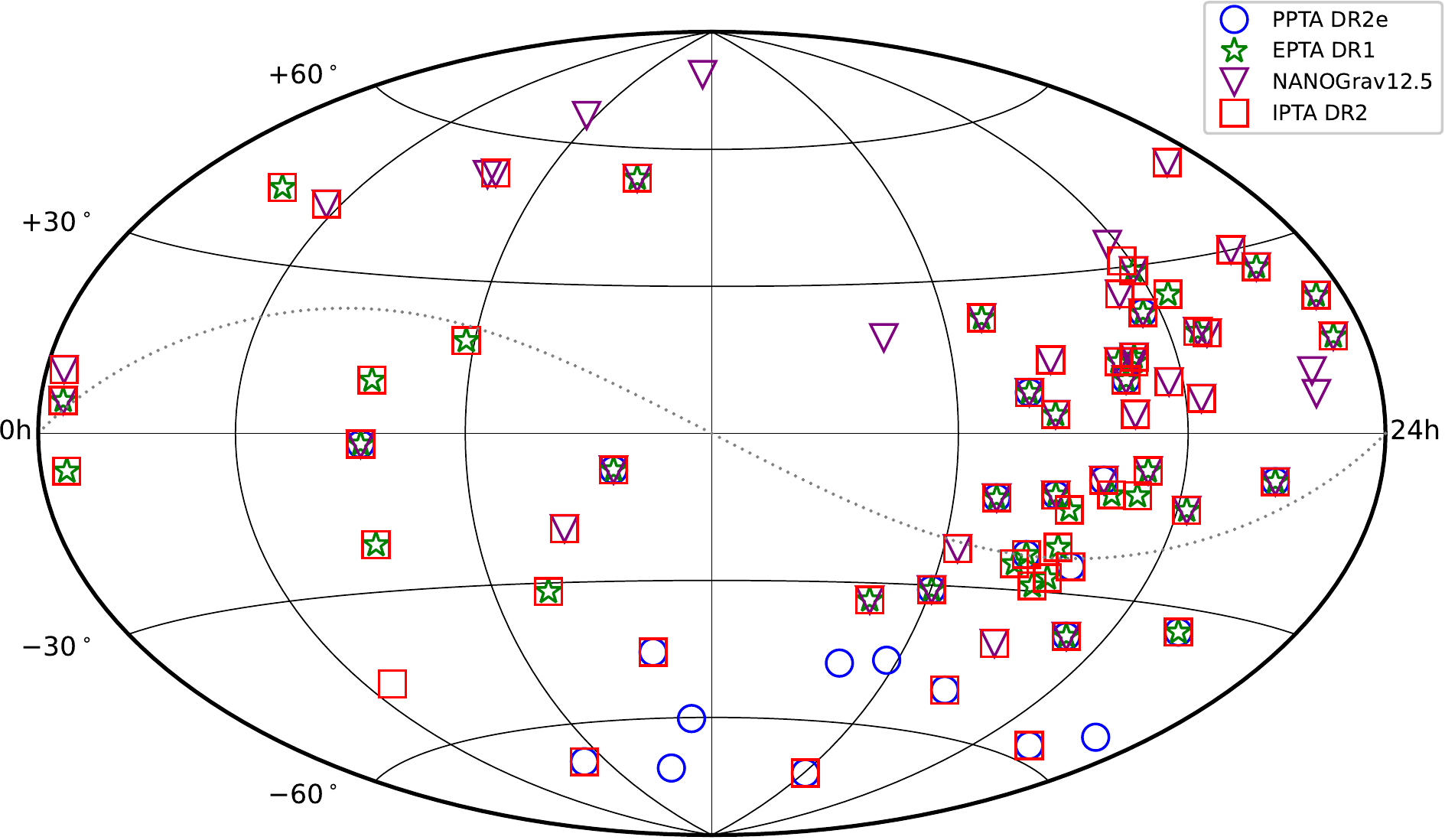}
    \caption[]{\label{fig:pta-psr-dist} %
    Sky distribution of the pulsars contained in different PTA data sets in the equatorial coordinate system.
    The dotted curve indicates the location of the ecliptic plane.
}
\end{figure}

%
\begin{table*}[htbp]
    \caption{\label{table:pta-astrometry-precision}
    Astrometric precision of the timing solutions based on different PTA data sets using DE440.
    }             
    \centering                          
    \begin{tabular}{c c c c c c c c}        
        \hline
        PTA &Subset &Pos. Epoch &Nb. PSR  & $\sigma_{\alpha^*}$ & $\sigma_{\delta}$ & $\sigma_{\mu_{\alpha^*}}$    & $\sigma_{\mu_\delta}$  \\ 
            &       &   &  &($\mathrm{\mu as}$)  &($\mathrm{\mu as}$)&($\mathrm{\mu as\,yr^{-1}}$)  &($\mathrm{\mu as\,yr^{-1}}$)  \\ 
        \hline                       
    PPTA DR2e                  &All                        &  2011.8  &    24  &      47  &      89  &      12  &      22    \\
                               &$|\beta|\!\le\!10^{\circ}$  &  2011.3  &     4  &     101  &    1469  &      29  &     427    \\
                               &$|\beta|\!>\!10^{\circ}$   &  2012.0  &    20  &      40  &      74  &       9  &      16    \\
    EPTA DR1                   &All                        &  2009.5  &    42  &     158  &     359  &      69  &     165    \\
                               &$|\beta|\!\le\!10^{\circ}$  &  2009.5  &    12  &     197  &    2255  &     120  &    1046    \\
                               &$|\beta|\!>\!10^{\circ}$   &  2009.5  &    30  &     140  &     302  &      67  &     136    \\
    NANOGrav12.5-WB            &All                        &  2013.6  &    47  &      82  &     156  &      44  &      89    \\
                               &$|\beta|\!\le\!10^{\circ}$  &  2012.5  &     6  &     212  &     977  &      84  &     245    \\
                               &$|\beta|\!>\!10^{\circ}$   &  2013.8  &    41  &      65  &     140  &      39  &      80    \\
    NANOGrav12.5-NB             &All                        &  2013.6  &    47  &      58  &     127  &      38  &      79    \\
                               &$|\beta|\!\le\!10^{\circ}$  &  2012.4  &     6  &     230  &     996  &      82  &     247    \\
                               &$|\beta|\!>\!10^{\circ}$   &  2013.8  &    41  &      58  &     108  &      30  &      58    \\
    IPTA DR2A                  &All                        &  2010.2  &    64  &     161  &     320  &      60  &     154    \\
                               &$|\beta|\!\le\!10^{\circ}$  &  2009.8  &    15  &     254  &    2171  &     119  &     924    \\
                               &$|\beta|\!>\!10^{\circ}$   &  2010.3  &    49  &     147  &     248  &      45  &      77    \\
    IPTA DR2B                  &All                        &  2010.2  &    64  &     145  &     314  &      59  &     136    \\
                               &$|\beta|\!\le\!10^{\circ}$  &  2009.8  &    15  &     281  &    1763  &     140  &     977    \\
                               &$|\beta|\!>\!10^{\circ}$   &  2010.3  &    49  &     109  &     173  &      44  &      70    \\
                               \hline
    Combination                &All                             &  2011.4  &    80  &     107  &     207  &      49  &     107    \\
                               &$|\beta|\!\le\!10^{\circ}$      &  2010.4  &    15  &     240  &    1465  &     102  &     581    \\
                               &$|\beta|\!>\!10^{\circ}$        &  2011.6  &    65  &      86  &     146  &      39  &      70    \\
    \hline                                   
    \end{tabular}
    \tablefoot{
    $\sigma_{\alpha^*}\!=\!\sigma_{\alpha}\cos\delta$ and $\sigma_{\mu_{\alpha^*}}\!=\!\sigma_{\mu_\alpha}\cos\delta$.
    The first four columns tabulate the PTA name, subset, mean position epoch, and the number of pulsars, followed by the typical precision (median) of the position and proper motion measurements.
    The last three rows represent the combination of all PTA data sets above.
    }
\end{table*}

    We retrieved the pulsar timing data sets from several pulsar timing arrays, including data release 1.0 of the European Pulsar Timing Array \citepads[EPTA DR1;][]{2016MNRAS.458.3341D}, the extended second data release of the Parkes Pulsar Timing Array \citepads[PPTA DR2e;][]{2021MNRAS.507.2137R}, the narrowband and wideband versions of the 12.5 yr data set of the North American Nanohertz Observatory for Gravitational Waves (NANOGrav12.5-NB and NANOGrav12.5-WB; \citeads{2021ApJS..252....4A}, \citeyearads{2021ApJS..252....5A}), and two versions (A and B) of the second data release of the International Pulsar Timing Array \citepads[IPTA DR2A and IPTA DR2B;][]{2019MNRAS.490.4666P}.
    All of these timing observations were made between 1986 and 2018.
    While the data set in the EPTA DR1, PPTA DR2e, and NANOGrav12.5 are obtained from independent observations of three regional PTAs in Europe, North America, and Australia, the IPTA combines the data sets from these individual PTAs to improve the sensitivity.
    The combination also leads to an increasing number of pulsars and better sky coverage compared to individual PTAs.
    For IPTA DR2, the input TOAs came from the previous data releases of these PTAs, which are EPTA DR1, the 9~yr data set from NANOGrav \citepads{2016ApJ...818...92M}, and the first data release of PPTA and its extended version \citepads{2016MNRAS.455.1751R}.
    The main difference between the two versions of IPTA DR2 (i.e., IPTA DR2A and IPTA DR2B) lies in the modeling of the dispersion measure and handling of the noise properties of pulsars.
    In this work, EPTA DR1, PPTA DR2e, and NANOGrav12.5 were used in addition to IPTA DR2 for two reasons.
    The first one was that PPTA DR2e and NANOGrav12.5 surpassed their previous data releases used to generate IPTA DR2 in terms of astrometric accuracy and sample size.
    There were five pulsars in PPTA DR2e and ten pulsars in NANAGrav12.5 that were not included in IPTA DR2.
    The other reason was that different PTAs with independent observations and analyses could be used to examine the consistency between results of planetary ephemeris comparison based on pulsar timing observations.
    The timing solutions from all the data sets were combined to derive the final results.

    Since the planetary ephemerides are mainly constrained by observations on the ecliptic plane, it would be interesting to use only pulsars right on or close to the ecliptic plane.
    However, the major axis of the error ellipse tends to align with the ecliptic longitude for timing astrometry.
    The timing astrometric error ellipse would become extremely elongated for pulsars right on the ecliptic, making the measurement of right ascension and declination highly correlated and leading to a less reliable timing astrometric solution.
    One example in our sample is PSR J1022+1001, whose position is usually provided in the ecliptic coordinate system in the timing solution. 
    We failed to reproduce timing solutions using several planetary ephemerides for this pulsar.
    This pulsar might not be suitable for our analyses and thus was removed from our sample.
    We believe that missing one pulsar should not affect our results much because we studied the common features yielded in the timing solutions of all pulsars.
    To compare the results based on pulsars near or far from the ecliptic plane, we divided the sample into three subsets: all pulsars, pulsars close to the ecliptic plane (i.e., ecliptic latitude $|\beta|\!\le\!10^{\circ}$), and pulsars far from the ecliptic plane (i.e., $|\beta|\!>\!10^{\circ}$).
    Figure~\ref{fig:pta-psr-dist} depicts the sky distribution of all pulsars in the equatorial coordinate system, from which we found that the sample was dominated by pulsars at high ecliptic latitudes.
    
    The timing models for each pulsar given in each PTA data set were directly used to rerun the timing analysis, except that different planetary ephemerides were adopted.
    We processed the published TOAs of all PTA data sets by \texttt{TEMPO2}\footnote{\url{https://bitbucket.org/psrsoft/tempo2/src/master/}} \citepads[version 2022.05.1;][]{2006MNRAS.369..655H}.
    Noting that the planetary masses used in \texttt{TEMPO2} (defined as constant variables) usually differed from those used in the planetary ephemerides, we manually modified the values of the related variables in \texttt{TEMPO2} to be consistent with each planetary ephemeris to be studied.
    The timing astrometric solutions of all pulsars in each PTA under the same planetary ephemeris formed a pulsar catalog, which was considered as an independent representation of the reference frame of the used ephemerides.
    
    Since the DE ephemerides were widely used in the pulsar timing analyses, we chose the timing solutions using DE440 as the reference solution for each PTA data set.
    Table~\ref{table:pta-astrometry-precision} presents a brief overview of these reference solutions.
    The typical timing position precision (median) was approximately 50\,${\rm \mu as}$--150\,$\mathrm{\mu as}$ in right ascension and 100\,$\mathrm{\mu as}$--350\,$\mathrm{\mu as}$ in declination.
    For the proper motion measurements, the typical precision was in the range of 10\,$\mathrm{\mu as\,yr^{-1}}$ to 70\,$\mathrm{\mu as\,yr^{-1}}$ in right ascension and twice as poor in declination.
    Comparing two samples of $|\beta|\!\le\!10^{\circ}$ and $|\beta|\!>\!10^{\circ}$, we found that the astrometric precision in either right ascension or declination for pulsars near the ecliptic plane was twice or more worse than those for pulsars away from the ecliptic plane, which is a typical feature of timing astrometry.
    
    Assuming the timing precision merely consists of the measurement noise, the sensitivity to the systematics of the ephemeris frames to be examined would be improved by a factor of the root square of the number of pulsars in the sample.
    Using the values presented in Table~\ref{table:pta-astrometry-precision}, we obtained 8\,${\rm \mu as}$ to 24\,$\mathrm{\mu as}$ for right ascension, 18\,${\rm \mu as}$ to 55\,$\mathrm{\mu as}$ for declination, 2\,$\mathrm{\mu as\,yr^{-1}}$ to 11\,$\mathrm{\mu as\,yr^{-1}}$ for proper motion in right ascension, and 4\,$\mathrm{\mu as\,yr^{-1}}$ to 26\,$\mathrm{\mu as\,yr^{-1}}$ for proper motion in declination.
    The detection sensitivity in the position system was thus well below the orientation accuracy of the planetary ephemeris reference frames (i.e., $\sim$0.25\,mas), enabling us to perform meaningful investigations into the possible systematics in the planetary ephemeris reference frames.


\subsection{VLBI solutions of pulsars}   \label{subsect:vlbi-data}

\begin{table*}[htbp]
    \caption{Overview of timing and VLBI solutions for seven pulsars used for evaluating the rotation between ephemerides frames and extragalactic frame.}              
    \label{table:timing-vs-vlbi-msp}      
    \centering          
    \begin{tabular}{c c c c c c c c c c c c}     
    \hline
    \multirow{2}{*}{Pulsar} &\multicolumn{5}{c}{Timing} &\multicolumn{5}{c}{VLBI} 
    &\multirow{2}{*}{Ref.} \\
    \cmidrule(r){2-6} \cmidrule(r){7-11}
    &  Epoch  & $\sigma_{\alpha^*}$ & $\sigma_{\delta}$  & $\sigma_{\mu_{\alpha^*}}$    & $\sigma_{\mu_\delta}$  &Eopch &$\sigma_{\alpha^*}$  & $\sigma_{\delta}$  & $\sigma_{\mu_{\alpha^*}}$    & $\sigma_{\mu_\delta}$  &  \\ 
    &  &($\mathrm{\mu as}$)  &($\mathrm{\mu as}$)  &($\mathrm{\mu as\,yr^{-1}}$)  &($\mathrm{\mu as\,yr^{-1}}$)  &  &($\mathrm{\mu as}$)  &($\mathrm{\mu as}$)  &($\mathrm{\mu as\,yr^{-1}}$)  &($\mathrm{\mu as\,yr^{-1}}$)  &     \\
    \hline                    
    J0437$-$4715  &  2009.5  &    5  &    5  &    1  &    2  &  2007.0  & 1015  &  994  &   50  &   90  &1 \\
    J1012+5307  &  2009.5  &   49  &   64  &   10  &   14  &  2016.9  &  900  & 1000  &   90  &  140  &2 \\
    J1640+2224  &  2009.5  &   46  &   65  &   13  &   19  &  2016.3  &  971  & 1000  &   80  &  140  &3 \\
    J1713+0747  &  2009.5  &    5  &   10  &    1  &    2  &  2002.0  & 1486  & 2000  &  170  &  160  &4 \\
    J2010$-$1323  &  2009.5  &  148  &  614  &   56  &  237  &  2012.2  & 1459  & 4000  &  329  &  303  &5 \\
    J2145$-$0750  &  2009.5  &   97  &  268  &   25  &   69  &  2012.2  & 1486  & 4000  &   52  &   90  &5 \\
    J2317+1439  &  2009.5  &   88  &  173  &   25  &   49  &  2012.2  & 1451  & 1000  &  465  &  704  &5 \\
    \hline                 
    \end{tabular}
    \tablefoot{The timing solutions were obtained through a timing analysis of the IPTA DR2B data set using DE440.
    The last column gives the reference to the VLBI position.
    }
    \tablebib{
(1)~\citetads{2009ApJ...701.1243D};
(2)~\citetads{2020ApJ...896...85D};
(3)~\citetads{2018ApJ...855..122V};
(4)~\citetads{2009ApJ...698..250C};
(5)~\citetads{2019ApJ...875..100D}.
}
\end{table*}
%

    We found the VLBI astrometric solutions for seven pulsars in our sample (PSR J1022+1001 excluded), which mostly came from the PSR$\pi$ and MSPSR$\pi$ campaigns \citepads{2019ApJ...875..100D}.
    These sources were all included in the IPTA DR2 sample.
    Among these seven pulsars, six sources were common to the EPTA DR1 and NANOGrav samples, and two pulsars were found in PPTA DR2e.
    There are two pulsars with $|\beta|\!\le\!10^{\circ}$, which are PSR J2010$-$1323 and PSR J2145$-$0750.
    
    We noted that the positional uncertainty in \citetads{2009ApJ...701.1243D} did not include the systematics due to the calibrator position uncertainty, core shift, and phase referencing errors from the primary calibrator to the in-beam calibrator.
    Therefore, we inflated the positional uncertainty therein using the same method as described in \citet{2023A&A...670A.173L}.
    That is, we added in quadrature to the positional uncertainty given in \citetads{2009ApJ...701.1243D} an empirical value of 0.8\,mas in each coordinate for the core-shift, 1.26\,mas in right ascension and 0.59\,mas in declination accounting for the phase referencing error, and the error in the absolute position of the calibrator.
    The positional precision of the timing and VLBI astrometric solutions is displayed in Table~\ref{table:timing-vs-vlbi-msp}.
    The VLBI positional uncertainty now reaches $\sim\mathrm{1\,mas}$, while the timing positional uncertainty is at least ten times smaller.
    The correlations between right ascension and declination in the VLBI solutions were not available and were assumed to be zero.
    
    The \textit{Gaia} data also provided the astrometric solutions for pulsars in the \textit{Gaia} reference frame \citepads[Gaia-CRF3;][]{2022A&A...667A.148G}.
    We followed the method given in \citetads{2021MNRAS.501.1116A} to search for pulsars in \textit{Gaia} Data Release 3 \citepads{2022arXiv220800211G}.
    We only found two pulsars, preventing us from performing a robust comparison between the planetary ephemeris frames and the \textit{Gaia} reference frame.


\subsection{Modeling of systematic differences in position and proper motion}  \label{subsect:model}

    The pulsar timing position and proper motion differences between using different ephemerides and using DE440 were computed as
    \begin{align}
            \Delta\alpha^* &= (\alpha_{\rm e} - \alpha_{\rm r}) \cos \delta_{\rm r}, \label{eq:ra_offset} \\
            \Delta\delta &= \delta_{\rm e} - \delta_{\rm r}, \label{eq:dec_offset} \\
            \Delta\mu_{\alpha^*} &= \mu_{\alpha^*,{\rm e}} - \mu_{\alpha^*,{\rm r}}, \label{eq:pmra_offset} \\
            \Delta\mu_{\delta} &= \mu_{\delta,{\rm e}} - \mu_{\delta,{\rm r}}, \label{eq:pmdec_offset} 
    \end{align}
    where the subscripts (as well as superscripts shown below) ``r'' and ``e'' represent the reference solution using DE440 and the timing solutions using ephemerides other than DE440, respectively.
    The notation $\mu_{\alpha^*}\!=\!\mu_\alpha\cos\delta$, together with other similar notations (e.g., $\sigma_{\alpha^*}\!=\!\sigma_{\alpha}\cos\delta$), was used throughout the paper.
    We used the four-dimensional vector in the form of 
    \begin{equation} \label{eq:4d-obs}
        \boldsymbol{y}=(\Delta\alpha^*,\, \Delta\delta,\, \Delta\mu_{\alpha^*},\, \Delta\mu_{\delta})^{\rm T}
    \end{equation}
    as the input observable. 
    The full covariance matrix of the observable was expressed as 
    \begin{equation} \label{eq:obs-covmat}
        \boldsymbol{C}_{\boldsymbol{y}}= \left[
            \begin{array}{cccc}
                \sigma^2_{\Delta\alpha^*} 
                & C_{\Delta\alpha^*,\Delta\delta}
                & C_{\Delta\alpha^*,\Delta\mu_{\alpha^*}}
                & C_{\Delta\alpha^*,\Delta\mu_{\delta}} \\
                C_{\Delta\alpha^*,\Delta\delta}
                & \sigma^2_{\Delta\delta} 
                & C_{\Delta\delta,\Delta\mu_{\alpha^*}}
                & C_{\Delta\delta,\Delta\mu_{\delta}} \\
                C_{\Delta\alpha^*,\Delta\mu_{\alpha^*}} 
                & C_{\Delta\delta,\Delta\mu_{\alpha^*}}
                & \sigma^2_{\Delta\mu_{\alpha^*}} 
                & C_{\Delta\mu_{\alpha^*},\Delta\mu_{\delta}} \\
                C_{\Delta\alpha^*,\Delta\mu_{\delta}}
                & C_{\Delta\delta,\Delta\mu_{\delta}} 
                & C_{\Delta\mu_{\alpha^*},\Delta\mu_{\delta}} 
                & \sigma^2_{\Delta\mu_{\delta}} 
            \end{array}
            \right],
    \end{equation}
    where
    \begin{align}
        \sigma_{\Delta\alpha^*}  &= \sqrt{\sigma^2_{\alpha^*,{\rm e}} + \sigma^2_{\alpha^*,{\rm r}}} ,  \\
        \sigma_{\Delta\delta}    &= \sqrt{\sigma^2_{\delta,{\rm e}} + \sigma^2_{\delta,{\rm r}}} ,     \\
        \sigma_{\Delta\mu_{\alpha^*}}  &= \sqrt{\sigma^2_{\mu_{\alpha^*},{\rm e}} + \sigma^2_{\mu_{\alpha^*},{\rm r}}} ,  \\
        \sigma_{\Delta\mu_{\delta}}    &= \sqrt{\sigma^2_{\mu_{\delta},{\rm e}} + \sigma^2_{\mu_{\delta},{\rm r}}} ,     \\
        C_{\Delta\alpha^*,\Delta\delta} &= \rho^{\rm e}_{\alpha,\delta} \sigma_{\alpha^*,{\rm e}}\sigma_{\delta,{\rm e}} + \rho^{\rm r}_{\alpha,\delta} \sigma_{\alpha^*,{\rm r}}\sigma_{\delta,{\rm r}}, \\
        C_{\Delta\alpha^*,\Delta\mu_{\alpha^*}} &= \rho^{\rm e}_{\alpha,\mu_{\alpha^*}} \sigma_{\alpha^*,{\rm e}}\sigma_{\mu_{\alpha^*},{\rm e}} + \rho^{\rm r}_{\alpha,\mu_{\alpha^*}} \sigma_{\alpha^*,{\rm r}}\sigma_{\mu_{\alpha^*},{\rm r}}, \\
        C_{\Delta\alpha^*,\Delta\mu_{\delta}} &= \rho^{\rm e}_{\alpha,\mu_{\delta}} \sigma_{\alpha^*,{\rm e}}\sigma_{\mu_{\delta},{\rm e}} + \rho^{\rm r}_{\alpha,\mu_{\delta}} \sigma_{\alpha^*,{\rm r}}\sigma_{\mu_{\delta},{\rm r}}, \\
        C_{\Delta\delta,\Delta\mu_{\alpha^*}} &= \rho^{\rm e}_{\delta,\mu_{\alpha^*}} \sigma_{\delta,{\rm e}}\sigma_{\mu_{\alpha^*},{\rm e}} + \rho^{\rm r}_{\delta,\mu_{\alpha^*}} \sigma_{\delta,{\rm r}}\sigma_{\mu_{\alpha^*},{\rm r}},\\
        C_{\Delta\delta,\Delta\mu_{\delta}} &= \rho^{\rm e}_{\delta,\mu_{\delta}} \sigma_{\delta,{\rm e}}\sigma_{\mu_{\delta},{\rm e}} + \rho^{\rm r}_{\delta, \mu_{\delta}} \sigma_{\delta,{\rm r}}\sigma_{\mu_{\delta},{\rm r}}, \\
        C_{\Delta\mu_{\alpha^*},\Delta\mu_{\delta}} &= \rho^{\rm e}_{\mu_{\alpha^*},\mu_{\delta}} \sigma_{\mu_{\alpha^*},{\rm e}}\sigma_{\mu_{\delta},{\rm e}} + \rho^{\rm r}_{\mu_{\alpha^*}, \mu_{\delta}} \sigma_{\mu_{\alpha^*},{\rm r}}\sigma_{\mu_{\delta},{\rm r}},.
    \end{align}
    We used $\sigma$ and $\rho$ to represent the formal uncertainty and correlation coefficient, respectively.
    To quantify the significance of the position and proper motion difference, we adopted a normalized quantity $X_{\boldsymbol{y}}$ defined as 
    \begin{equation} \label{eq:X-y}
        X_{\boldsymbol{y}}^{2}=\left[\begin{array}{llll}
                \Delta\alpha^* &\Delta\delta  &\Delta\mu_{\alpha^*} &\Delta\mu_{\delta}
                \end{array}\right] 
                \boldsymbol{C}_{\boldsymbol{y}}^{-1}
                \left[\begin{array}{l}
                \Delta\alpha^* \\
                \Delta\delta  \\
                \Delta\mu_{\alpha^*} \\
                \Delta\mu_{\delta}
            \end{array}\right].
    \end{equation}
  
    We considered the global differences between both the position and proper motion systems of the planetary ephemerides, which would manifest themselves in the systematic changes of the pulsar timing positions and proper motions when different planetary ephemerides were used in the timing analyses. 
    These global differences were modeled by the VSH of the first degree, including a rotation vector and a glide vector.
    
    For the position system, the rotation vector $\boldsymbol{R}\!=\!(R_{\rm X},R_{\rm Y},R_{\rm Z})^{\rm T}$ mainly represents the orientation offsets between the planetary ephemeris reference frames at the coordinate epoch (i.e., J2000).
    On the other hand, the glide vector $\boldsymbol{G}\!=\!(G_{\rm X},G_{\rm Y},G_{\rm Z})^{\rm T}$ models a dipolar positional offset field that can be caused by the large-scale deformation in the planetary ephemeris reference frames and relative displacements and motions of the Solar System Barycenter (SSB) defined implicitly by the planetary ephemerides.
    The pulsar positional offsets at J2000 modeled by these two vectors can be expressed as 
    \begin{align}
        \Delta{\alpha^{*}_0} = &-R_{\rm X} \cos \alpha_{\rm r} \sin \delta_{\rm r} -R_{\rm Y} \sin \alpha_{\rm r} \sin \delta_{\rm r} +R_{\rm Z} \cos \delta_{\rm r}  \nonumber \\
        &-G_{\rm X} \sin \alpha_{\rm r} +G_{\rm Y} \cos \alpha_{\rm r}, \label{eq:pos-oft-ra-vsh01} \\
        \Delta{\delta}_0 = &+R_{\rm X} \sin \alpha_{\rm r} -R_{\rm Y} \cos \alpha_{\rm r} \nonumber\\
        &-G_{\rm X} \cos \alpha_{\rm r} \sin \delta_{\rm r} -G_{\rm Y} \sin \alpha_{\rm r} \sin \delta_{\rm r} +G_{\rm Z} \cos \delta_{\rm r}. \label{eq:pos-oft-dec-vsh01}
    \end{align}

    For the proper motion system, the rotation vector $\boldsymbol{r}\!=\!(r_{\rm X},r_{\rm Y},r_{\rm Z})^{\rm T}$ characterizes the relative spin of the planetary ephemeris reference frames, which can be attributed to the mean motion difference in the orbital determinations of the Earth.
    Meanwhile, the glide vector $\boldsymbol{g}\!=\!(g_{\rm X},g_{\rm Y},g_{\rm Z})^{\rm T}$ can be related to the relative accelerations of the SSB defined by different planetary ephemerides (if they exist).
    Similar to Eqs.(\ref{eq:pos-oft-ra-vsh01})--(\ref{eq:pos-oft-dec-vsh01}), we obtained the pulsar proper motion offsets as 
    \begin{align}
        \Delta\mu_{\alpha^{*},{\rm C}} = &-r_{\rm X}\cos\alpha_{\rm r}\sin\delta_{\rm r} - r_{\rm Y} \sin \alpha_{\rm r} \sin \delta_{\rm r} + r_{\rm Z} \cos \delta_{\rm r} \nonumber \\
        &-g_{\rm X} \sin \alpha_{\rm r} +g_{\rm Y} \cos \alpha_{\rm r},  \label{eq:dpmra_c} \\
        \Delta\mu_{\delta,{\rm C}} = &+r_{\rm X} \sin \alpha_{\rm r} -r_{\rm Y} \cos \alpha_{\rm r} \nonumber\\
        &-g_{\rm X} \cos \alpha_{\rm r} \sin \delta_{\rm r} -g_{\rm Y} \sin \alpha_{\rm r} \sin \delta_{\rm r} +g_{\rm Z} \cos \delta_{\rm r}.  \label{eq:dpmdec_c} 
    \end{align}
    In addition, the resulting proper motion offsets can also lead to a positional displacement if the reference epoch $t$ of the pulsar position differs from the coordinate epoch $t_0$ of the planetary ephemeris frame (i.e., J2000).
    Therefore, the total positional offsets due to the global differences of the aforementioned planetary ephemeris reference frame can be written as
    \begin{align}
        \Delta\alpha^*_C &= \Delta\alpha^*_0 + \Delta\mu_{\alpha^*,C} \left( t - t_0 \right), \label{eq:dra_c} \\
        \Delta\delta_C   &= \Delta\delta_0 + \Delta\mu_{\delta,C} \left( t - t_0 \right). \label{eq:ddec_c}    
    \end{align}

    The VSH parameters (components in vectors $\boldsymbol{R}$, $\boldsymbol{G}$, $\boldsymbol{r}$, and $\boldsymbol{g}$) were determined by a least-squares fit of the model given by Eqs.~(\ref{eq:dpmra_c})--(\ref{eq:ddec_c}) to the observables described by Eqs.~(\ref{eq:ra_offset})--(\ref{eq:pmdec_offset}).
    The observables were weighted by the inverse of the covariance matrix $\boldsymbol{C}_{\boldsymbol{y}}$ given in Eq.~(\ref{eq:obs-covmat}).
    To improve the robustness of the fitting against the potential outliers, we computed the normalized difference $X_{\boldsymbol{y}}$ based on the postfit residuals following Eq.~(\ref{eq:X-y}).
    The sources were considered outliers when the normalized differences exceeded the median value of the normalized differences for all sources by a chosen factor $\kappa$ (called the clip limit), as used in \citetads{2021A&A...649A...9G}.
    The fit was iterated until no new outlier was detected.
    We set $\kappa\!=\!3$ in this work.
    Different values of $\kappa$ were examined: the estimates of the VSH parameters changed little.
    We also adopted a bootstrap fit to cross-check the results of the least-squares fit.
    We generated 1000 trial samples, for which we randomly selected the same number of pulsars from the input sample with replacements.
    The least-squares fit was performed for each trial sample. 
    The median value and the median absolute deviation (MAD) multiplied by a factor of 1.4826 (i.e., normalized MAD, which is equivalent to 1-sigma in a Gaussian distribution) were used as the final estimate and the associated uncertainty for each unknown.

    For the comparison between timing and VLBI astrometry, we first propagated the VLBI positions from their own epoch to those of the timing positions using the VLBI proper motions and then computed the offsets of the timing positions with respect to the propagated VLBI positions.
    The uncertainties of these positional offsets in either right ascension or declination were computed as the square root of the quadratic sum of both uncertainties in VLBI and timing positions as well as the propagation errors induced by the formal error in VLBI proper motion.  
    Since there were a limited number of pulsars in common, we only modeled these offsets with the orientation offset, that is, the contribution from the vector $\boldsymbol{R}$, as given in Eqs.~(\ref{eq:pos-oft-ra-vsh01})--(\ref{eq:pos-oft-dec-vsh01}).    

    We performed the fit separately for each PTA data set to check their consistencies.
    The final estimates, together with their uncertainties, were obtained by a least-squares fit to the combination of the timing solutions from all PTAs.

\section{Results} \label{sect:results}

\subsection{Offsets of the pulsar timing position and proper motion due to different ephemerides} \label{subsect:timing-vs-timing}
    
\begin{figure*}[htbp]
  \centering
  \includegraphics[width=\columnwidth]{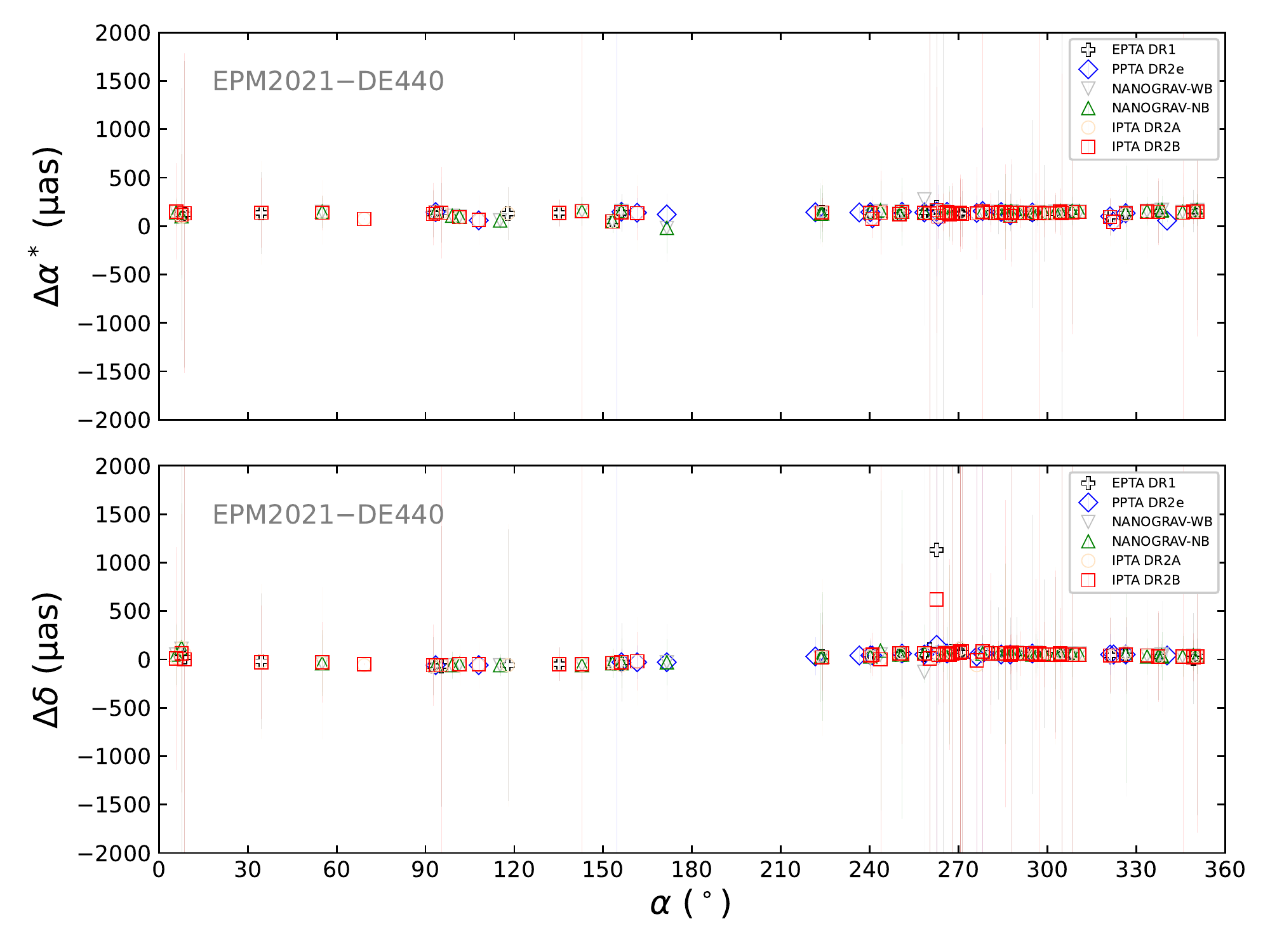}
  \includegraphics[width=\columnwidth]{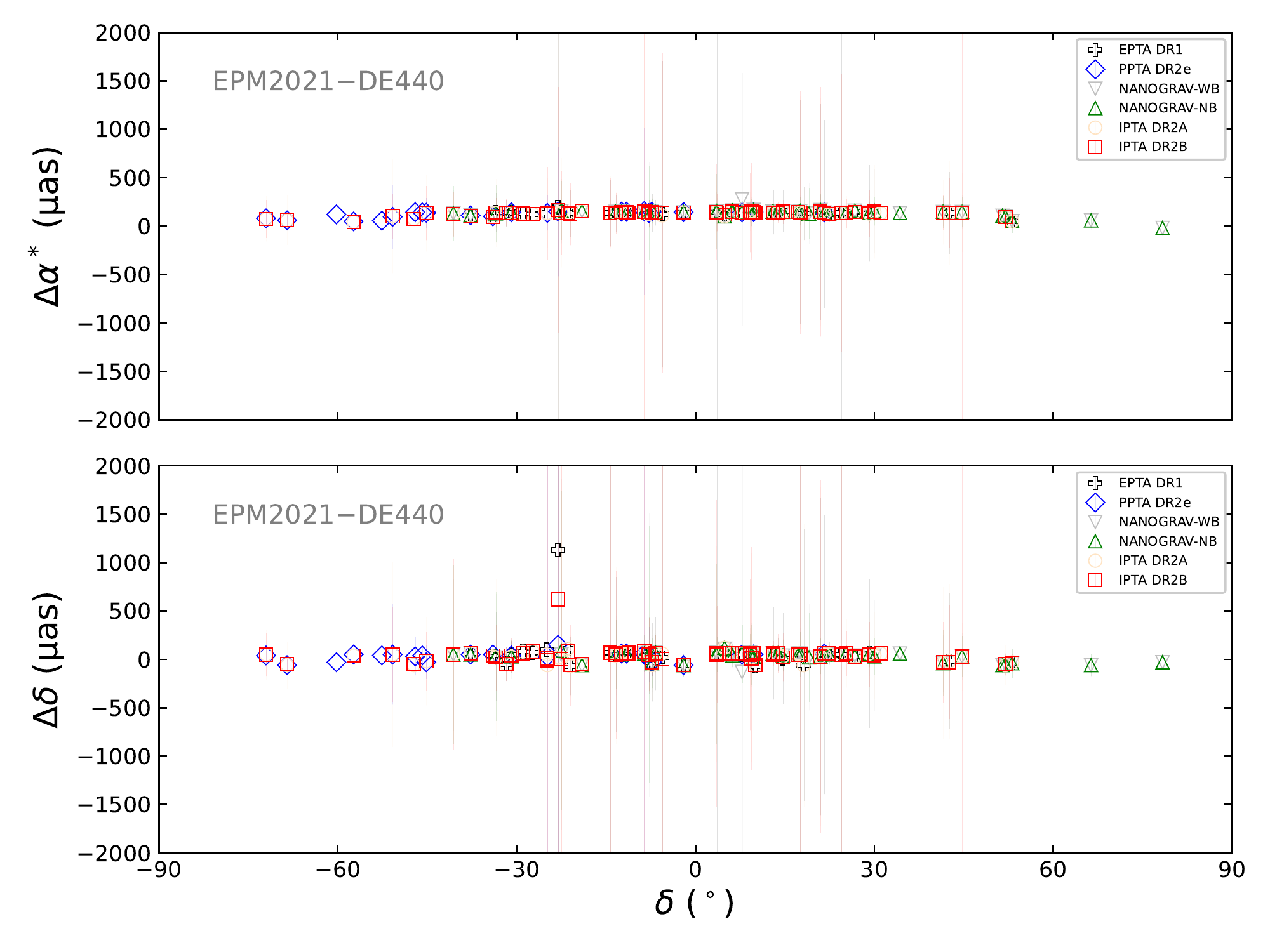}
  \caption[]{\label{fig:equ-pos-epm2021-vs-de440} %
  Offsets of the pulsar timing positions in the EPM2021 frame with respect to those in the DE440 frame as a function of right ascension (left) and declination (right).
}
\end{figure*}

\begin{figure*}[htbp]
  \centering
  \includegraphics[width=\columnwidth]{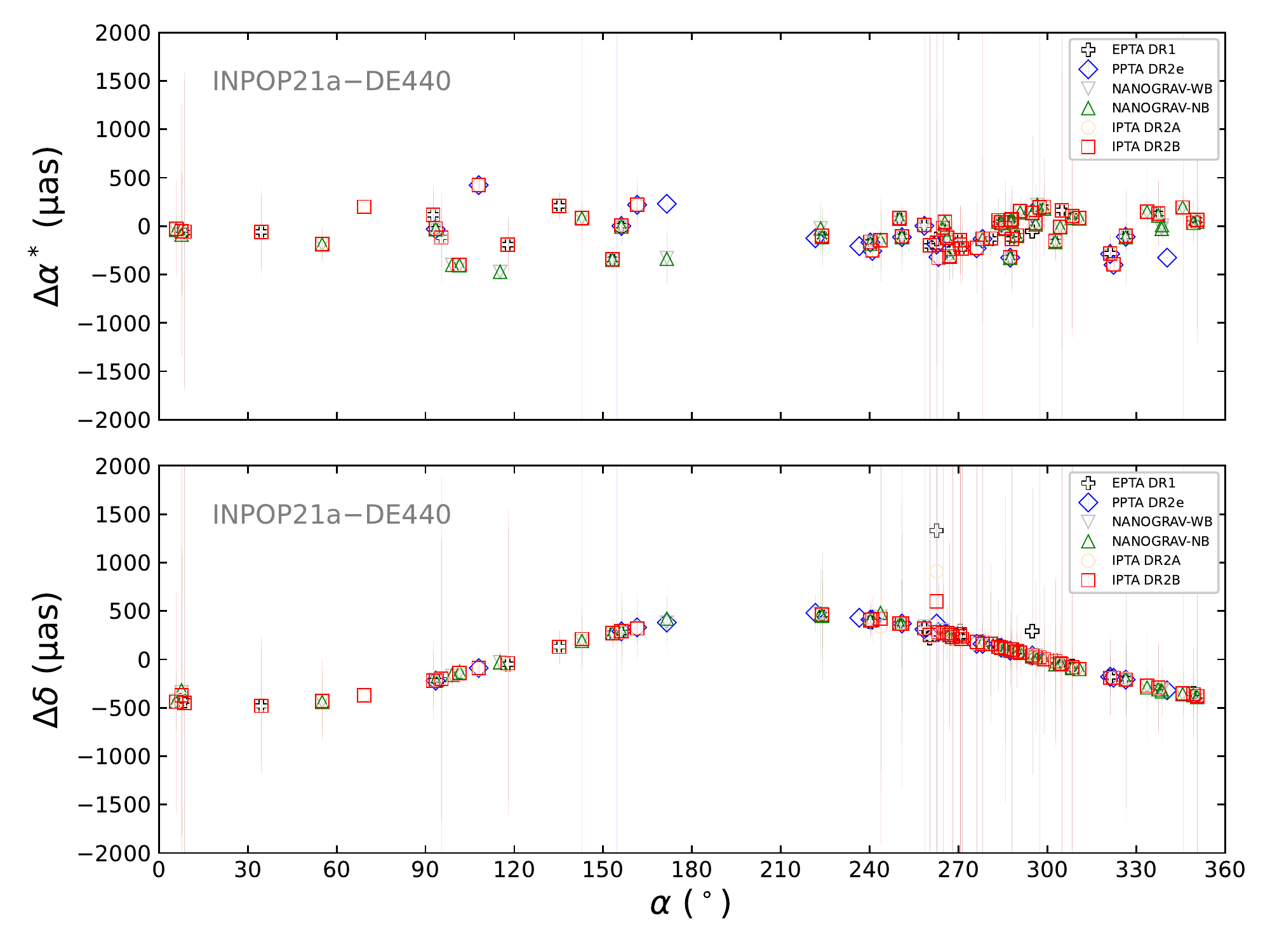}
  \includegraphics[width=\columnwidth]{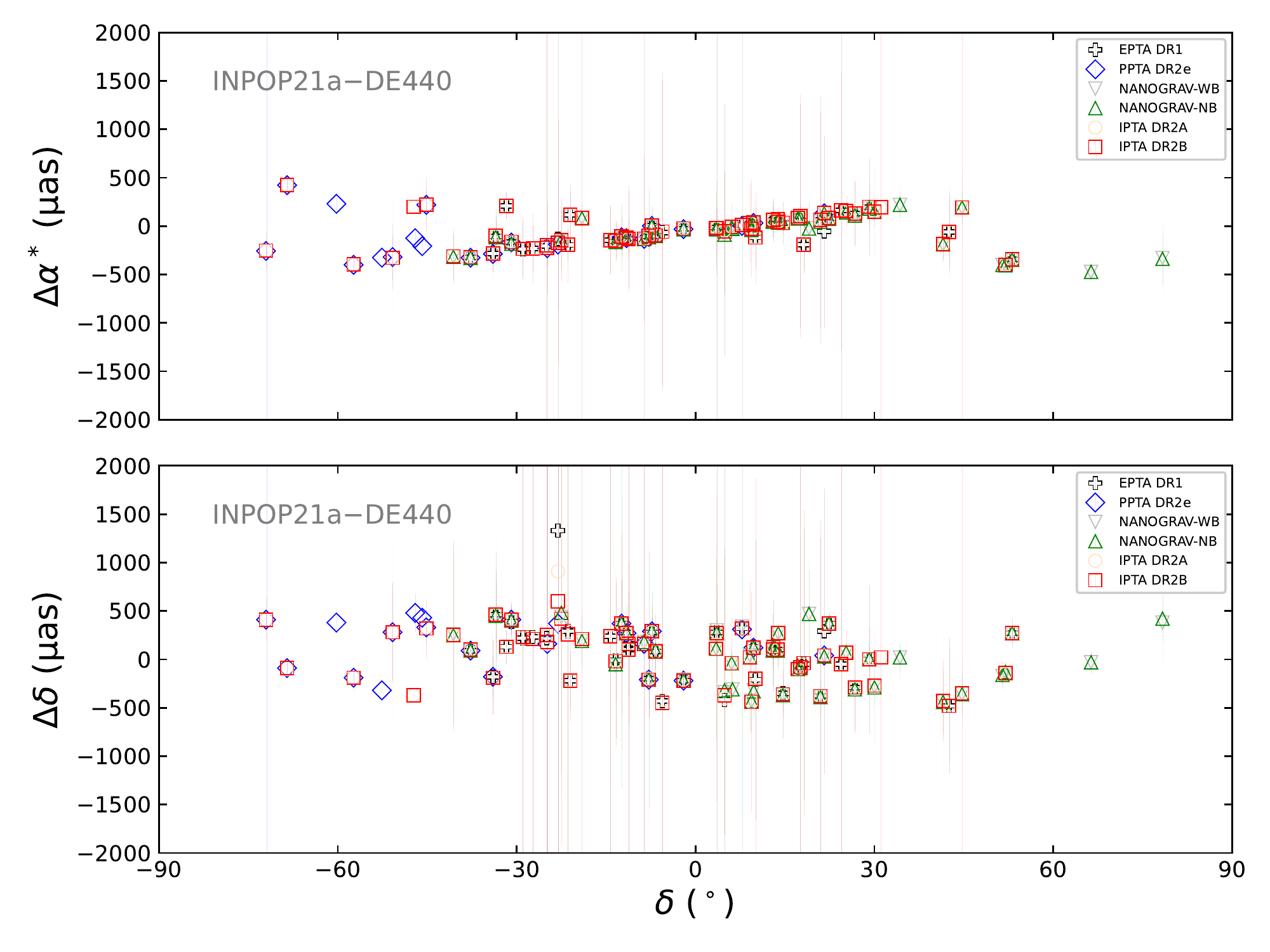}
  \caption[]{\label{fig:equ-pos-inpop21a-vs-de440} %
  Offsets of the pulsar timing positions in the INPOP21a frame with respect to those in the DE440 frame as a function of right ascension (upper) and declination (lower).
}
\end{figure*}

\begin{figure*}[htbp]
  \includegraphics[width=\columnwidth]{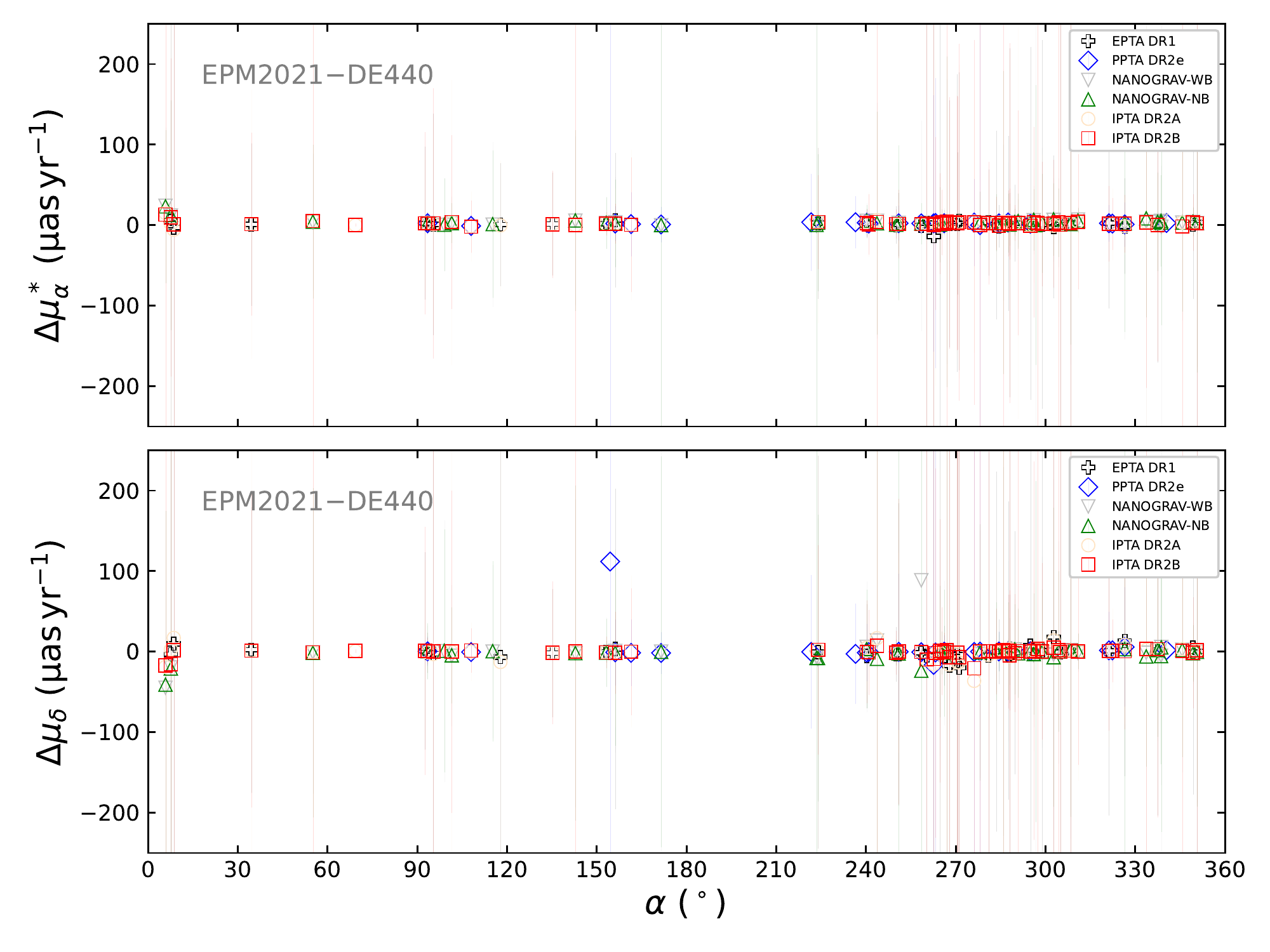}
  \includegraphics[width=\columnwidth]{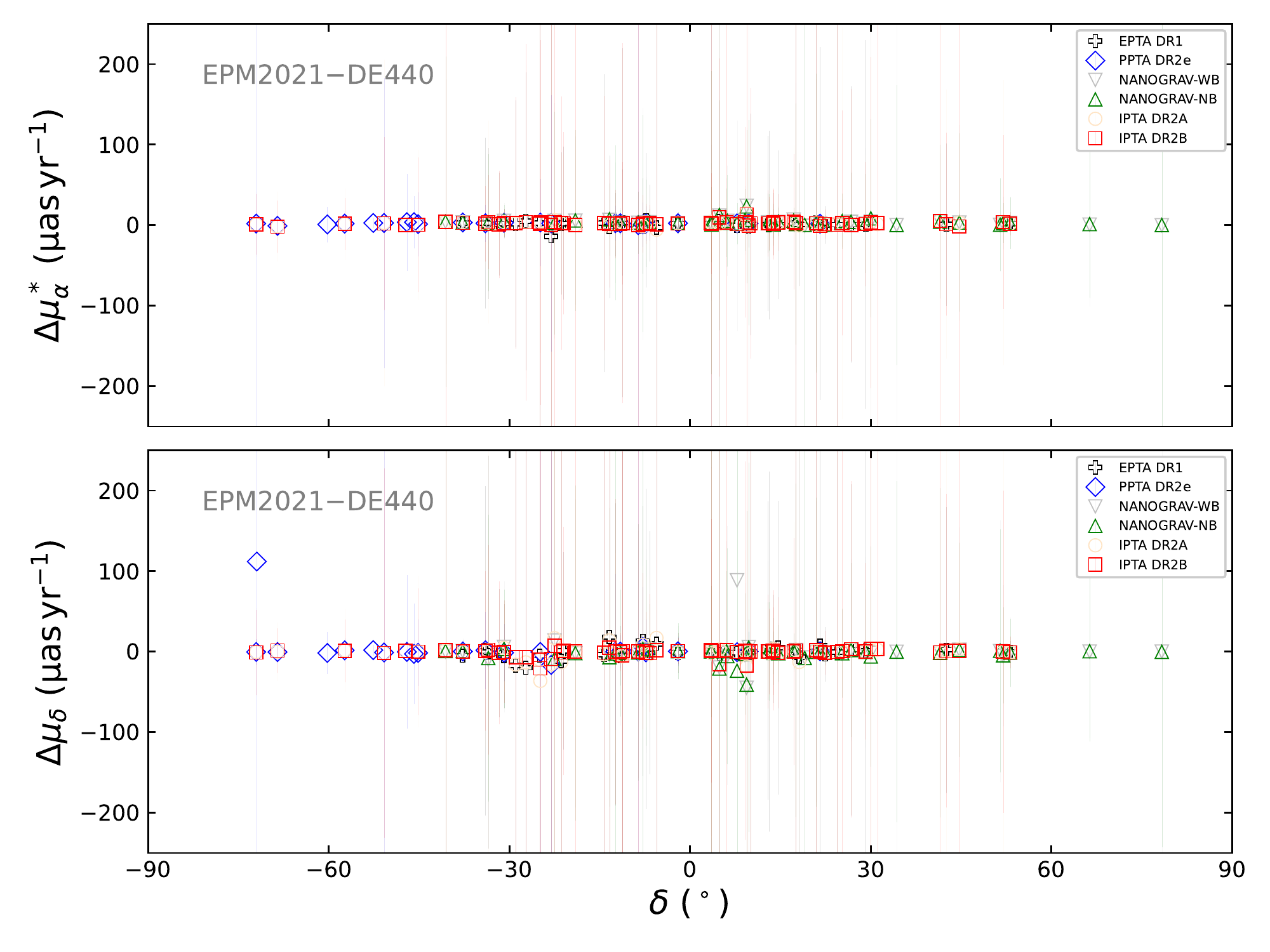}
  \caption[]{\label{fig:equ-pm-epm2021-vs-de440} %
  Offsets of the pulsar timing proper motions in the EPM2021 frame with respect to those in the DE440 frame as a function of right ascension (upper) and declination (lower).
}
\end{figure*}

\begin{figure*}[htbp]
  \includegraphics[width=\columnwidth]{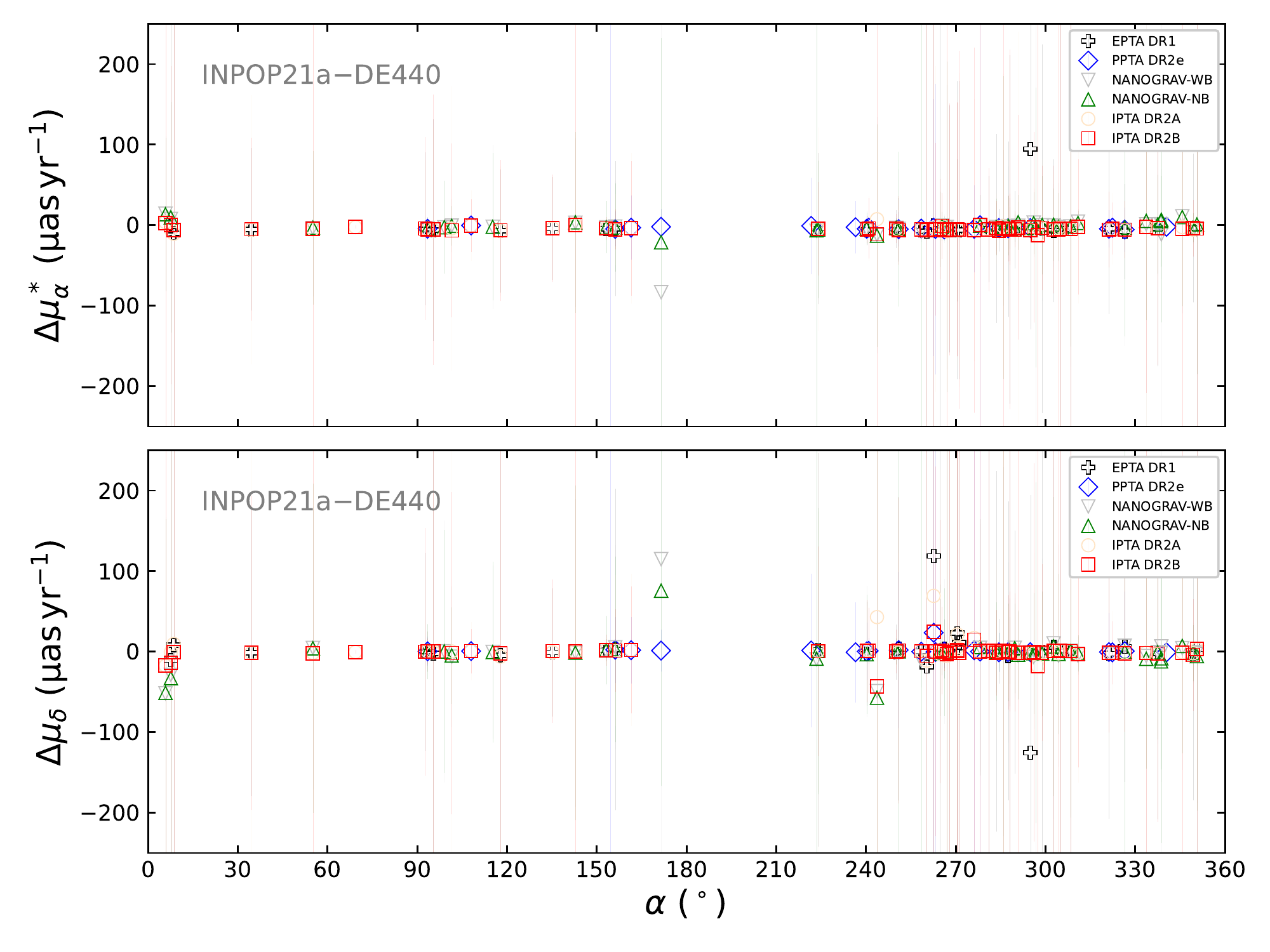}
  \includegraphics[width=\columnwidth]{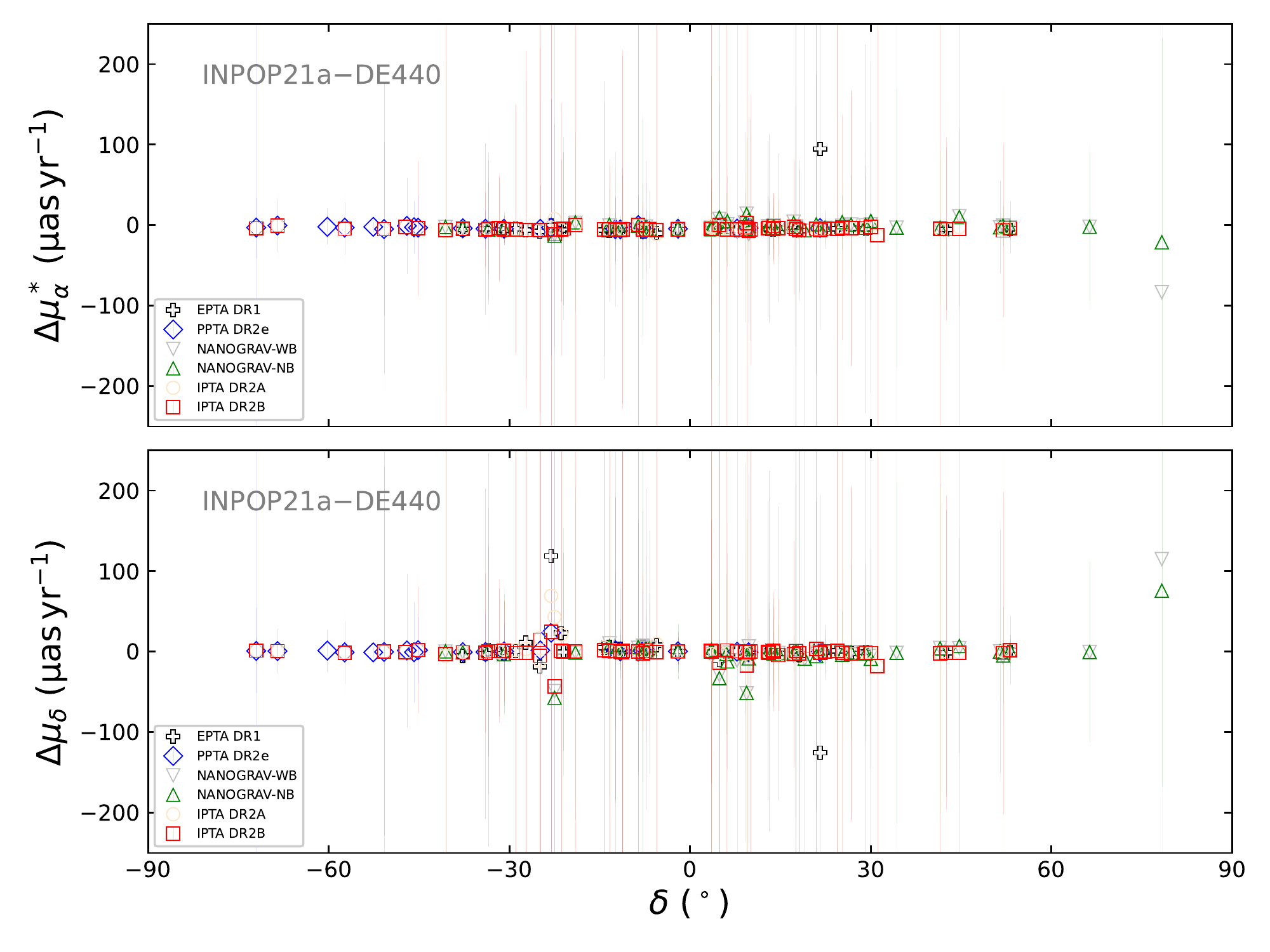}
  \caption[]{\label{fig:equ-pm-inpop21a-vs-de440} %
  Offsets of the pulsar timing proper motions in the INPOP21a frame with respect to those in the DE440 frame as a function of right ascension (upper) and declination (lower).
}
\end{figure*}
    
    Figures~\ref{fig:equ-pos-epm2021-vs-de440}--\ref{fig:equ-pos-inpop21a-vs-de440} depict the offsets of the timing positions derived using EPM2021 and INPOP21a with respect to those using DE440 at the original timing position reference epochs for each pulsar. 
    The comparisons of the pulsar timing positions between other ephemerides and DE440 are shown in Appendix~\ref{sect:appendix-pos-offset}.
    The differences in the right ascension are approximately $\mathrm{100\,\mu as}$ between EPM2021 and DE440 and $\mathrm{200\,\mu as}$ between INPOP21a and DE440, while the respective offsets in the declination are $\mathrm{50\,\mu as}$ and $\mathrm{300\,\mu as}$.
    The weighted root-mean-squares (WRMS) in right ascension and declination are $\mathrm{118\,\mu as}$ and $\mathrm{55\,\mu as}$ for EPM2021 versus DE440 and $\mathrm{232\,\mu as}$ and $\mathrm{305\,\mu as}$ for INPOP21a versus DE440.  
    These positional offsets are not significant over 1-sigma compared to their uncertainties for approximately half of the pulsars.
    When using older planetary ephemerides, the positional offsets become more pronounced and significant; they are almost all confident at 1-sigma for DE200 and DE405.
    In addition, we found obvious patterns in these plots, especially in the plots of $\Delta\alpha^*$ versus $\delta$ and $\Delta\delta$ versus $\alpha$, which implied clear dependencies of the positional offsets on the right ascension and declination.  
    The amplitude of these signatures exceeded the detection sensitivity of the PTA data sets mentioned in Sect.~\ref{subsect:timing-data}, suggesting that the global difference in the position system between planetary ephemerides could be determined with the timing solutions.
    
    Similarly, the offsets of the pulsar timing proper motions using EPM2021 and INPOP21a with respect to those using DE440 are presented in Figs.~\ref{fig:equ-pm-epm2021-vs-de440}--\ref{fig:equ-pm-inpop21a-vs-de440}, while the related plots for the remaining ephemerides can be found in Appendix~\ref{sect:appendix-pm-offset}.  
    For most cases, we found a dominant proper motion bias in the right ascension, while the proper motion in the declination was much less affected.
    Except for a small fraction of pulsars, the differences in the proper motion in the right ascension are approximately $\mathrm{5\,\mu as\,yr^{-1}}$ for INPOP21a and EPM2021; they are only a few $\mathrm{\,\mu as\,yr^{-1}}$ in declination.
    These proper motion differences, however, are at least one order-of-magnitude smaller than their uncertainties and thus not statistically significant.
    We observed similar results between other planetary ephemerides and DE440, except those between DE200 and DE440 (Fig.~\ref{fig:equ-pm-de200-vs-de440}).
    
    We also compared the timing astrometric solutions using the same ephemerides but based on different PTA data sets.
    These timing solutions based on the IPTA DR2B data set were used as the reference solutions.
    The positional agreements (WRMS) are between $\mathrm{30\,\mu as}$ and $\mathrm{60\,\mu as}$ in right ascension and between $\mathrm{60\,\mu as}$ and $\mathrm{180\,\mu as}$ in declination.
    For proper motion measurements, the agreements are $\mathrm{\leq 20\,\mu as\,yr^{-1}}$ for right ascension and between $\mathrm{20\,\mu as\,yr^{-1}}$ and $\mathrm{100\,\mu as\,yr^{-1}}$ for declination.

\subsection{Modeling the differences}

\begin{figure*}[htbp]
  \centering
  \includegraphics[width=17cm]{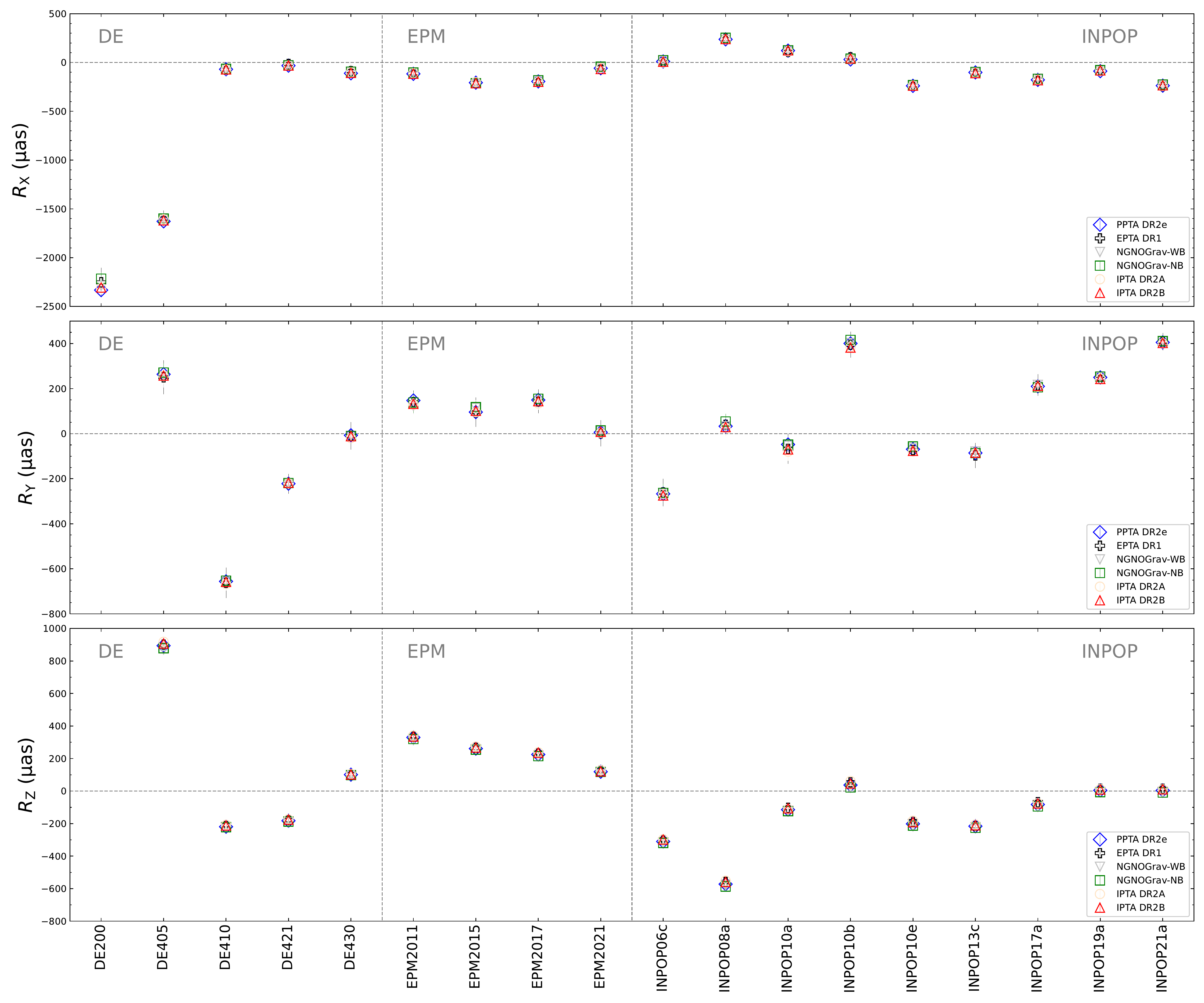}
  \caption[]{\label{fig:pos-vsh01-rot} %
    Rotation parameters around the $X$-, $Y$-, and $Z$-axes in the position system of the planetary ephemeris frames with respect to DE440 in the equatorial coordinate system (from top to bottom).
   $R_{\rm Z}$ between DE200 and DE440 is approximately $-12$\,mas, which is beyond the range of the vertical axis.
}
\end{figure*}

\begin{figure*}[htbp]
  \centering
  \includegraphics[width=17cm]{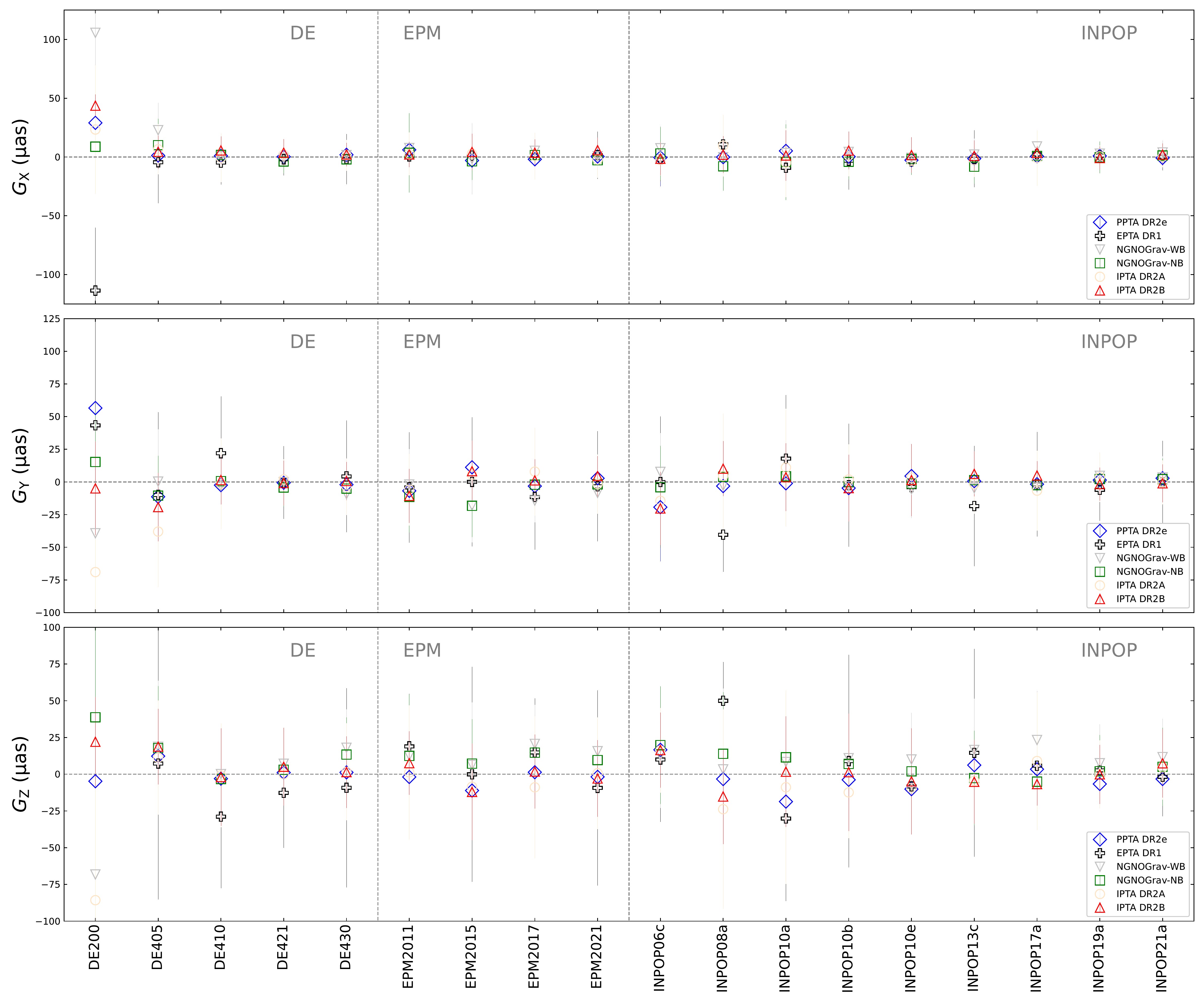}
  \caption[]{\label{fig:pos-vsh01-gli} %
    Glide parameters around the $X$-, $Y$-, and $Z$-axes in the position system of the planetary ephemeris frames with respect to DE440 in the equatorial coordinate system (from top to bottom).
}
\end{figure*}
    
\begin{figure*}[htbp]
  \centering
  \includegraphics[width=17cm]{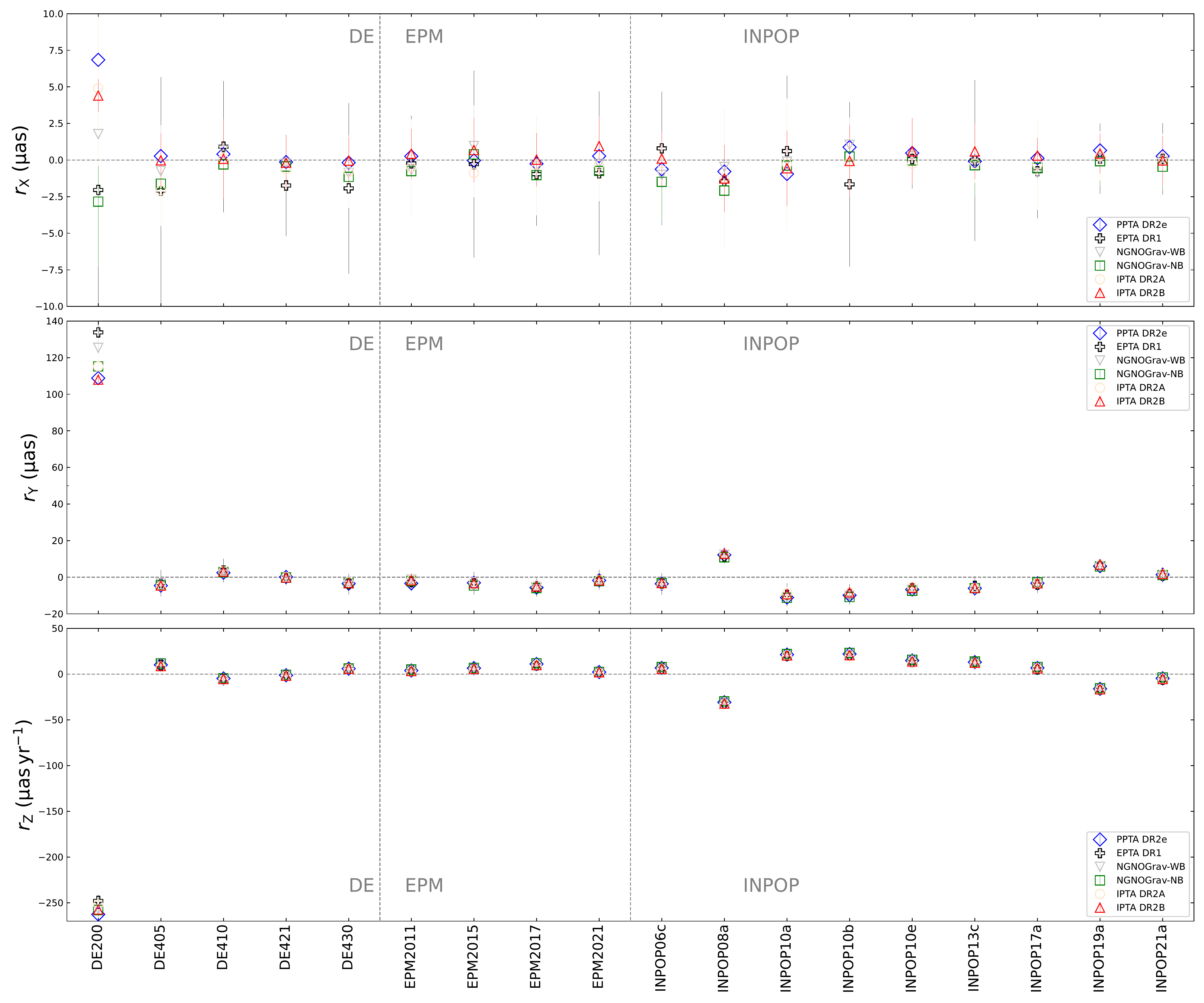}
  \caption[]{\label{fig:pm-vsh01-rot} %
    Rotation parameters around the $X$-, $Y$-, and $Z$-axes in the proper motion system of the planetary ephemeris frames with respect to DE440 in the equatorial coordinate system (from top to bottom).
}
\end{figure*}

\begin{figure*}[htbp]
  \centering
  \includegraphics[width=17cm]{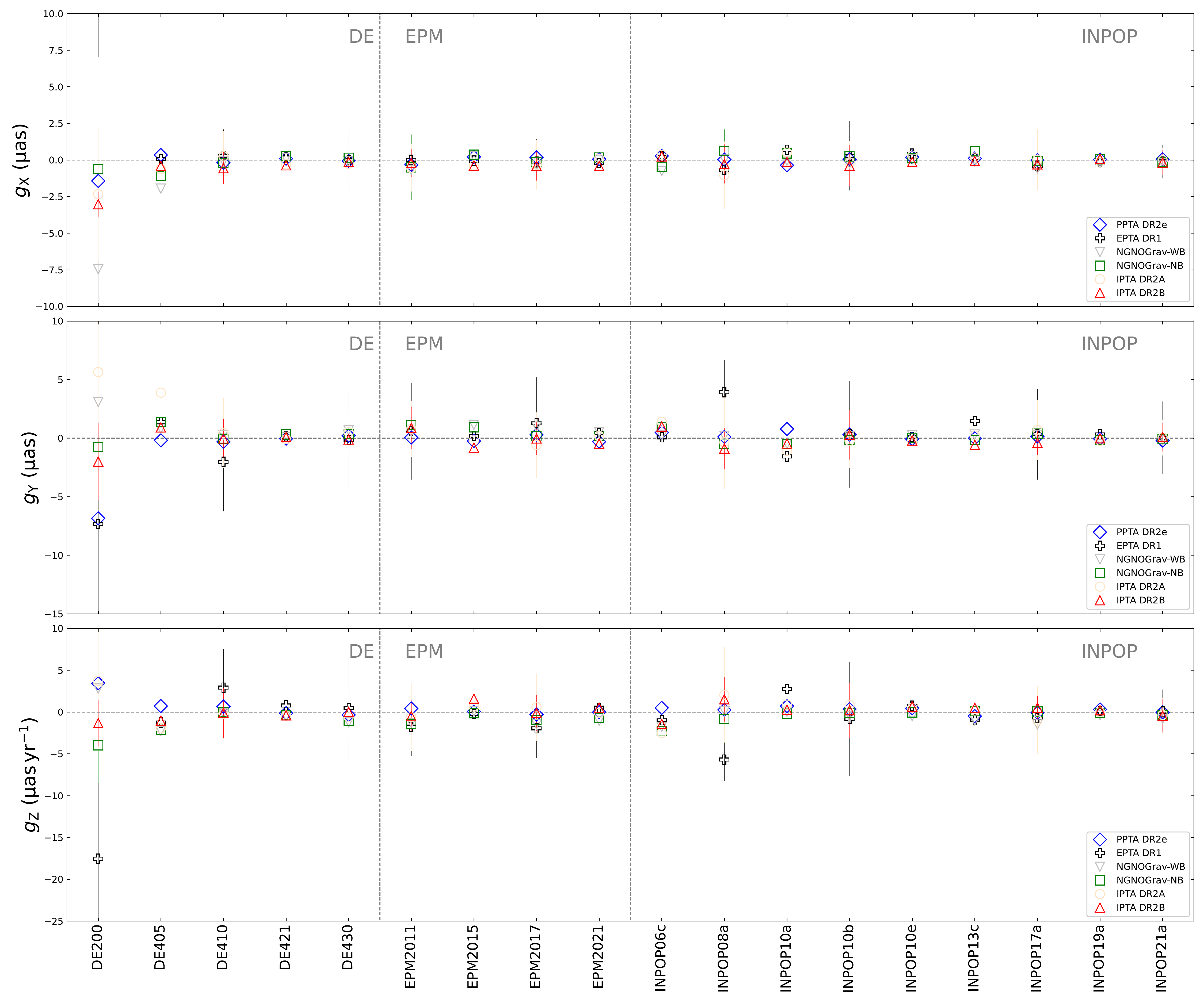}
  \caption[]{\label{fig:pm-vsh01-gli} %
    Glide parameters around the $X$-, $Y$-, and $Z$-axes  in the proper motion system of the planetary ephemeris frames with respect to DE440 in the equatorial coordinate system (from top to bottom).  
}
\end{figure*}

\begin{figure*}[htbp]
  \includegraphics[width=\columnwidth]{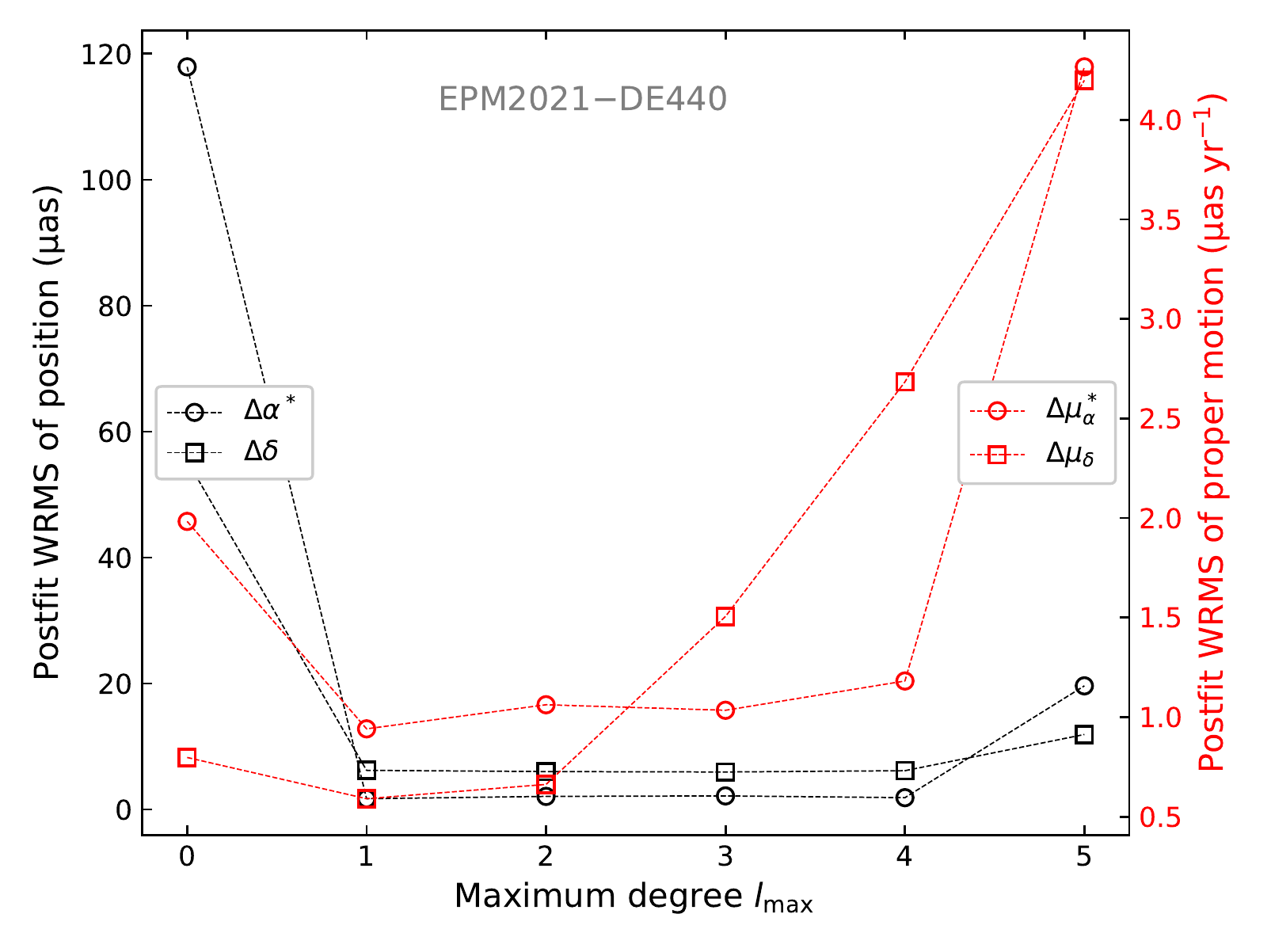}
  \includegraphics[width=\columnwidth]{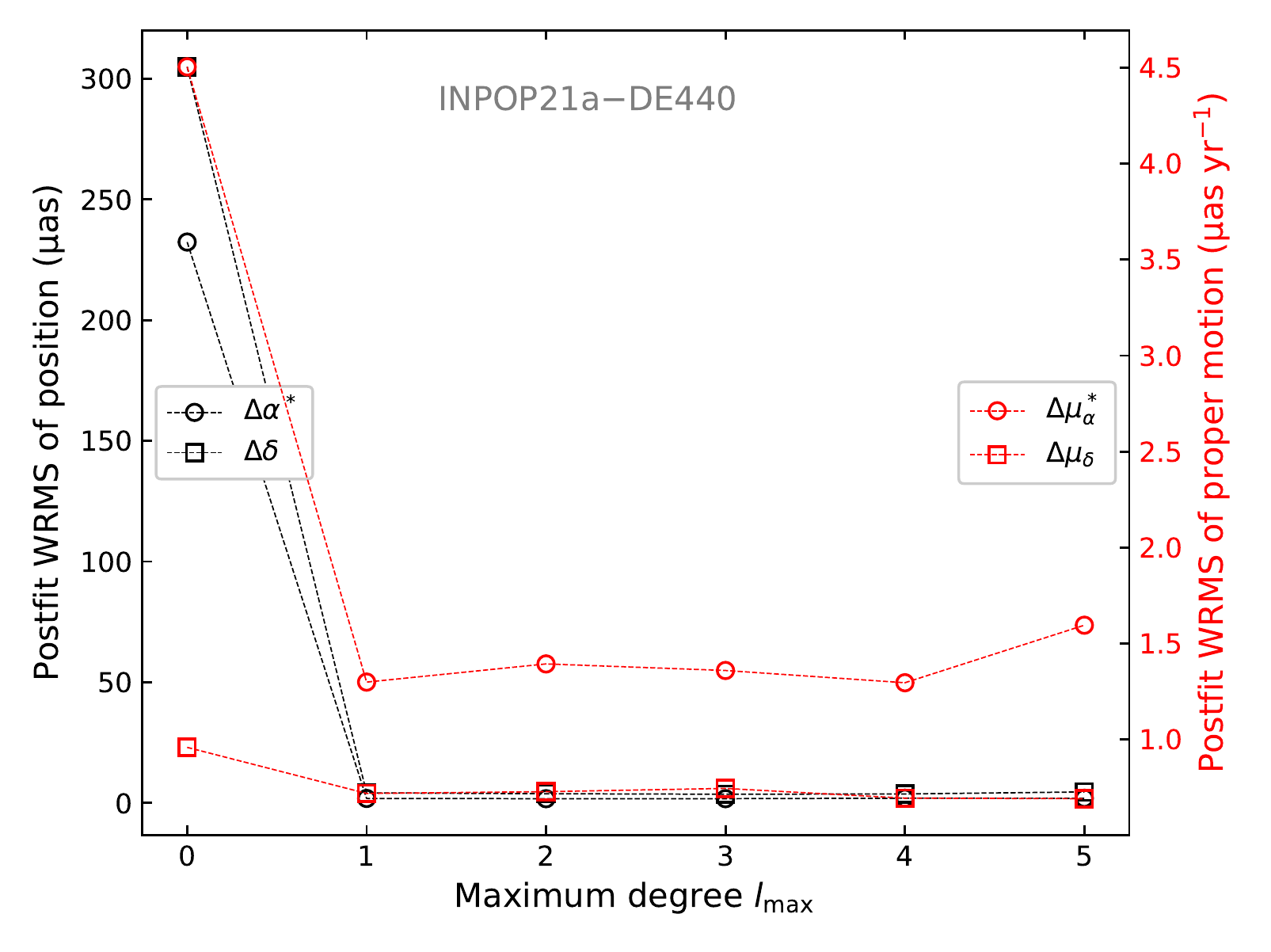}
  \caption[]{\label{fig:postfit-wrms-vs-lmax} %
  WRMS of the postfit residuals as a function of the maximum degree of the VSH model.
  Left: EPM2021 versus DE440; Right: INPOP21a versus DE440.
}
\end{figure*}

%
\begin{table*}
    \caption{\label{table:pos-vsh}   
        Rotation and glide of the position system of ephemerides with respect to that of DE440 in the equatorial coordinate system.} 
    \centering          
    \begin{tabular}{c r r r r r r}     
    \hline   
    Ephemeris &\multicolumn{1}{c}{$R_{\rm X}$} &\multicolumn{1}{c}{$R_{\rm Y}$}  &\multicolumn{1}{c}{$R_{\rm Z}$}  &\multicolumn{1}{c}{$G_{\rm X}$} &\multicolumn{1}{c}{$G_{\rm Y}$}  &\multicolumn{1}{c}{$G_{\rm Z}$}  \\ 
    &\multicolumn{1}{c}{($\mathrm{\mu as}$)}  &\multicolumn{1}{c}{($\mathrm{\mu as}$)}  &\multicolumn{1}{c}{($\mathrm{\mu as}$)}  &\multicolumn{1}{c}{($\mathrm{\mu as}$)}  &\multicolumn{1}{c}{($\mathrm{\mu as}$)}  &\multicolumn{1}{c}{($\mathrm{\mu as}$)}   \\
    \hline                    
    DE200     &$-2291$(22)  &$-12432$(19)  &$-12428$(13)  &$8$(14)  &$-1$(17)  &$-17$(25)  \\
    DE403     &$-1838$(13)  &$597$(13)  &$-2309$(8)  &$10$(9)  &$-17$(12)  &$-6$(16)  \\
    DE405     &$-1623$(13)  &$248$(12)  &$891$(9)  &$6$(9)  &$-8$(11)  &$-5$(16)  \\
    DE410     &$-66$(9)  &$-665$(8)  &$-219$(5)  &$2$(5)  &$-1$(7)  &$-1$(11)  \\
    DE421     &$-33$(7)  &$-227$(7)  &$-184$(4)  &$-2$(3)  &$-3$(5)  &$-0$(8)  \\
    DE430     &$-106$(7)  &$-14$(7)  &$101$(3)  &$1$(3)  &$-1$(6)  &$2$(9)  \\
    \hline
    EPM2011   &$-113$(8)  &$132$(8)  &$327$(4)  &$-1$(4)  &$-6$(7)  &$2$(9)  \\
    EPM2015   &$-204$(8)  &$97$(7)  &$258$(5)  &$-2$(5)  &$-2$(7)  &$-1$(10)  \\
    EPM2017   &$-192$(8)  &$140$(8)  &$223$(4)  &$-1$(4)  &$-4$(7)  &$0$(10)  \\
    EPM2021   &$-58$(6)  &$8$(6)  &$120$(3)  &$1$(3)  &$-1$(5)  &$2$(7)  \\
    \hline
    INPOP06c  &$3$(11)  &$-275$(10)  &$-308$(7)  &$1$(8)  &$-3$(10)  &$-5$(13)  \\
    INPOP08a  &$244$(9)  &$32$(9)  &$-572$(5)  &$2$(4)  &$-6$(7)  &$0$(10)  \\
    INPOP10a  &$116$(10)  &$-66$(8)  &$-116$(5)  &$-3$(6)  &$6$(9)  &$-9$(12)  \\
    INPOP10b  &$38$(12)  &$388$(10)  &$29$(7)  &$2$(7)  &$-5$(10)  &$1$(16)  \\
    INPOP10e  &$-229$(8)  &$-74$(7)  &$-202$(5)  &$0$(5)  &$-5$(6)  &$-1$(10)  \\
    INPOP13c  &$-107$(7)  &$-86$(6)  &$-218$(4)  &$-3$(4)  &$1$(6)  &$0$(8)  \\
    INPOP17a  &$-176$(9)  &$208$(10)  &$-81$(4)  &$1$(4)  &$-4$(7)  &$1$(10)  \\
    INPOP19a  &$-83$(6)  &$246$(6)  &$4$(3)  &$-2$(3)  &$-2$(5)  &$-1$(7)  \\
    INPOP21a  &$-237$(4)  &$405$(5)  &$3$(3)  &$-3$(3)  &$-2$(5)  &$2$(6)  \\
    \hline                 
    \end{tabular}
    \tablefoot{
    The uncertainties for the estimates given in parentheses were obtained from the least-squares fit and could be underestimated several times, as suggested by the bootstrap fit.}
\end{table*}
%

%
\begin{table*}[htbp]
\caption{\label{table:pm-vsh}   
Rotation and glide of the proper motion system of ephemerides with respect to that of DE440 in the equatorial coordinate system.}             
\centering          
\begin{tabular}{c r r r r r r }     
\hline\hline       
Ephemeris &\multicolumn{1}{c}{$r_{\rm X}$} &\multicolumn{1}{c}{$r_{\rm Y}$}  &\multicolumn{1}{c}{$r_{\rm Z}$}  &\multicolumn{1}{c}{$g_{\rm X}$} &\multicolumn{1}{c}{$g_{\rm Y}$}  &\multicolumn{1}{c}{$g_{\rm Z}$}  \\ 
    &\multicolumn{1}{c}{($\mathrm{\mu as\,yr^{-1}}$)}   &\multicolumn{1}{c}{($\mathrm{\mu as\,yr^{-1}}$)}   &\multicolumn{1}{c}{($\mathrm{\mu as\,yr^{-1}}$)}  &\multicolumn{1}{c}{($\mathrm{\mu as\,yr^{-1}}$)}   &\multicolumn{1}{c}{($\mathrm{\mu as\,yr^{-1}}$)}   &\multicolumn{1}{c}{($\mathrm{\mu as\,yr^{-1}}$)}  \\
    \hline                    
    DE200     &$2.4$(1.6)  &$114.6$(1.5)  &$-259.0$(1.0)  &$-0.7$(1.0)  &$0.4$(1.3)  &$-0.3$(1.9)  \\
    DE403     &$2.1$(1.0)  &$-4.8$(1.0)  &$20.1$(0.6)  &$-1.0$(0.7)  &$1.9$(0.9)  &$-0.7$(1.2)  \\
    DE405     &$-0.7$(1.0)  &$-3.2$(0.9)  &$10.8$(0.7)  &$-0.8$(0.7)  &$1.1$(0.8)  &$-1.1$(1.2)  \\
    DE410     &$-0.2$(0.7)  &$3.5$(0.6)  &$-4.7$(0.4)  &$-0.2$(0.4)  &$0.1$(0.5)  &$-0.1$(0.8)  \\
    DE421     &$-0.2$(0.5)  &$0.3$(0.5)  &$-1.0$(0.3)  &$0.2$(0.3)  &$0.3$(0.4)  &$0.0$(0.6)  \\
    DE430     &$-0.4$(0.6)  &$-3.2$(0.5)  &$6.1$(0.3)  &$-0.1$(0.3)  &$0.1$(0.4)  &$-0.3$(0.6)  \\
    \hline 
    EPM2011   &$-0.1$(0.6)  &$-1.9$(0.6)  &$4.3$(0.3)  &$-0.0$(0.3)  &$0.7$(0.5)  &$-0.6$(0.7)  \\
    EPM2015   &$0.0$(0.6)  &$-3.3$(0.6)  &$6.6$(0.4)  &$0.2$(0.4)  &$0.1$(0.6)  &$0.3$(0.8)  \\
    EPM2017   &$-0.5$(0.6)  &$-5.1$(0.6)  &$11.2$(0.3)  &$0.1$(0.3)  &$0.4$(0.5)  &$-0.1$(0.7)  \\
    EPM2021   &$0.0$(0.5)  &$-1.9$(0.5)  &$2.3$(0.2)  &$0.0$(0.3)  &$0.1$(0.4)  &$-0.2$(0.6)  \\
    \hline 
    INPOP06c  &$-0.5$(0.8)  &$-2.7$(0.8)  &$7.0$(0.5)  &$-0.3$(0.6)  &$0.7$(0.7)  &$-0.9$(1.0)  \\
    INPOP08a  &$-1.6$(0.7)  &$12.3$(0.6)  &$-30.9$(0.4)  &$-0.1$(0.3)  &$0.5$(0.5)  &$-0.2$(0.8)  \\
    INPOP10a  &$0.1$(0.7)  &$-10.0$(0.6)  &$21.5$(0.4)  &$0.3$(0.4)  &$-0.5$(0.6)  &$1.0$(0.9)  \\
    INPOP10b  &$0.3$(0.9)  &$-9.0$(0.7)  &$22.5$(0.5)  &$-0.1$(0.5)  &$0.4$(0.7)  &$0.1$(1.1)  \\
    INPOP10e  &$-0.1$(0.6)  &$-6.1$(0.5)  &$15.0$(0.4)  &$-0.0$(0.4)  &$0.4$(0.5)  &$0.2$(0.7)  \\
    INPOP13c  &$0.4$(0.5)  &$-5.9$(0.5)  &$13.3$(0.3)  &$0.2$(0.3)  &$-0.0$(0.4)  &$0.1$(0.6)  \\
    INPOP17a  &$-0.1$(0.7)  &$-3.0$(0.7)  &$6.8$(0.3)  &$-0.0$(0.3)  &$0.3$(0.6)  &$-0.2$(0.8)  \\
    INPOP19a  &$0.1$(0.5)  &$6.5$(0.5)  &$-16.0$(0.3)  &$0.2$(0.3)  &$0.1$(0.4)  &$0.1$(0.5)  \\
    INPOP21a  &$-0.0$(0.3)  &$1.5$(0.4)  &$-4.6$(0.2)  &$0.2$(0.2)  &$0.2$(0.4)  &$-0.2$(0.5)  \\
    \hline                  
\end{tabular}
\tablefoot{The uncertainties for the estimate given in parentheses were obtained from the least-squares fit and could be underestimated several times, as suggested by the bootstrap fit.}
\end{table*}
%


    We modeled the pulsar timing position and proper motion offsets seen in the previous section following methods introduced in Sect.~\ref{subsect:model}.
    The least-squares fits converged after two iterations or less.
    The WRMS of postfit residuals was reduced to a few $\mathrm{\mu as}$ for the position and a few $\mathrm{\mu as\,yr^{-1}}$ for the proper motion.
    This suggested that the variations in the timing position and proper motion were well accounted for by our models.
    
    However, we found a few large deviations from the bulk of the sample for a few pulsars (e.g., PSR J1730--2304 and PSR J1949+3106), for which the precision of the timing astrometric solutions was relatively low compared to the rest of the pulsars (approximately 10~mas for position and $\mathrm{5\,mas\,yr^{-1}}$ for proper motions).
    This implied that the systematics in the planetary ephemeris reference frames could be evinced in the pulsar timing data set only when the timing astrometric precision is sufficiently high.
    
    The estimates of the VSH parameters based on various PTA data sets are depicted in Figs.~\ref{fig:pos-vsh01-rot}--\ref{fig:pm-vsh01-gli}, which show good agreement among different PTAs.
    We observed a general trend of declining to zero for the successive DE ephemerides, supporting a better global agreement between newer editions of the DE ephemerides. 
    Similar results can be found for the successive EPM and INPOP ephemerides if one takes the results of the latest versions of each series (i.e., EPM2021 and INPOP21a) as the reference.
    The positional glide parameters (components in $\boldsymbol{G}$) were estimated to be $\mathrm{10\,\mu as}$ to $\mathrm{20\,\mu as}$ for most cases, obviously smaller than the corresponding positional rotation parameters (components in $\boldsymbol{R}$).
    This comment is also valid for the VSH parameters of the proper motion system.
    We also noticed that the $X$-components of the rotation vector $\boldsymbol{r}$ ($\sim \mathrm{1\,\mu as\,yr^{-1}}$) are much less pronounced than the $Y$- and $Z$-components.
    
    Table~\ref{table:pos-vsh} presents estimates of the rotation and glide parameters of the position system among the planetary ephemerides based on the combination of the timing solutions from all PTAs.
    Except for earlier DE planetary ephemerides (DE200 and DE405), the orientation offsets were generally less than $\mathrm{500\,\mu as}$, and almost no relative dipolar deformation was effectively detected.

    We tabulated the determinations of the VSH parameters among the planetary ephemeris proper motion systems in Table~\ref{table:pm-vsh}. 
    The spin around the $X$-axis is much smaller than those around the $Y$- and $Z$-axes.
    We converted the spin vectors from the equatorial coordinate system to the ecliptic coordinate system and found that the spins were mainly in the direction of the latitude, suggesting that the spin was likely due to the differences in the mean orbital motion of Earth.
    The glide parameters were less than $\mathrm{1\,\mu as\,yr^{-1}}$ and were statistically insignificant for most cases.

    The VSH fittings were also applied to the subsets of pulsars at high and low ecliptic latitudes.
    The results of pulsars near the ecliptic plane showed some discrepancies compared to the results based on subsets of all pulsars and only pulsars far from the ecliptic plane, which usually occurred for parameters on the Z-axis.
    The most pronounced differences occurred for the comparison between DE200 and DE440, for which the sample of pulsars at low ecliptic latitudes yielded a solution of $G_{\rm Z}\!=\!-634\,\pm\,375\,{\rm \mu as}$ and $g_{\rm Z}\!=\!36\,\pm\,32\,{\rm \mu as\,yr^{-1}}$.
    These discrepancies, however, were not statistically significant with respect to the uncertainties of the estimates for most cases.
    
    The bootstrap fit gave consistent results with those from the least-squares fit (i.e., reported in Tables~\ref{table:pos-vsh} and \ref{table:pm-vsh}), while the formal uncertainty was usually several times larger than the formal uncertainty of the latter.
    To cross-check the results, we processed the NANOGrav12.5 data set using \texttt{PINT}\footnote{\url{https://pypi.org/project/pint-pulsar/}} \citepads[version 0.9.1;][]{2021ApJ...911...45L} developed by the NANOGrav team under the DE planetary ephemeris frames, from which the results were consistent with those given by \texttt{TEMPO2}.
    
    It was also possible to extend our VSH model to high degrees, for example, to degree two to include the quadrupolar terms.
    However, we noticed that the postfit WRMS in the timing positions and proper motions were already small.
    Including higher orders of VSHs might not improve the fit much.
    The quadrupolar signals, if they exist, should not exceed $5\,\mathrm{\mu as}$ in the positions and $1\,\mathrm{\mu as\,yr^{-1}}$ in the proper motions. 
    Comparisons among timing solutions using DE440, EPM2021, and INPOP21a were used to test this speculation.
    The VSH model was extended to different maximum degrees $l_{\rm max}$ until $l_{\rm max}\!=\!5$, to which the differences between timing solutions were fitted.
    We found that the estimates of the rotation and glide parameters varied slightly when $l_{\rm max}\!\leq\!4$.
    Figure~\ref{fig:postfit-wrms-vs-lmax} shows the evolution of the WRMS of postfit residuals as $l_{\rm max}$ increases, which changes little compared to that of $l_{\rm max}\!=\!1$.
    The results of this experiment suggest that the VSH model of the first degree is sufficient to account for the global differences in position and proper motions between planetary ephemerides. 


\subsection{Comparison between timing and VLBI solutions}   \label{subsect:timing-vs-vlbi}

\begin{figure*}[htbp]
  \includegraphics[width=\columnwidth]{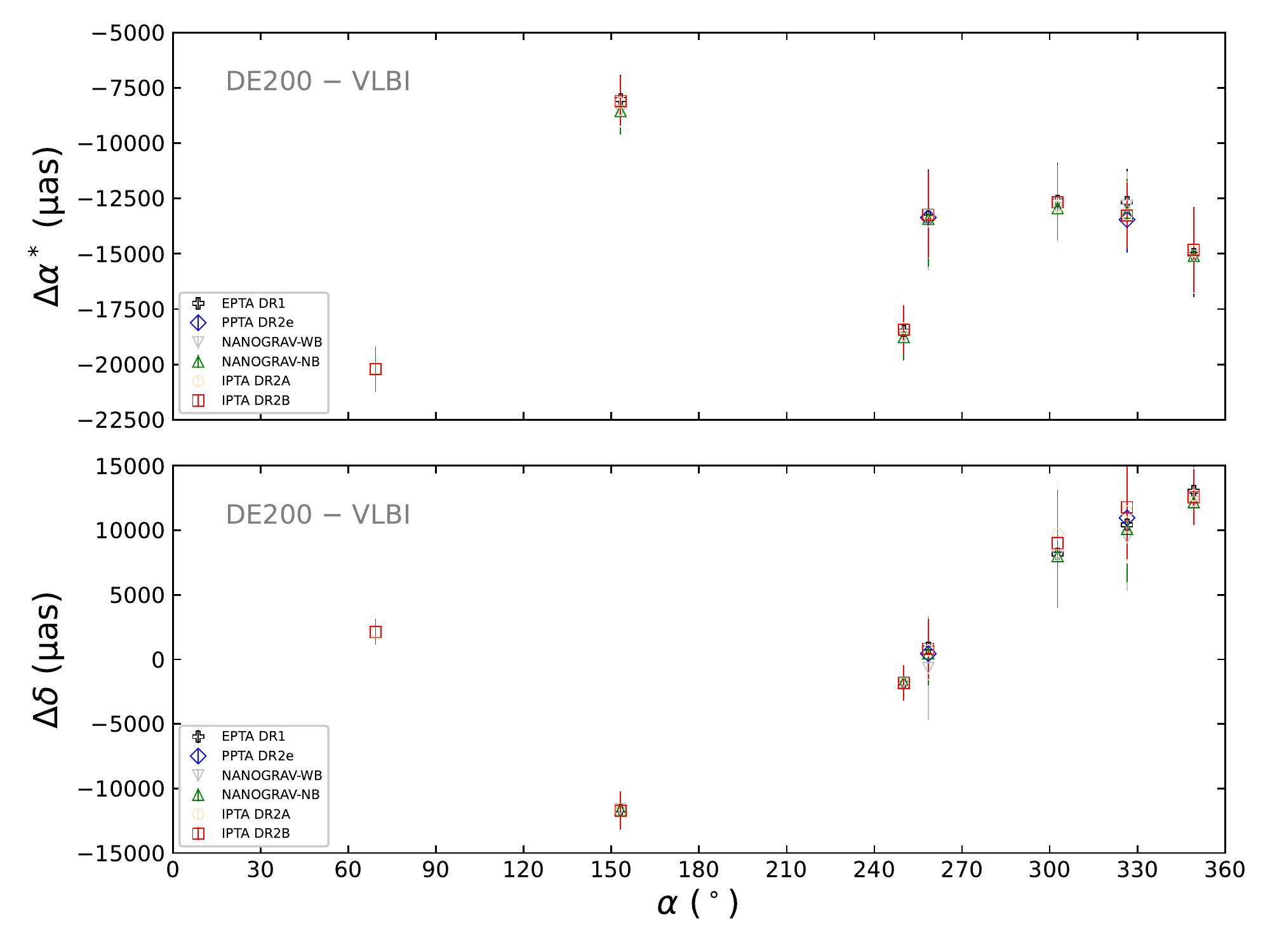}
  \includegraphics[width=\columnwidth]{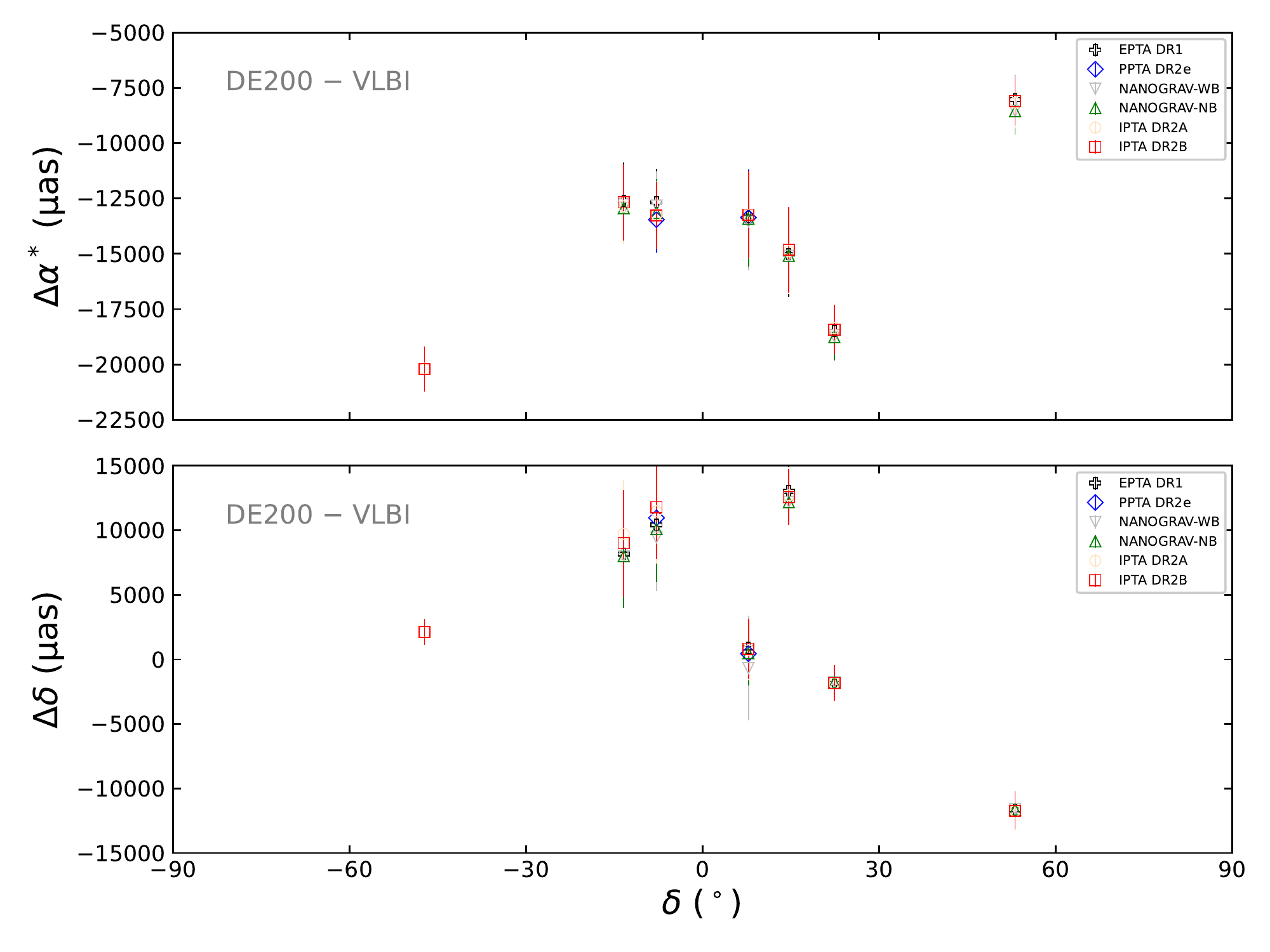}
  \caption[]{\label{fig:de200-vs-vlbi} %
   Offsets of the pulsar timing positions in the DE200 frame with respect to those in the VLBI frame as a function of right ascension (left) and declination (right).
}
\end{figure*}

\begin{figure*}[htbp]
  \includegraphics[width=\columnwidth]{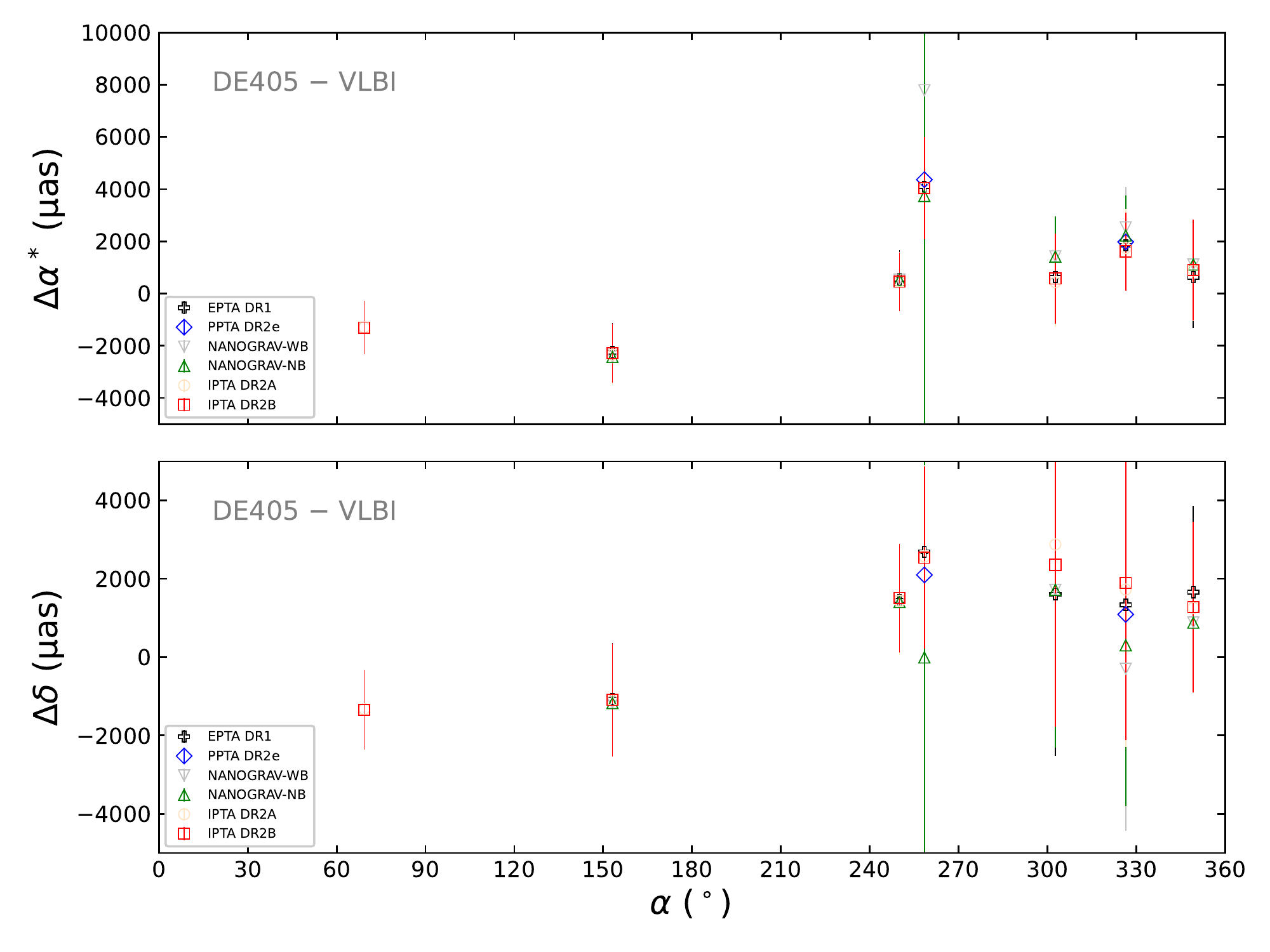}
  \includegraphics[width=\columnwidth]{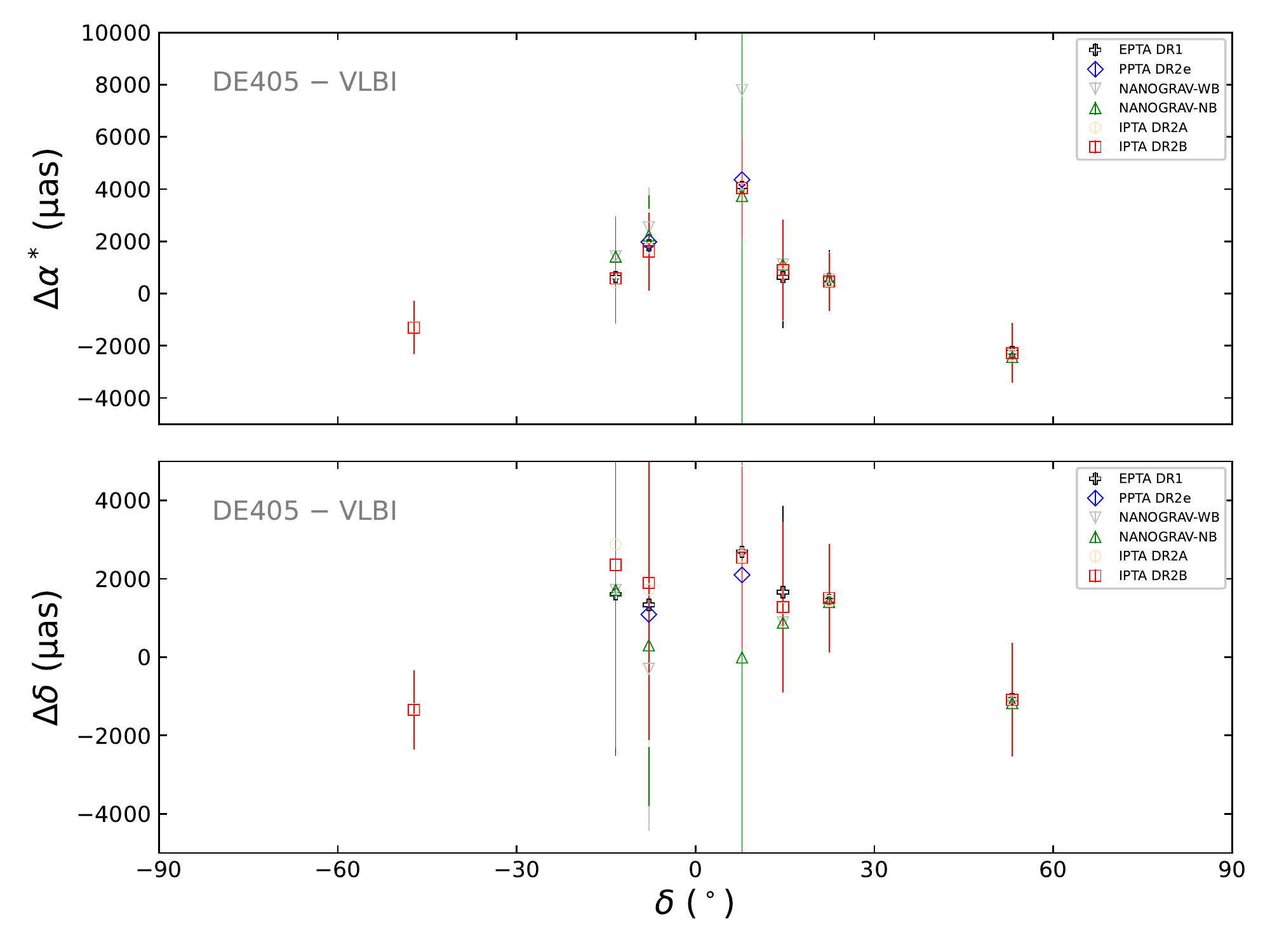}
  \caption[]{\label{fig:de405-vs-vlbi} %
   Offsets of the pulsar timing positions in the DE405 frame with respect to those in the VLBI frame as a function of right ascension (left) and declination (right).
}
\end{figure*}

\begin{figure*}[htbp]
  \includegraphics[width=\columnwidth]{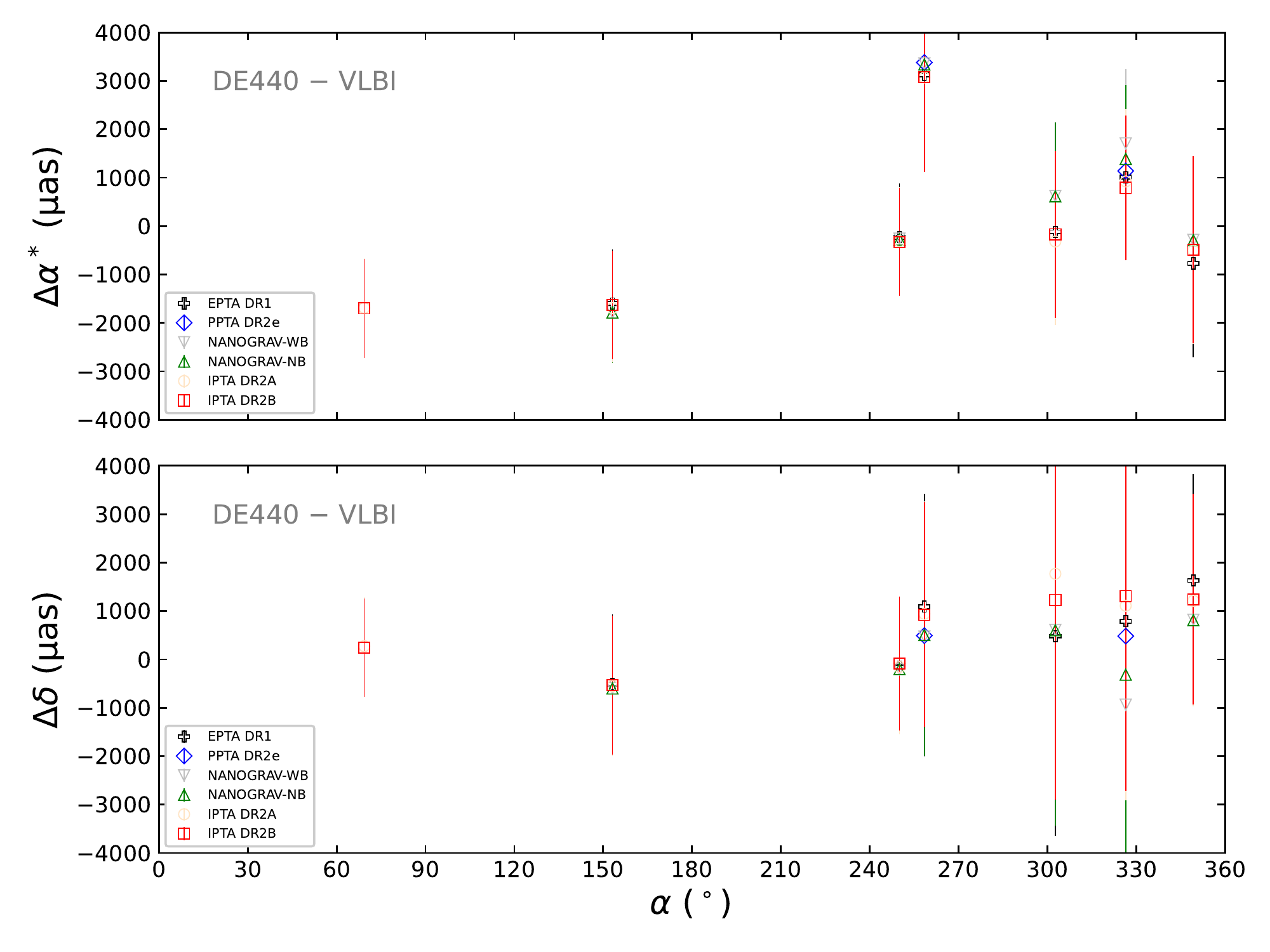}
  \includegraphics[width=\columnwidth]{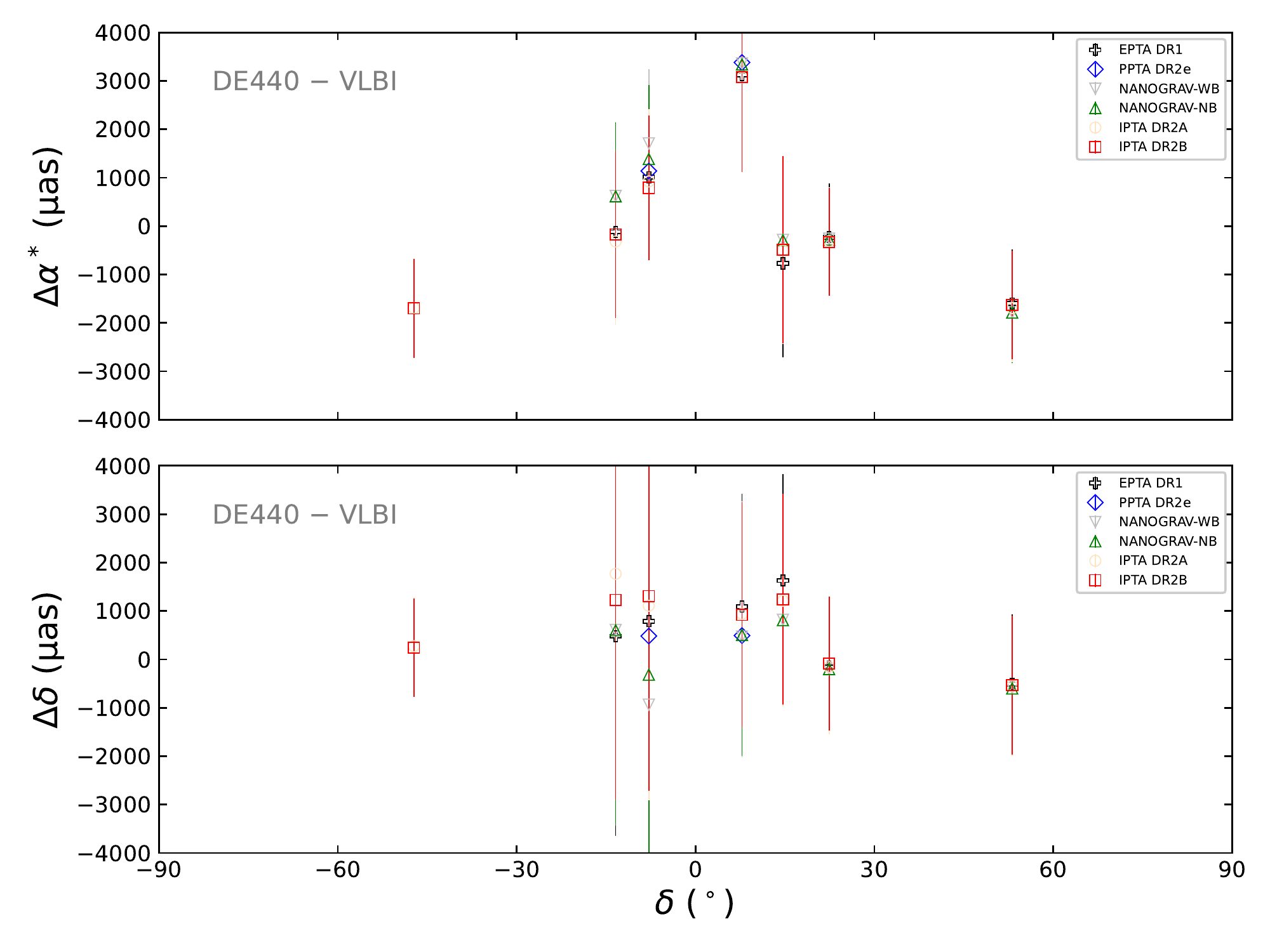}
  \caption[]{\label{fig:de440-vs-vlbi} %
   Offsets of the pulsar timing positions in the DE440 frame with respect to those in the VLBI frame as a function of right ascension (left) and declination (right).
}
\end{figure*}

%
\begin{table}[htbp]
    \caption{\label{table:pos-vsh-vlbi}  
        Orientation offsets of the planetary ephemeris frames with respect to the VLBI frame.}              
    \begin{tabular}{c r r r }     
    \hline
    Ephemeris &\multicolumn{1}{c}{$R_{\rm X}$} &\multicolumn{1}{c}{$R_{\rm Y}$}  &\multicolumn{1}{c}{$R_{\rm Z}$}    \\ 
    &\multicolumn{1}{c}{($\mathrm{\mu as}$)}  &\multicolumn{1}{c}{($\mathrm{\mu as}$)}  &\multicolumn{1}{c}{($\mathrm{\mu as}$)}     \\
    \hline                    
    DE200     &$ -3141(   221)$  &$-11806(   216)$  &$-14803(   189)$  \\
    DE405     &$ -2478(   219)$  &$  -372(   210)$  &$  1200(   187)$  \\
    DE410     &$  -929(   212)$  &$ -1233(   204)$  &$  -152(   183)$  \\
    DE421     &$  -904(   218)$  &$  -820(   211)$  &$     4(   187)$  \\
    DE430     &$ -1011(   220)$  &$  -637(   216)$  &$   431(   189)$  \\
    DE440     &$  -896(   221)$  &$  -587(   217)$  &$   292(   189)$  \\
    \hline 
    EPM2011   &$ -1013(   221)$  &$  -472(   217)$  &$   668(   190)$  \\
    EPM2015   &$ -1106(   221)$  &$  -523(   217)$  &$   619(   189)$  \\
    EPM2017   &$ -1070(   218)$  &$  -505(   212)$  &$   529(   188)$  \\
    EPM2021   &$  -971(   221)$  &$  -600(   216)$  &$   412(   189)$  \\
    \hline 
    INPOP06c  &$  -893(   222)$  &$  -882(   218)$  &$    61(   190)$  \\
    INPOP08a  &$  -655(   217)$  &$  -429(   211)$  &$  -683(   187)$  \\
    INPOP10a  &$  -769(   217)$  &$  -763(   211)$  &$   311(   187)$  \\
    INPOP10b  &$  -900(   221)$  &$  -286(   216)$  &$   534(   189)$  \\
    INPOP10e  &$ -1124(   223)$  &$  -726(   217)$  &$   242(   190)$  \\
    INPOP13c  &$ -1002(   221)$  &$  -732(   217)$  &$   210(   190)$  \\
    INPOP17a  &$ -1076(   221)$  &$  -413(   217)$  &$   281(   190)$  \\
    INPOP19a  &$  -978(   221)$  &$  -276(   217)$  &$   130(   189)$  \\
    INPOP21a  &$ -1170(   221)$  &$  -164(   216)$  &$   225(   189)$  \\
    \hline                 
    \end{tabular}
    \tablefoot{The uncertainties of these estimates given in parentheses were obtained from the least-square fitting and could be underestimated several times.
    }
\end{table}
%

    The positional offsets between the timing and VLBI solutions are presented in Figs.~\ref{fig:de200-vs-vlbi}--\ref{fig:de440-vs-vlbi}.
    Only the results of timing solutions using DE200, DE405, and DE440 are shown here; the results of using other planetary ephemerides are similar to those of using DE440 and thus not plotted.
    The positional offsets between DE200 and VLBI reach $-$20\,mas for some pulsars, while they decrease to approximately 4\,mas or less between DE440 and VLBI, with uncertainties of 1\,mas to 4\,mas in either right ascension or declination.
    The results based on different PTA data sets are consistent within their astrometric precision.
    
    The fittings of a frame rotation were performed to the positional offsets using each PTA data set separately except PPTA DR2e (due to the small sample size of pulsars in common) as well as their combination.
    The estimates of rotation parameters based on the IPTA DR2 data sets always differed by 0.3\,mas to 0.6\,mas from those based on the data sets from regional PTAs, although they still agreed with each other with their uncertainties returned by the least-squares fitting.
    These differences indicate possible systematics of the same level in the input VLBI and/or timing astrometric solutions.
    
    Table~\ref{table:pos-vsh-vlbi} reports the fitting results based on the combination of all data sets.
    We found an orientation offset of $\mathrm{\sim -3\,mas}$ on the $X$-axis and of $\mathrm{<\,-10\,mas}$ on the $Y$- and $Z$-axes in the DE200 frame.
    We also detected an orientation offset of $\mathrm{\sim -2.5\,mas}$ on the $X$-axis for the DE405 frame.
    The rest of the planetary ephemerides yield a common orientation offset of $\mathrm{\sim -1.0\,mas}$ on the $X$-axis.

    Since there were only two pulsars near the ecliptic plane, we could not use this subset to test the contribution of pulsars near the ecliptic plane to the determination of the frame rotation parameters.
    Instead, we excluded these two pulsars, leaving only pulsars with high ecliptic latitudes in the sample, and reran the solution.
    We found excellent agreements in the estimates of $R_{\rm X}$ and $R_{\rm Y}$, while there was an offset of $\mathrm{\sim -0.2\,mas}$ in the estimate of $R_{\rm Z}$ with respect to the solution using all pulsars.
    We speculated that this offset was caused by the removal of PSR J2145--0750, which yielded an offset of $\mathrm{\sim +0.8\,mas}$ in right ascension between timing and VLBI.
    An increase in the sample size might lead to a more robust fit of the frame rotation.

\section{Discussion}    


\subsection{Comparison with previous results}    

    \subsubsection{Intercomparison among planetary ephemerides} \label{subsubsect:disscusion-ephemeris-comparison}

    \citetads{2004A&A...417.1165S} compared DE200 and DE405 with DE409 (the newest edition of the DE series at that time) to approximate the accuracy of DE200 and DE405.
    Based on the heliocentric coordinates of the Earth-Moon barycenter, the author gave the rotation relations between DE200 and DE409 and between DE405 and DE409 at J2000.
    The rotation relation can be rewritten as 
    \begin{equation} \label{eq:standish-de200-de409}
        \bm{R}_{\rm DE200} = \left[\begin{array}{c}
        -2200 \\
        -11700 \\
        -12300
        \end{array}\right]\,{\rm \mu as}, ~
        \bm{r}_{\rm DE200} = \left[\begin{array}{c}
        +3 \\
        +108 \\
        -249
        \end{array}\right]\,{\rm \mu as\,yr^{-1}},
    \end{equation}
    for DE200 and 
    \begin{equation} \label{eq:standish-de405-de409}
        \bm{R}_{\rm DE405} = \left[\begin{array}{c}
        -1500 \\
        +1000 \\
        +1100
        \end{array}\right]\,{\rm \mu as}, ~
        \bm{r}_{\rm DE405} = \left[\begin{array}{c}
        +1 \\
        -7 \\
        +15
        \end{array}\right]\,{\rm \mu as\,yr^{-1}},
    \end{equation}
    for DE405 by converting to the same units used in this work.
    We computed the relative orientation offsets and spins of DE200 and DE405 with respect to DE410 by comparing the rotation parameters for DE200 and DE405 with those for DE410 reported in Tables~\ref{table:pos-vsh}--\ref{table:pm-vsh}.
    Assuming that DE409 was very close to DE410, we obtained consistent results with those of \citetads{2004A&A...417.1165S} given in Eqs.~(\ref{eq:standish-de200-de409})--(\ref{eq:standish-de405-de409}), which validated the effectiveness of our method.
    The relative spin found in the DE200 frame mainly reflects the mean motion errors of DE200 for the Earth-Moon barycenter (EMB), which was caused by the relatively short time-span coverage of the input ranging data used for deriving DE200 and the lack of asteroids in modeling their perturbations, as explained in \citetads{2004A&A...417.1165S}.
    The relative spin between DE405 and DE410 suggested a difference in the mean motion of Earth, as reported in \citet{standish2003de410}, which was mainly caused by the differences in the modeling of perturbation from the largest 20 asteroids and more accurate fitting of Mars' orbit in DE410 compared to DE405.

    Adopting a different computation of the asteroid perturbation compared to DE405 (considering perturbations of all 300 asteroids on all the planets and using an asteroid ring), the accuracy obtained with INPOP06 was improved to be comparable to the accuracy of DE414 \citepads{2008A&A...477..315F}.
    We observed in Table~\ref{table:pos-vsh} that the rotation and glide estimates for INPOP06c were much smaller than those of DE405 with respect to DE440.
    The accuracy of the Earth's orbit of DE421 was continuously improved by new VLBI tracking observations of spacecraft (\citeads{standish2004de414}; \citeads{standish2007de418}, \citeyearads{2009IPNPR.178C...1F}).
    In addition, the asteroid perturbation was modeled as the effects of 67 asteroids with individual masses and 276 asteroids whose masses were defined by different density classes.
    These improvements were reflected in the smaller values of the rotation and glide parameters between DE421 and DE440 compared to those between DE410 and DE440.
    For INPOP08a, the ring model and asteroid selection were revised. 
    \citetads{2009A&A...507.1675F} showed that the heliocentric longitudes of the EMB in the ecliptic coordinate computed using INPOP08 differed from those computed using DE421 or INPOP06 by a general linear drift, resulting in an ecliptic longitude offset of $\mathrm{-3\,mas}$ over 100 years. 
    It corresponded roughly to a difference of $\mathrm{-30\,\mu as\,yr^{-1}}$ in the spin around the $Z$-axis in the ecliptic coordinate, which agreed with the spin reported in Table~\ref{table:pm-vsh}.

    The main differences between INPOP08a and INPOP10a were the input data sets and the implementation of the fit.
    \citetads{2011CeMDA.111..363F} used timing astrometric solutions of 18 millisecond pulsars with a precision of better than 10\,mas to estimate the relative orientation offset between DE200, DE405, DE414, DE421, INPOP08, and INPOP10a.
    The reported results agreed with those obtained in this work within the formal uncertainties.
    Thanks to the increasing numbers of millisecond pulsars, more timing observations accumulated over more than 10 years and from more radio telescopes, and improvements achieved in the timing observing system, our results were several times more precise than those in \citetads{2011CeMDA.111..363F}, reaching a precision of several tens of $\mathrm{\mu as}$.  
    EPM2011 was the first edition of the EPM series included in this work.
    \citetads{2013SoSyR..47..386P} reported that the maximum differences in the heliocentric positions of Mars between EPM2011 and DE424 were approximately 0.7\,mas in right ascension and 0.5\,mas in declination.
    Our fitting suggested that the frame rotation between EPM2011 and DE421 was approximately 0.4\,mas around the $Y$-axis and 0.5\,mas around the $Z$-axis. 

    No ring model accounting for the perturbation of the Main belt objects was considered in the INPOP series since INPOP13c.
    The comparison of the geocentric coordinates of Mars between INPOP13c and DE430 provided by \citetads{2014arXiv1405.0484F} showed that the differences in right ascension and declination were all below 0.5\,mas.
    In this work, the agreement in the orientation of each axis between INPOP13c and DE430 was found to be approximately 0.3\,mas or better, which was thus consistent with the results in \citetads{2014arXiv1405.0484F}.
    The asteroid perturbation models were slightly modified in EPM2015 and EPM2017 compared to EPM2011.
    We found an orientation agreement of $\sim$0.1\,mas among these planetary ephemeris reference frames.
    The dynamical modeling of INPOP17a was almost the same as that of INPOP13c.
    The differences in the heliocentric positions of Mars between INPOP17a and DE430 given by the comparison in \citetads{2017NSTIM.108.....V} were approximately 0.2\,mas in both right ascension and declination, close to the differences in the rotation parameters between these two planetary ephemeris reference frames.
    A new model for the Trans-Neptunian objects was introduced in INPOP19a in addition to the increase in the amount of main-belt asteroids in comparison to INPOP17a, leading to a difference of $\sim$ 100\,km in the barycentric position of the EMB that was mainly in the direction perpendicular to the ecliptic \citepads{2019NSTIM.109.....F}.
    It seemed that, however, these improvements did not greatly affect the orientation of the INPOP reference frame.
    \citetads{2021NSTIM.110.....F} compared the heliocentric position of Earth between DE440 and INPOP21a, which gave an agreement of 0.2\,mas in right ascension and 0.4\,mas in declination.
    These values agreed roughly with the orientation offset between DE440 and INPOP21a frames determined from the timing astrometry.

    Comparisons with previous results based on other data sets or methods have validated the effectiveness of our method.
    Therefore, the transformation relation between other planetary ephemerides that are not included in this study might also be derived following the same procedure as described in Sect.~\ref{subsect:model}.
    
    \subsubsection{Comparison between planetary ephemerides and extragalactic frames} \label{subsubsect:disscusion-ephemeris-vlbi-comparison}

    DE200 was constructed on its own dynamical equinox of J2000.0 with an accuracy of 1 arcsecond \citepads{1982A&A...114..297S}.
    \citetads{1994A&A...287..279F} estimated the orientation offset between the DE200 frame and extragalactic frame to be approximately $\mathrm{-2\,mas}$, $\mathrm{-12\,mas}$, and $\mathrm{-6\,mas}$ around the $X$-, $Y$-, and $Z$-axes in 1988 by a joint analysis of the lunar laser ranging and VLBI measurements.
    These values roughly match our determination of the frame rotation between the DE200 frame and the VLBI frame at J2000.0, plus the propagation of the frame spin.

    Since DE403, the tie between all the planetary ephemeris reference frames and the ICRF is achieved through the adjustment of the planetary ephemerides to the VLBI observations of the spacecraft relative to the extragalactic sources.
    For the orbits of inner planets, this frame tie is dominated by VLBI measurements of Mars-orbiting spacecraft.
    Therefore, the accuracy of the frame tie between the planetary ephemeris reference frames and the ICRF is on the same level as the accuracy of these VLBI measurements.

    The DE405 frame was tied to ICRF1 \citepads{1998AJ....116..516M} with an uncertainty of $\sim\mathrm{1\,mas}$ \citep{standish1998de405} based on VLBI observations of the Magellan spacecraft in orbit around Venus.
    Later, in DE410, the frame tie was improved by VLBI points of Mars-orbiting spacecraft \citep{standish2003de410}.
    The improvement of frame-tie from DE405 to DE410 can be easily seen in Table~\ref{table:pos-vsh-vlbi}, although our estimations suggest larger orientation offsets between these two ephemeris frames and the VLBI frame.
    
    The DE421 frame was tied to the ICRF1 with an accuracy better than $\mathrm{1\,mas}$ \citepads{2009IPNPR.178C...1F}.
    \citetads{2013arXiv1301.1510F} reported that the link between the INPOP10e reference frame and the ICRF was maintained with an accuracy of $\mathrm{\sim 1\,mas}$ over 10 years.
    \citetads{2013SoSyR..47..386P} determined the orientation offset of the EPM2011 frame with respect to the ICRF1, which depended strongly on the time span of the input VLBI observations of various spacecraft. 
    Nevertheless, the frame tie between EPM2011 and ICRF1 is most likely to also be better than $\mathrm{1\,mas}$.
    Our comparisons between the timing and VLBI solutions agree with these results within the quoted uncertainty.
    
    The DE430 frame was aligned with ICRF2 \citepads{2015AJ....150...58F} with an accuracy of $\mathrm{0.2\,mas}$ (\citeads{2014IPNPR.196C...1F}; \citeads{2015HiA....16..219F}).
    \citetads{2015AJ....150..121P} also determined the orientation of the DE430 frame relative to the ICRF by using VLBA measurements of Mars with an uncertainty of 0.23 mas.
    The link accuracy of the INPOP13c frame to ICRF2 was improved to be better than $\mathrm{0.5\,mas}$ compared to INPOP10e \citepads{2014arXiv1405.0484F}.
    The EPM2015 and EPM2017 frames were oriented to the ICRF2 with an accuracy better than 0.2 mas at 3-sigma (\citeads{pitjeva2017}; \citeads{2018CeMDA.130...57P}).
    A recent study presented by \citetads{2022CeMDA.134...32D} used the observations of asteroids in the \textit{Gaia} Data Release 2 \citepads{2018A&A...616A...1G} to determine the frame rotation between the INPOP19a reference frame and \textit{Gaia}-CRF2 \citepads[][]{2018A&A...616A..14G}.
    The misalignment between these two reference frames was found to be approximately $\mathrm{0.25\,mas}$; it decreased to a few $\mathrm{\mu as}$ when only very accurate observations of inner planets were considered.
    Considering that the agreement between \textit{Gaia}-CRF2 and ICRF3 is on the level of $\mathrm{20\,\mu as}$--$\mathrm{30\,\mu as}$ (\citeads{2020A&A...644A.159C}; \citeads{2018A&A...616A..14G}; \citeads{2020A&A...634A..28L}), the tie between the INPOP19a frame and ICRF3 should be accurate at $\mathrm{0.25\,mas}$ or much better.
    The average accuracy of the orientation of the inner planet orbits for DE440 tied to ICRF3 is approximately 0.2\,mas \citepads{2021AJ....161..105P}.
    For these planetary ephemeris frames, however, we detected a common frame rotation $R_{\rm X}\!\simeq\!\mathrm{-1\,mas}$ that is confident at 4-sigma or more; this frame rotation would decrease if only an individual PTA data set was used.
    Therefore, we suspected that the common rotation was likely related to the systematics in the input timing data sets.

    \citetads{2011CeMDA.111..363F} found a significant rotation of $\mathrm{\sim 10\,mas}$ in the planetary ephemerides with respect to the VLBI frame based on four pulsars with both VLBI and timing astrometric measurements, which diminishes in our results with improved timing and VLBI astrometric data.
    Therefore, the rotation reported in \citetads{2011CeMDA.111..363F} is likely caused by the unmodeled systematics in the input timing or VLBI positions used therein.
    \citetads{2017MNRAS.469..425W} studied the rotation between VLBI positions and timing positions using different ephemerides (DE405, DE414, DE421, DE430, DE432, and DE435) for five millisecond pulsars and reported a statistically significant rotation angle of $\mathrm{\sim 2\,mas}$ around the $X$-axis for DE405.
    Similar results were obtained in our comparison (see Table~\ref{table:pos-vsh-vlbi}).

    The orientation offsets of the DE440, EPM2021, and INPOP21a frames with respect to the VLBI frame range from $\mathrm{0.2\,mas}$ to $\mathrm{1.2\,mas}$ in an absolute sense.
    Taking the formal uncertainty of $\mathrm{\sim 0.2\,mas}$ of the rotation parameters and the possible systematics of $\mathrm{0.3\,mas}$--$\mathrm{0.6\,mas}$ (Sect.~\ref{subsect:timing-vs-vlbi}) into consideration, the nonzero orientation offset of the ephemeride frame is approximately $\mathrm{0.4\,mas}$.
    This deviation is similar to the accuracy of the input VLBI measurements used for constructing the planetary ephemerides.

    We noted that the formal uncertainty of the frame rotation based on pulsar astrometry was close to those based on the VLBA measurements, even though we used only a small fraction of the sample from the MSPSR$\pi$ campaign.
    More precise frame-tie between the planetary ephemeris frame and the VLBI frame via pulsars can be expected with the increased sample size and reduced systematics in the absolute VLBI positions in the future data release of this campaign.
    

\subsection{Accuracy of state-of-the-art planetary ephemeris reference frames}   \label{subsect:eph-accuracy}

    The orientation offset and global spin of the ephemerides generally decrease with respect to the latest ephemerides in each series (i.e., DE440, EPM2021, and INPOP21a).
    These results can be expected and validate the improvements in the error control for the successive ephemerides in the DE, EPM, and INPOP series from the aspect of the pulsar timing astrometry.

    According to \citetads{2004A&A...417.1165S}, there are three main factors determining the accuracy of an ephemeris, including (i) the correctness and completeness of the equations of motion, (ii) the method and algorithm of integration, and (iii) the input observational data to create the ephemeris.
    The third factor contributed mainly to the improvements of the ephemerides in the DE series since DE405.
    Here, we use the INPOP series as an example to discuss the effect of these factors on our results, for which INPOP21a is used as the reference ephemeris.
    There is no modification in the dynamical modeling of INPOP10a compared with INPOP08, but new observations, especially the normal points deduced from flybys of the Messenger spacecraft around Mercury, were used to create INPOP10a \citepads{2011CeMDA.111..363F}.
    This may explain the improvements of $\mathrm{0.4\,mas}$ in the orientation of the INPOP10a frame compared with the INPOP08a frame.
    INPOP10a and INPOP10b were fitted on the same data sample, while the modeling of the asteroid perturbations on planet orbits for INPOP10b was improved and determined with an advanced algorithm \citep{fienga2012complementary}.
    Our comparison suggests that the orientation offset of the INPOP10b frame was reduced by $\mathrm{0.4\,mas}$ in the $Y$-axis compared to the INPOP10a frame, which can be considered as a contribution from the first two factors.
    
    The orientation agreement was found to be better than 0.15\,mas between DE440 and EPM2021 frames and 0.40\,mas between DE440 and INPOP21a frames (Sect.~\ref{subsect:timing-vs-timing}).
    In addition, the discrepancy of orientation between the ephemerides frames and VLBI frame was estimated to be $\mathrm{\sim 0.4\,mas}$ (Sect.~\ref{subsubsect:disscusion-ephemeris-vlbi-comparison}).
    From these results, we concluded that the orientation of the state-of-art planetary ephemeris frames is accurate at 0.4\,mas.
    The relative spin is approximately $\mathrm{2\,\mu as\,yr^{-1}}$ between DE440 and EPM2021 and $\mathrm{5\,\mu as\,yr^{-1}}$ between DE440 and INPOP21a, leading to the conclusion that the nonrotating of the planetary frame is no greater than $\mathrm{5\,\mu as\,yr^{-1}}$.
    In addition, there is no significant dipolar deformation in the planetary frames at the level of $\mathrm{1\,\mu as}$ in the position and $\mathrm{0.1\,\mu as\,yr^{-1}}$ in the proper motion.


\section{Conclusions}

    In this work, we used the up-to-date pulsar timing observations from the data releases of several pulsar timing arrays and astrometric VLBI measurements for pulsars to investigate the systematics in the planetary ephemeris reference frames.
    We reported the rotation and glide in the position and proper motion systems of successive editions of the DE series, EPM series, and INPOP series with respect to those of DE440, which is as precise as several $\mathrm{\mu as}$ in position and $\mathrm{0.1\,\mu as\,yr^{-1}}$ in proper motion.
    Depending on the sky location, the variations in the pulsar timing position and proper motion due to the different choices are on the same order of magnitude as the derived rotation and glide parameters.
    The orientation offset between the ephemerides frames and VLBI frame is estimated to be approximately $\mathrm{0.4\,mas}$ when considering the formal uncertainty.
    Our determinations of rotation parameters are improved compared to previous studies thanks to the precise pulsar timing observations of more than 60 millisecond pulsars, which were not available before.
    Based on the results mentioned above, we conclude that the misalignment of the latest planetary reference frames is no greater than 0.4\,mas with a nonrotating of better than $\mathrm{5\,\mu as\,yr^{-1}}$.

\begin{acknowledgements}
    We sincerely thank the referee, Dr.~Agnès Fienga, for the constructive comments and useful suggestions, which have greatly improve the work.
    We wish to thank Prof.~Xue-Mei Deng for her critical reading and comments on the first draft.
    We also thank Dr.~Jing Luo for his kind instructions on \texttt{PINT}.
    N. Liu and Z. Zhu were supported by the National Natural Science Foundation of China (NSFC) under grants Nos~11833004 and 12103026.
    N. Liu was also supported by the China Postdoctoral Science Foundation (Grant Number: 2021M691530) and the Yuxiu Postdoctoral Institute at Nanjing University (``Yuxiu Young Scholars Program'').
    J. Antoniadis was supported by the Stavros Niarchos Foundation (SNF) and the Hellenic Foundation for Research and Innovation (H.F.R.I.) under the 2nd Call of  ``Science and Society'' Action Always strive for excellence -- ``Theodoros Papazoglou'' (Project Number: 01431).
    This research also made use of Numpy \citepads{2011CSE....13b..22V}, IPython \citepads{2007CSE.....9c..21P}, Astropy\footnote{\url{http://www.astropy.org}} \citepads{2018AJ....156..123A}, the Python 2D plotting library Matplotlib \citepads{2007CSE.....9...90H}, and NASA's Astrophysics Data System.
    All the results shown in this work are also presented in a series of \textsc{Python Jupyter Notebooks} posted and publicly accessible online\footnote{\url{https://git.nju.edu.cn/neo/pulsar-ephem-crf}}.
\end{acknowledgements}

\bibliographystyle{aa} 
\bibliography{references} 

\begin{appendix} 

\section{Pulsar timing position differences using different ephemerides} \label{sect:appendix-pos-offset}

    Figures~\ref{fig:equ-pos-de200-vs-de440}--\ref{fig:equ-pos-inpop19a-vs-de440} presents the offsets of the pulsar position from the timing analyses using successive versions of planetary ephemerides from the DE series, EPM series, and INPOP series with respect to those from the timing solution using DE440.

\input{pos-diff-plot}

\section{Pulsar timing proper motion differences using different ephemerides} \label{sect:appendix-pm-offset}

    Figures~\ref{fig:equ-pm-de200-vs-de440}--\ref{fig:equ-pm-inpop19a-vs-de440} presents the offsets of the pulsar proper motion from the timing solutions using successive versions of planetary ephemerides from the DE series, EPM series, and INPOP series with respect to those from the timing solution using DE440.

\input{pm-diff-plot}


\end{appendix}

\end{document}

%% file: pos-diff-plot.tex
\begin{figure*} 
  \includegraphics[width=\columnwidth]{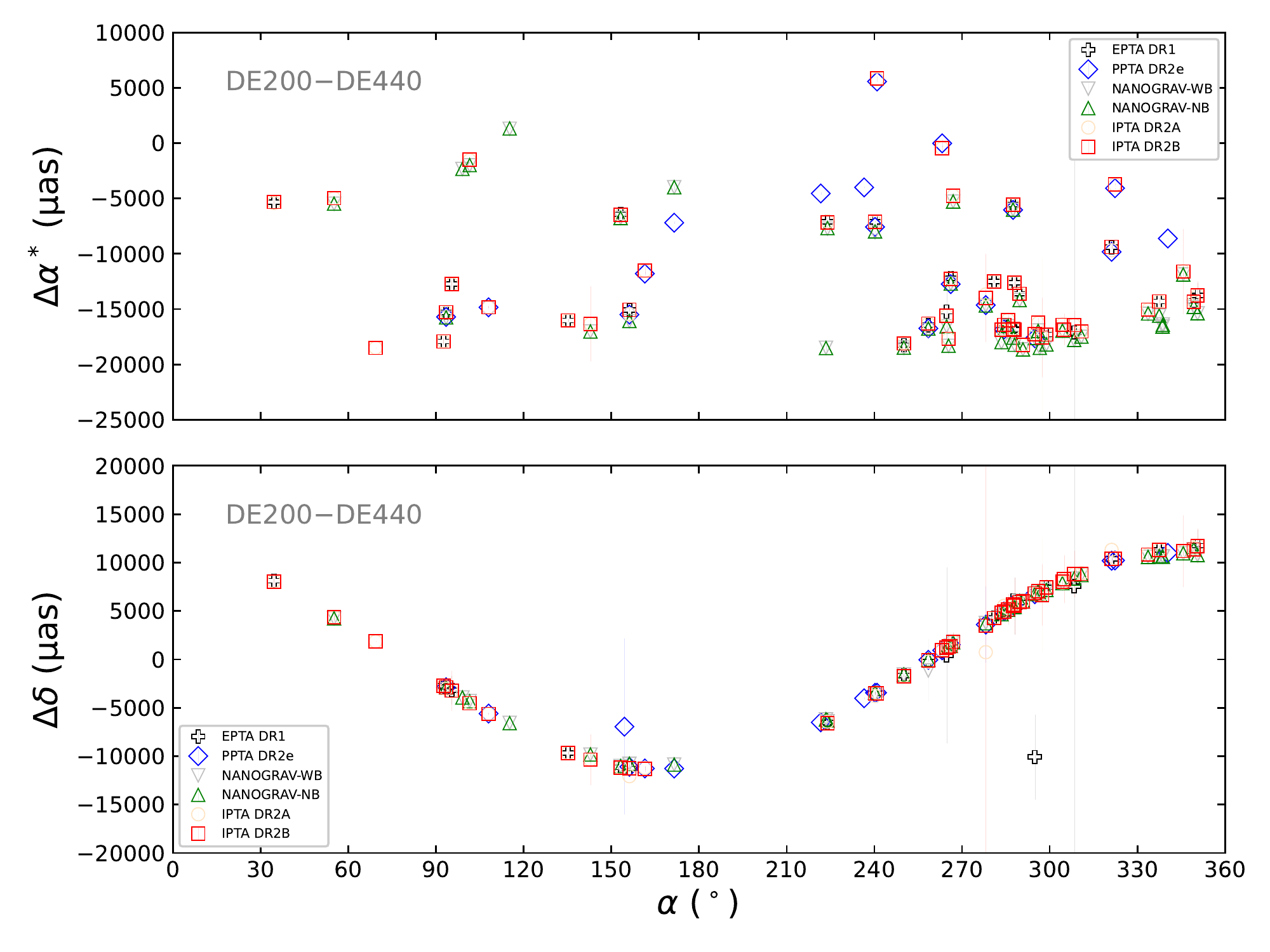}
  \includegraphics[width=\columnwidth]{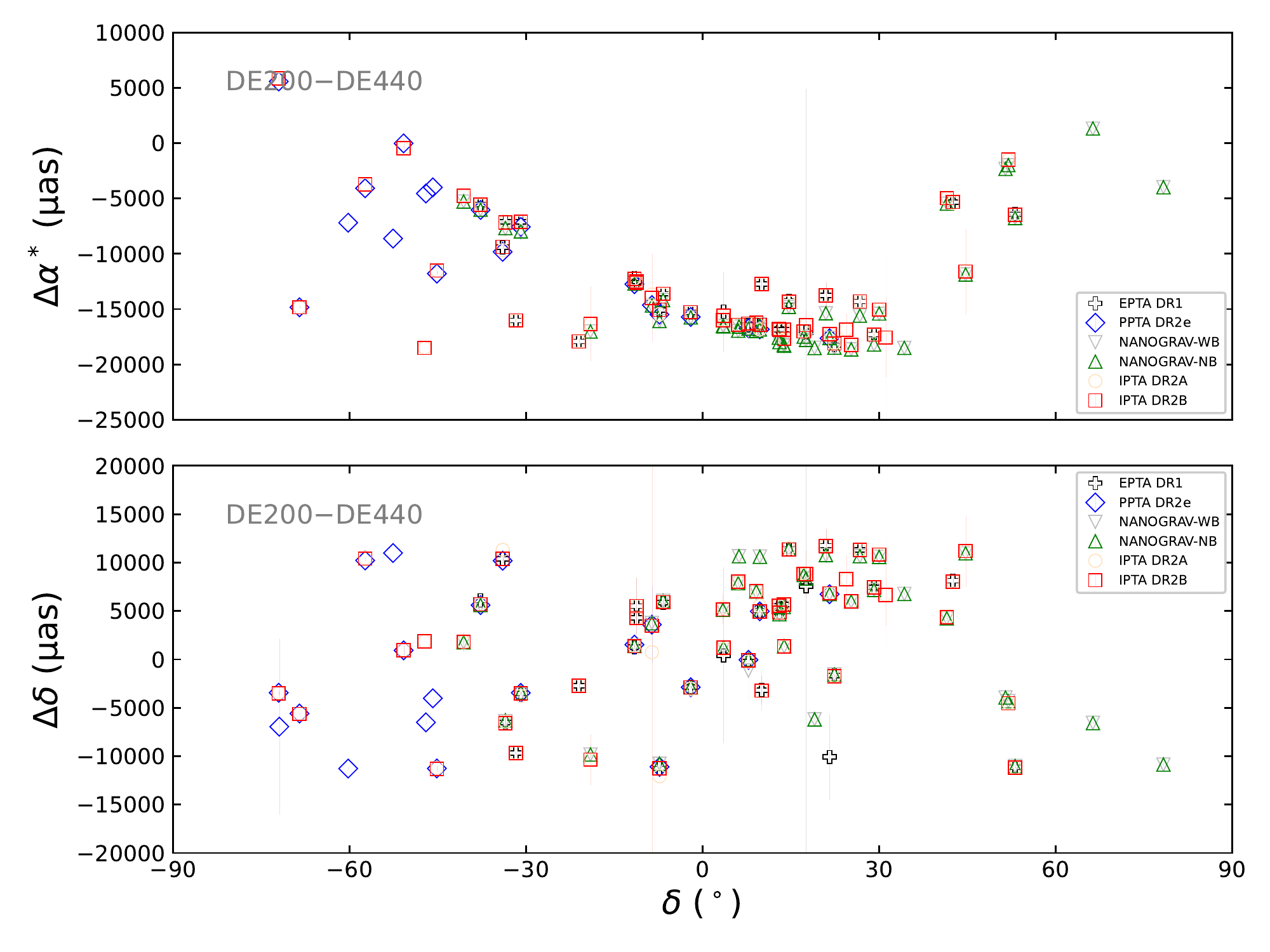}
  \caption[]{\label{fig:equ-pos-de200-vs-de440} %
  Offsets of the pulsar timing positions in the DE200 frame with referred to those in the DE440 frame as a function of the right ascension (left) and declination (right).
}
\end{figure*}

\begin{figure*} 
  \includegraphics[width=\columnwidth]{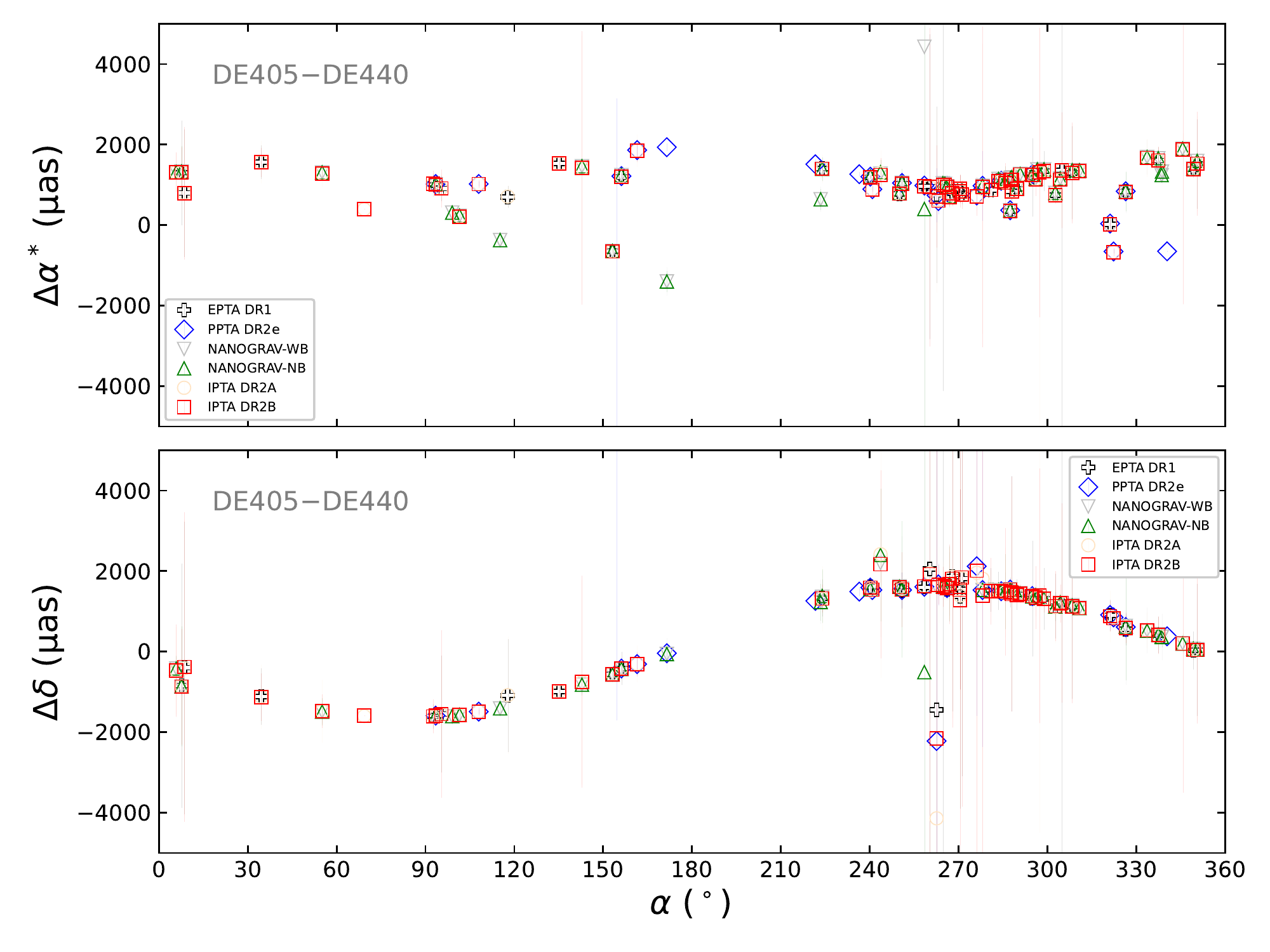}
  \includegraphics[width=\columnwidth]{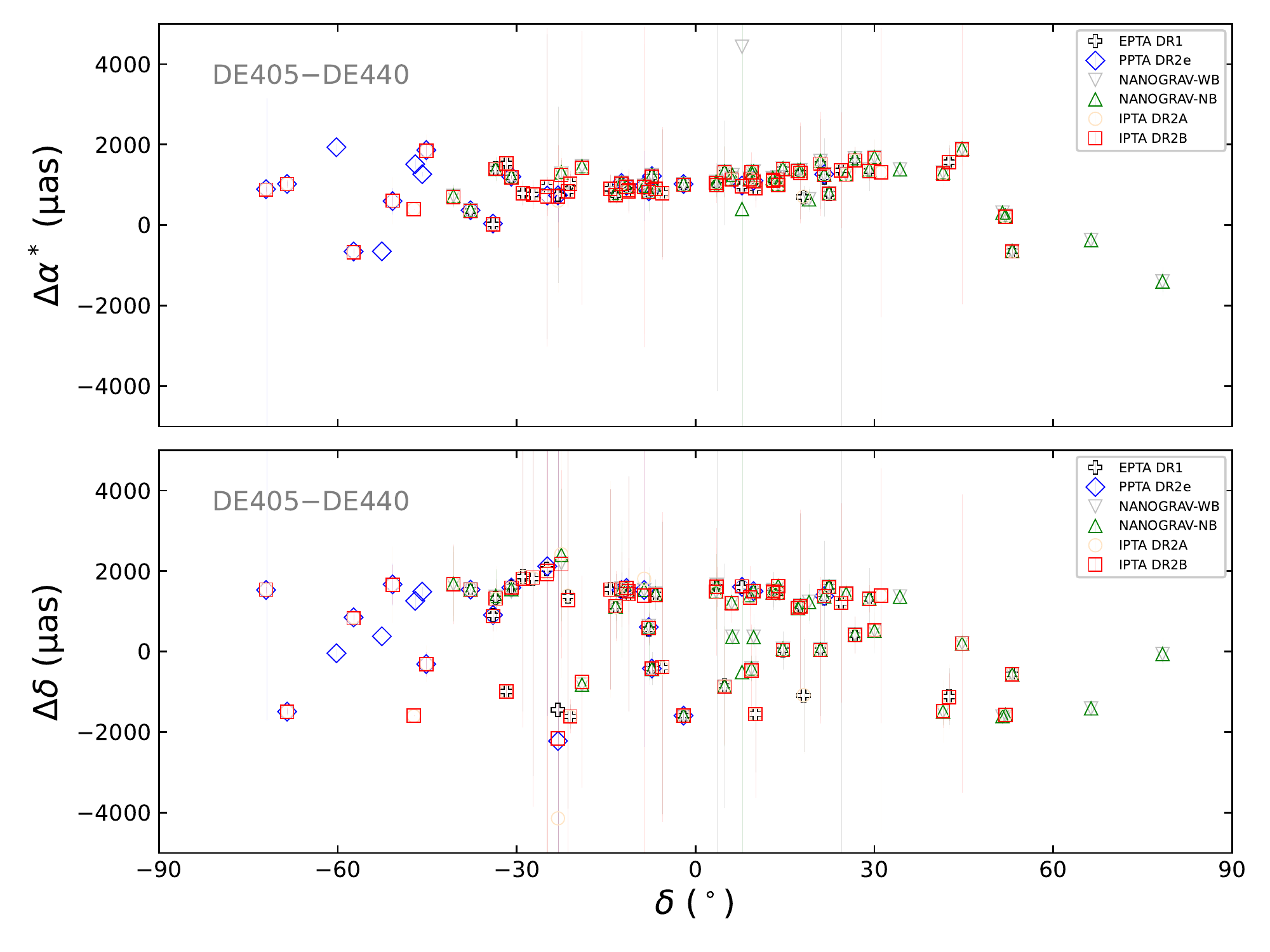}
  \caption[]{\label{fig:equ-pos-de405-vs-de440} %
  Offsets of the pulsar timing positions in the DE405 frame with referred to those in the DE440 frame as a function of the right ascension (left) and declination (right).
}
\end{figure*}

\begin{figure*} 
  \includegraphics[width=\columnwidth]{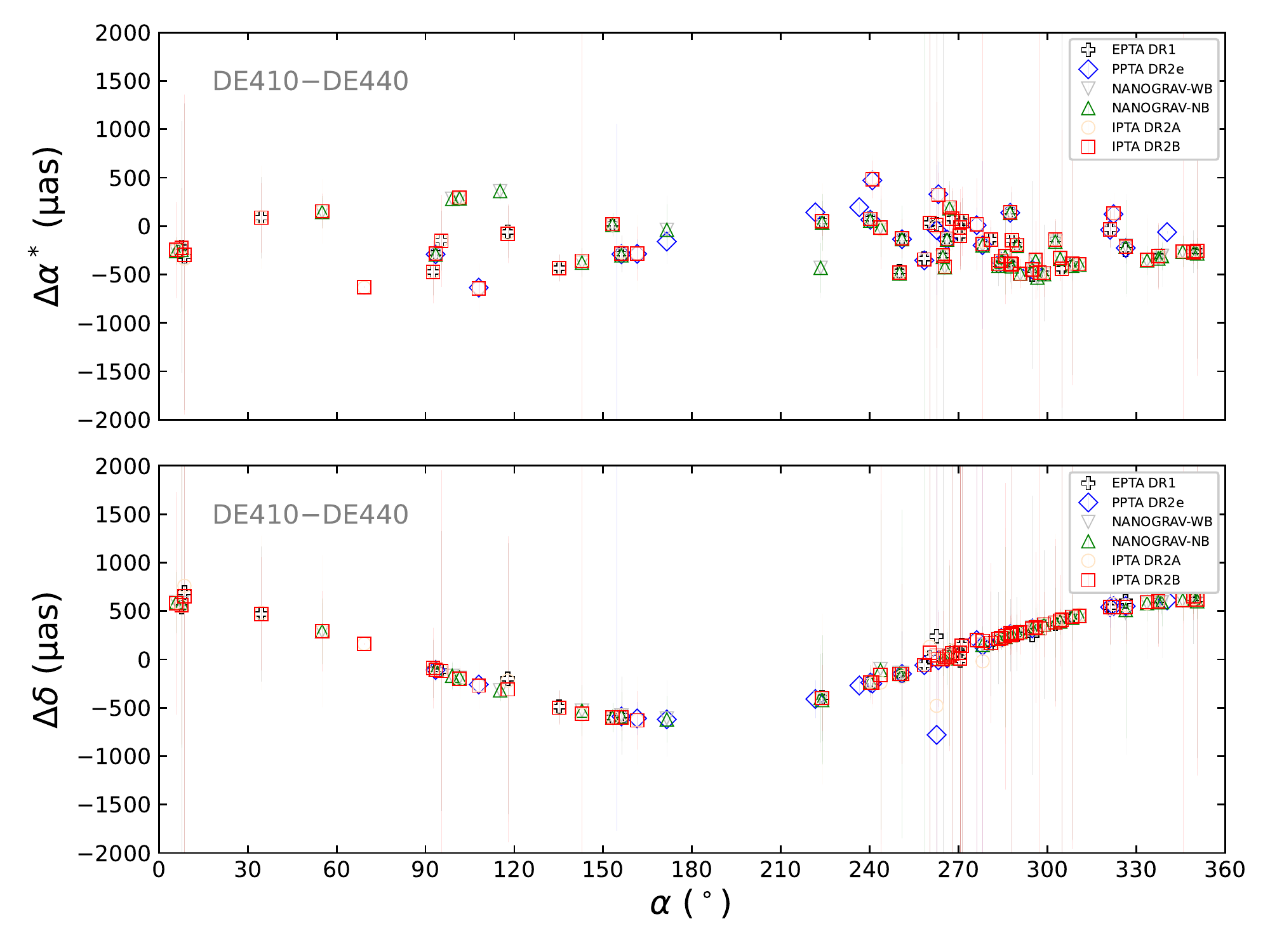}
  \includegraphics[width=\columnwidth]{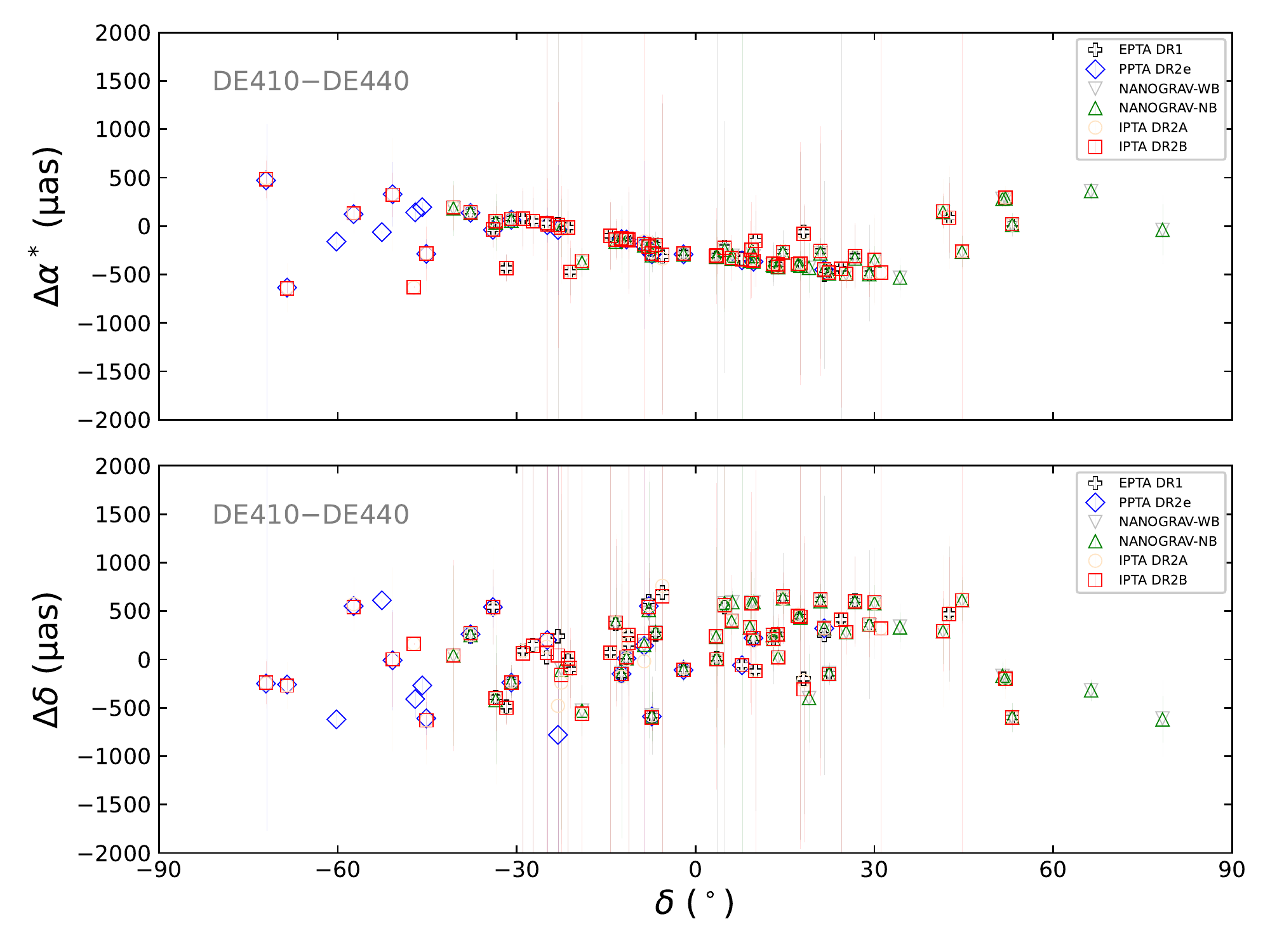}
  \caption[]{\label{fig:equ-pos-de410-vs-de440} %
  Offsets of the pulsar timing positions in the DE410 frame with referred to those in the DE440 frame as a function of the right ascension (left) and declination (right).
}
\end{figure*}

\FloatBarrier

\begin{figure*} 
  \includegraphics[width=\columnwidth]{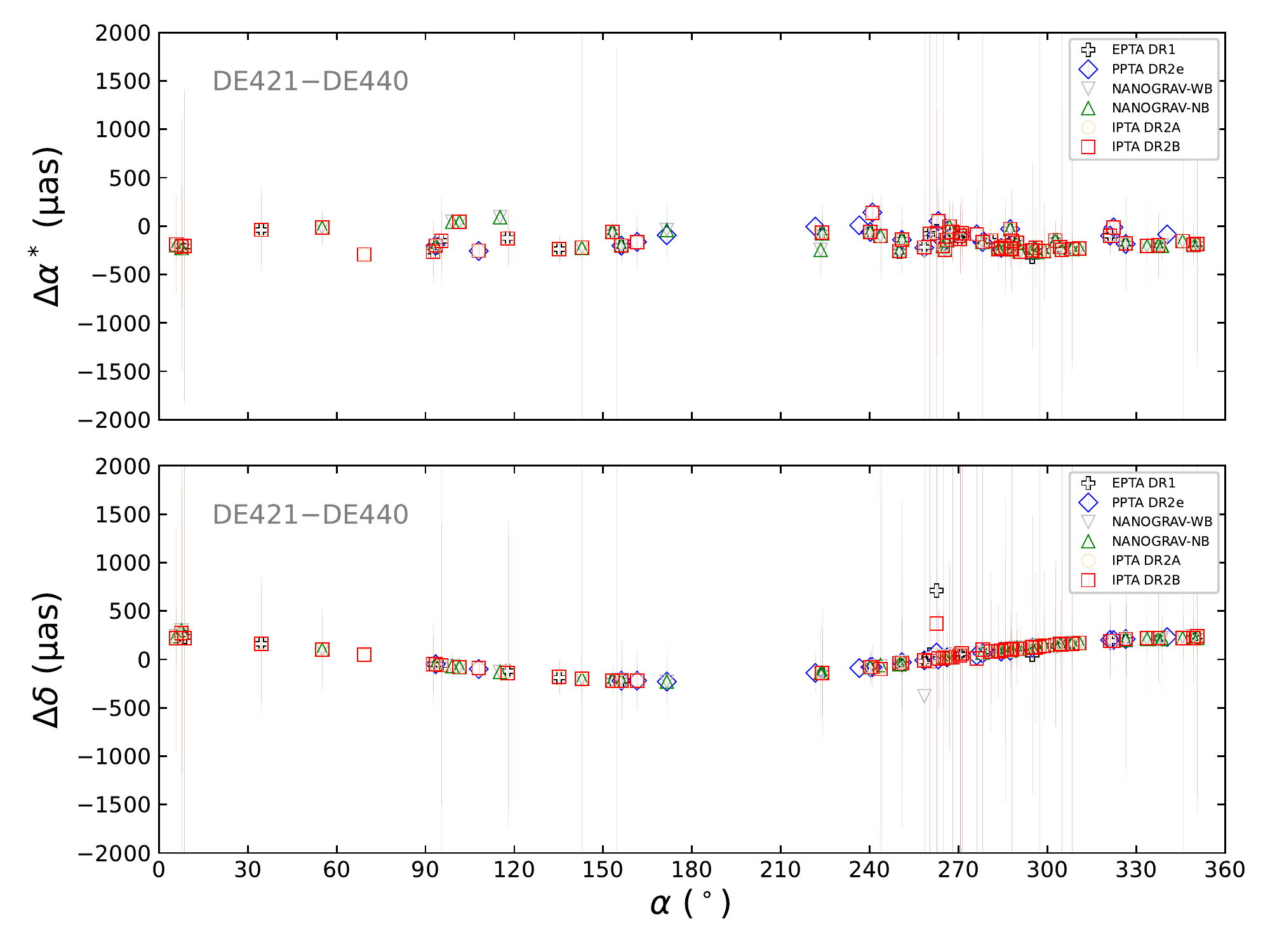}
  \includegraphics[width=\columnwidth]{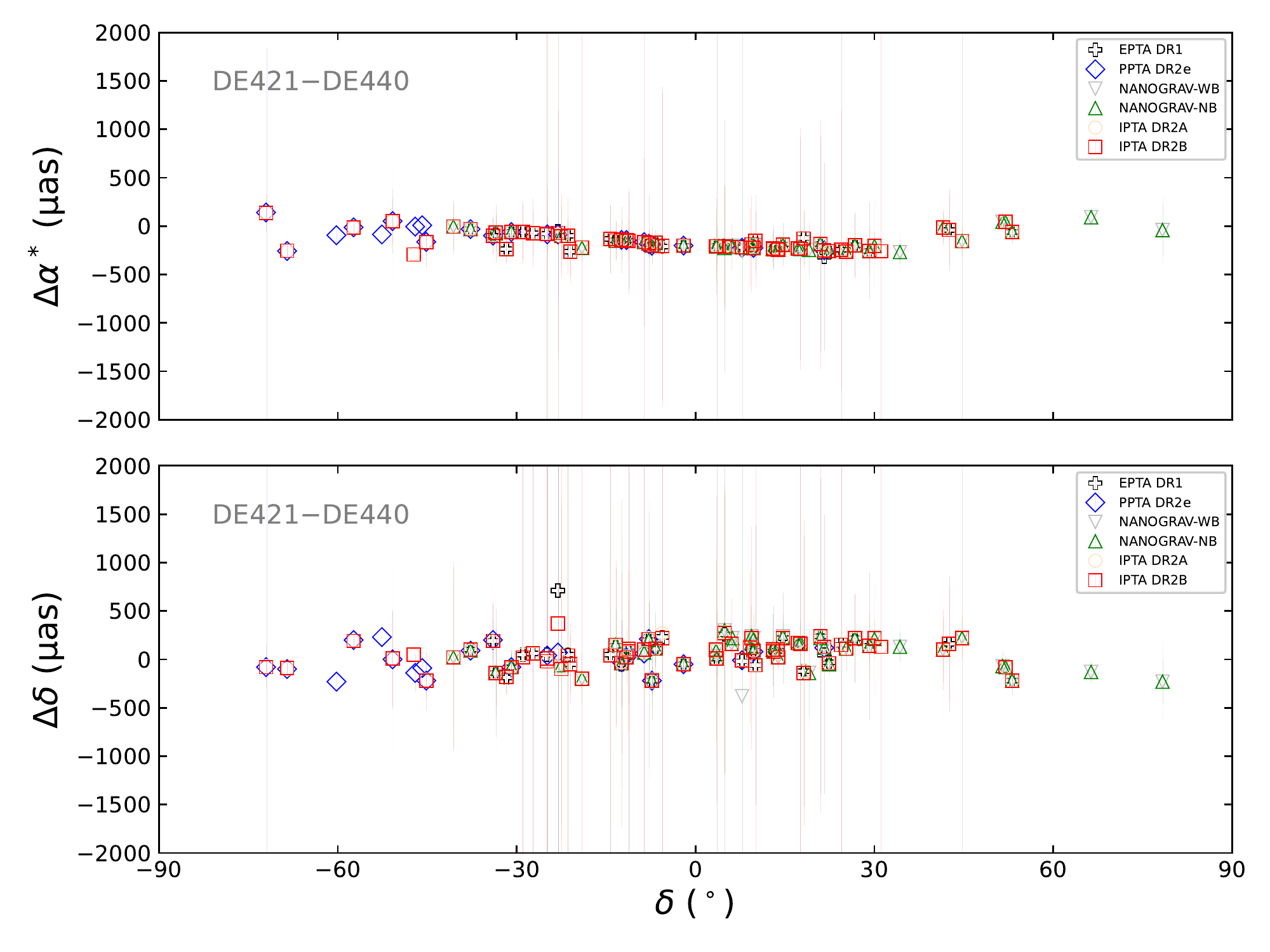}
  \caption[]{\label{fig:equ-pos-de421-vs-de440} %
  Offsets of the pulsar timing positions in the DE421 frame with referred to those in the DE440 frame as a function of the right ascension (left) and declination (right).
}
\end{figure*}

\begin{figure*} 
  \includegraphics[width=\columnwidth]{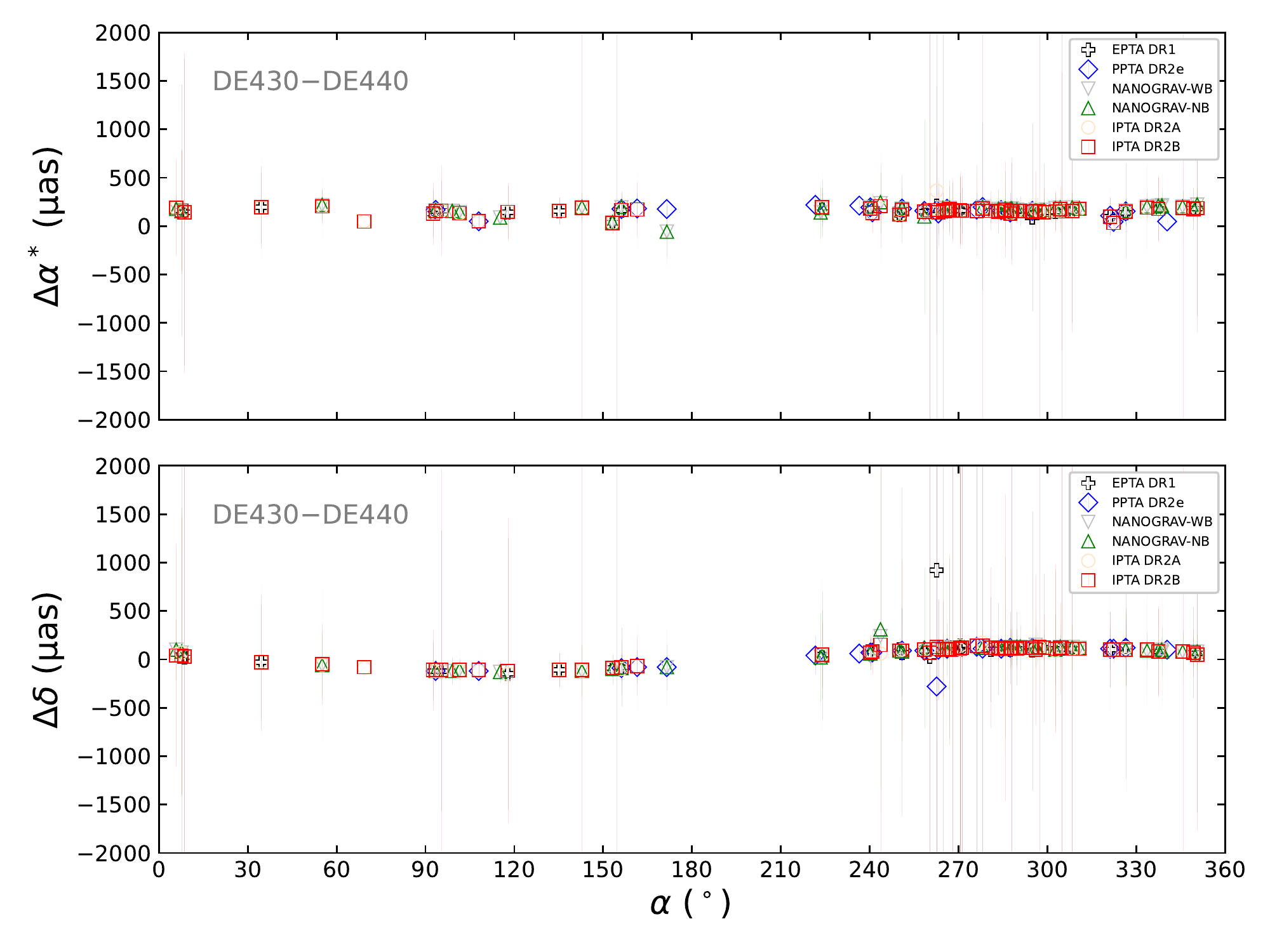}
  \includegraphics[width=\columnwidth]{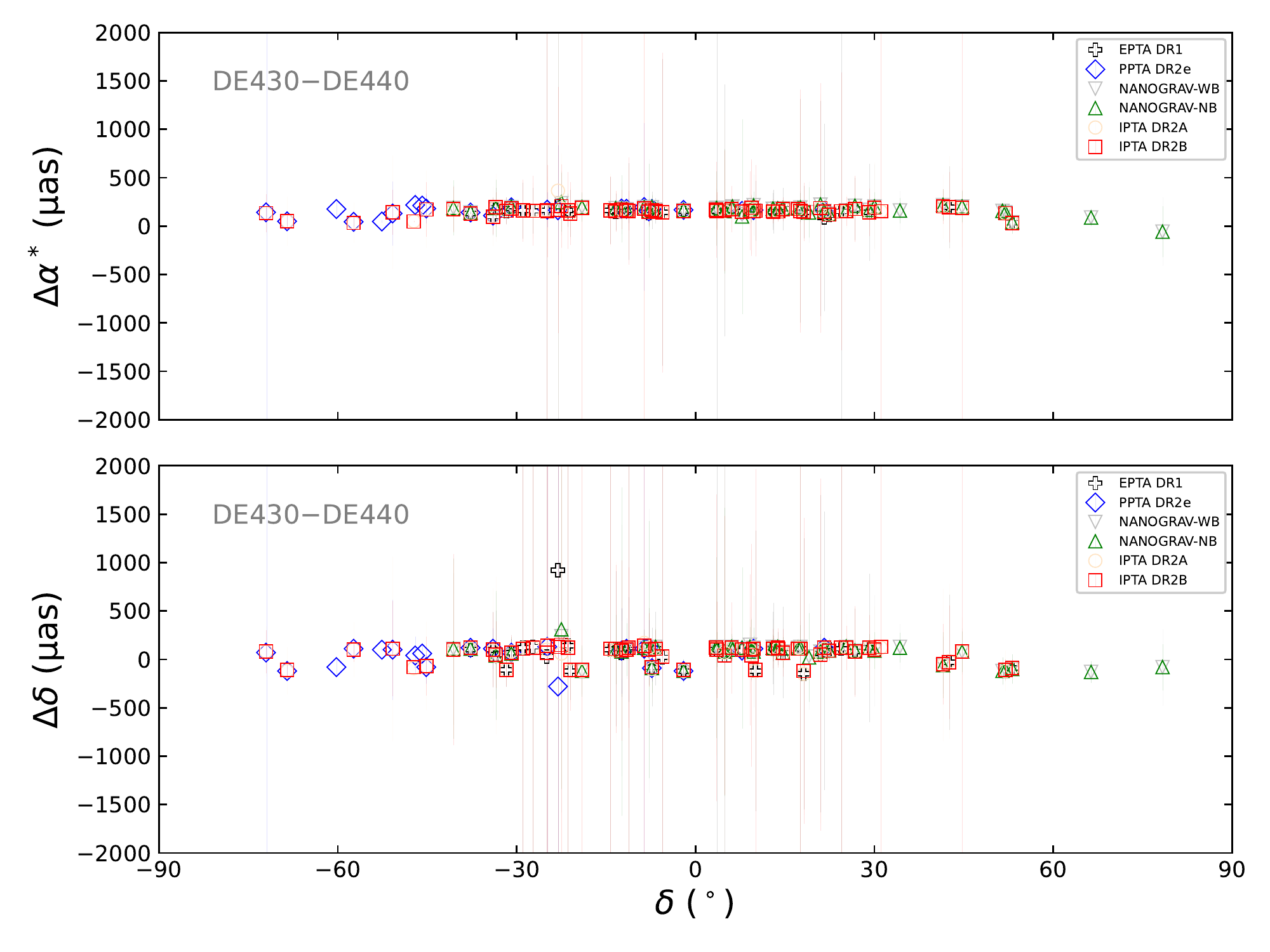}
  \caption[]{\label{fig:equ-pos-de430-vs-de440} %
  Offsets of the pulsar timing positions in the DE430 frame with referred to those in the DE440 frame as a function of the right ascension (left) and declination (right).
}
\end{figure*}

\begin{figure*} 
  \includegraphics[width=\columnwidth]{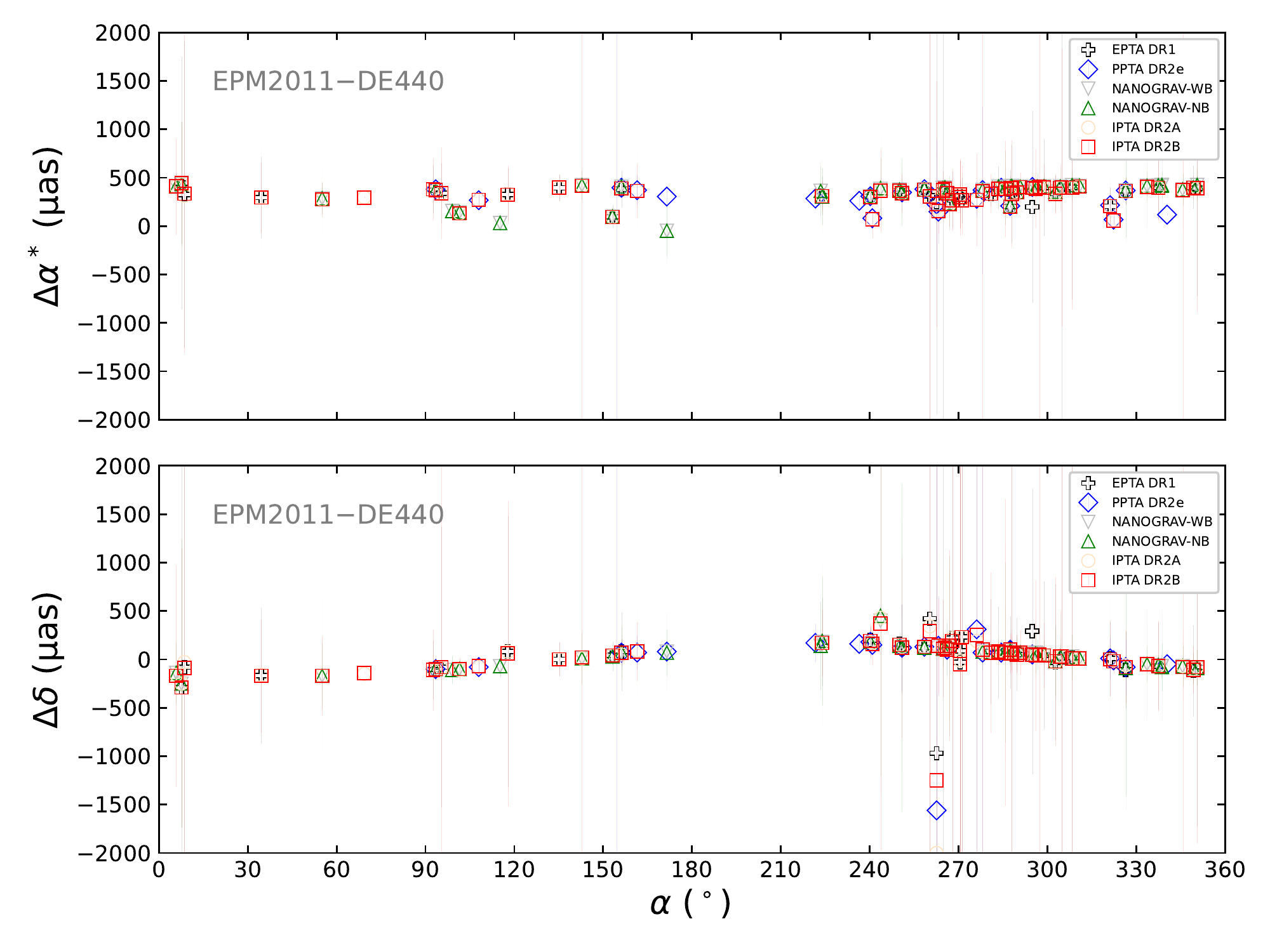}
  \includegraphics[width=\columnwidth]{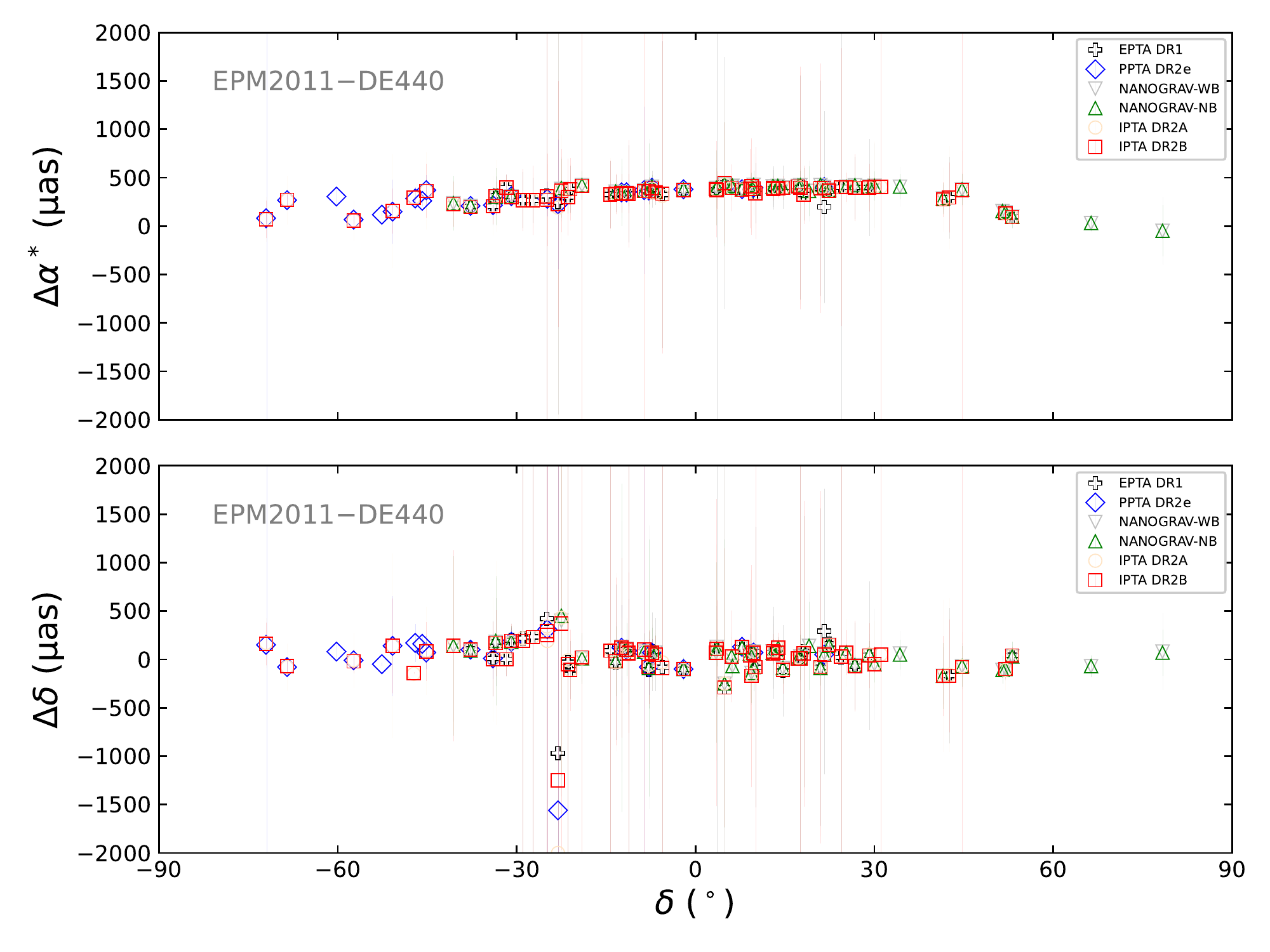}
  \caption[]{\label{fig:equ-pos-epm2011-vs-de440} %
  Offsets of the pulsar timing positions in the EPM2011 frame with referred to those in the DE440 frame as a function of the right ascension (left) and declination (right).
}
\end{figure*}

\FloatBarrier

\begin{figure*} 
  \includegraphics[width=\columnwidth]{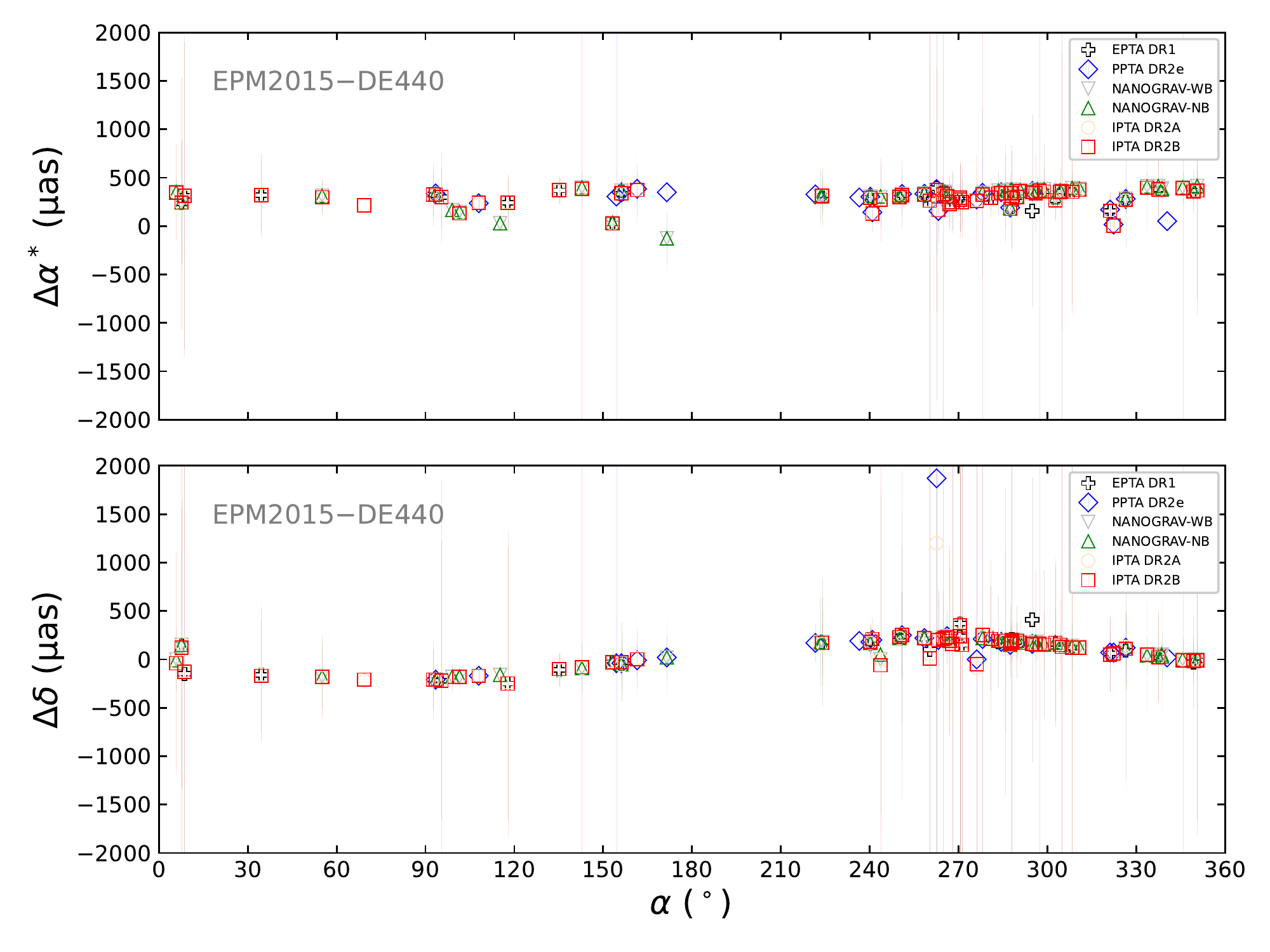}
  \includegraphics[width=\columnwidth]{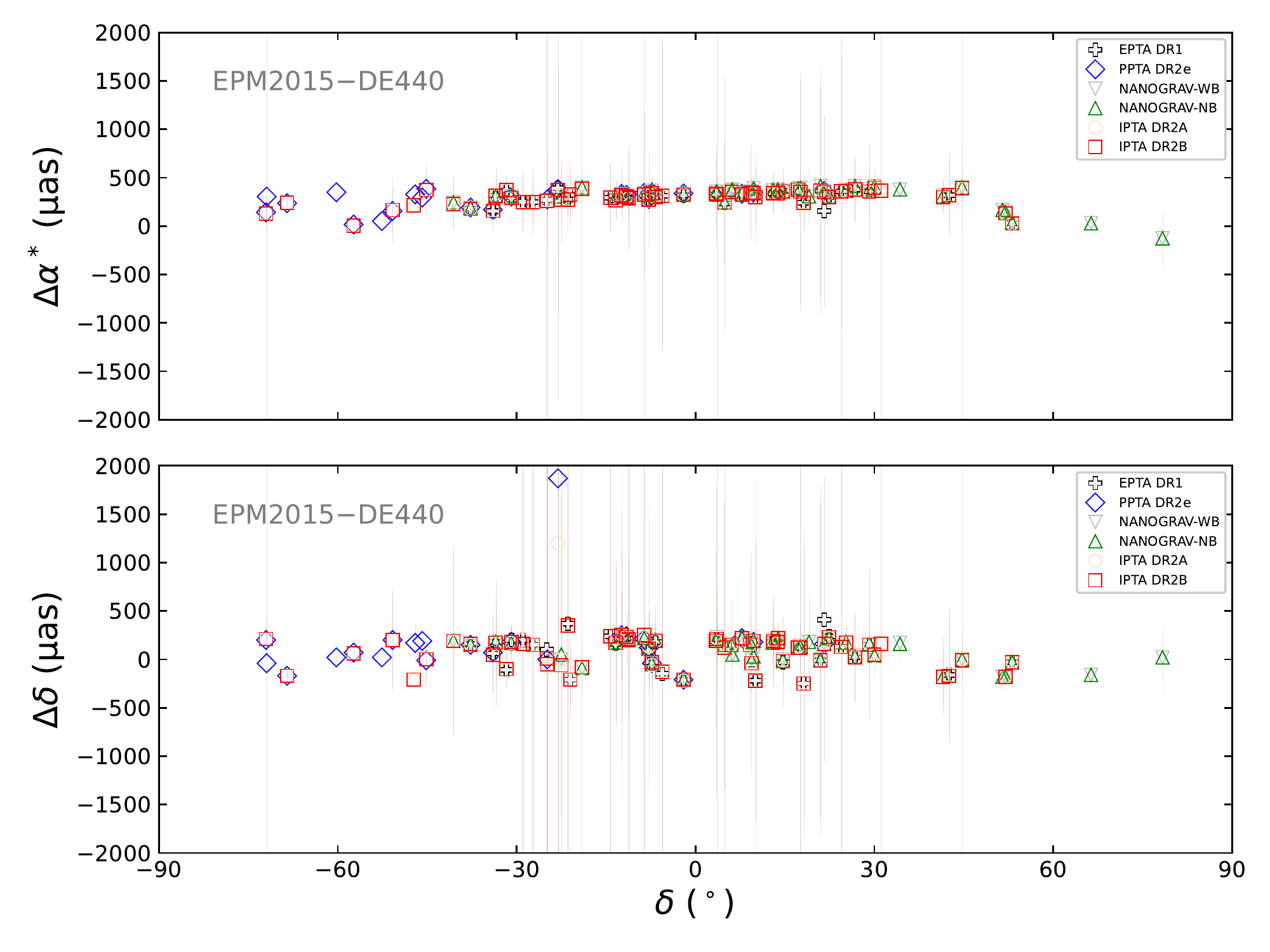}
  \caption[]{\label{fig:equ-pos-epm2015-vs-de440} %
  Offsets of the pulsar timing positions in the EPM2015 frame with referred to those in the DE440 frame as a function of the right ascension (left) and declination (right).
}
\end{figure*}

\begin{figure*} 
  \includegraphics[width=\columnwidth]{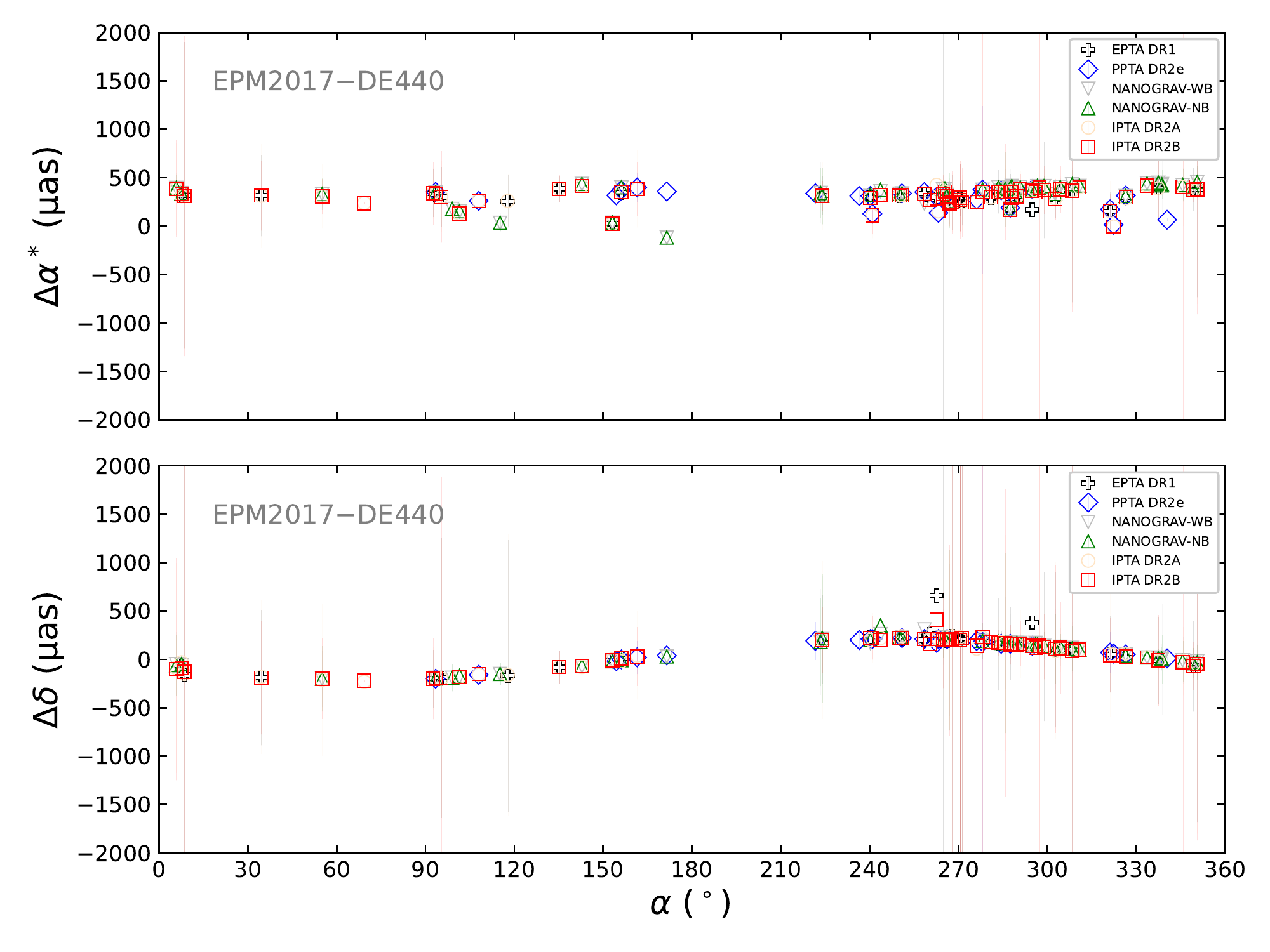}
  \includegraphics[width=\columnwidth]{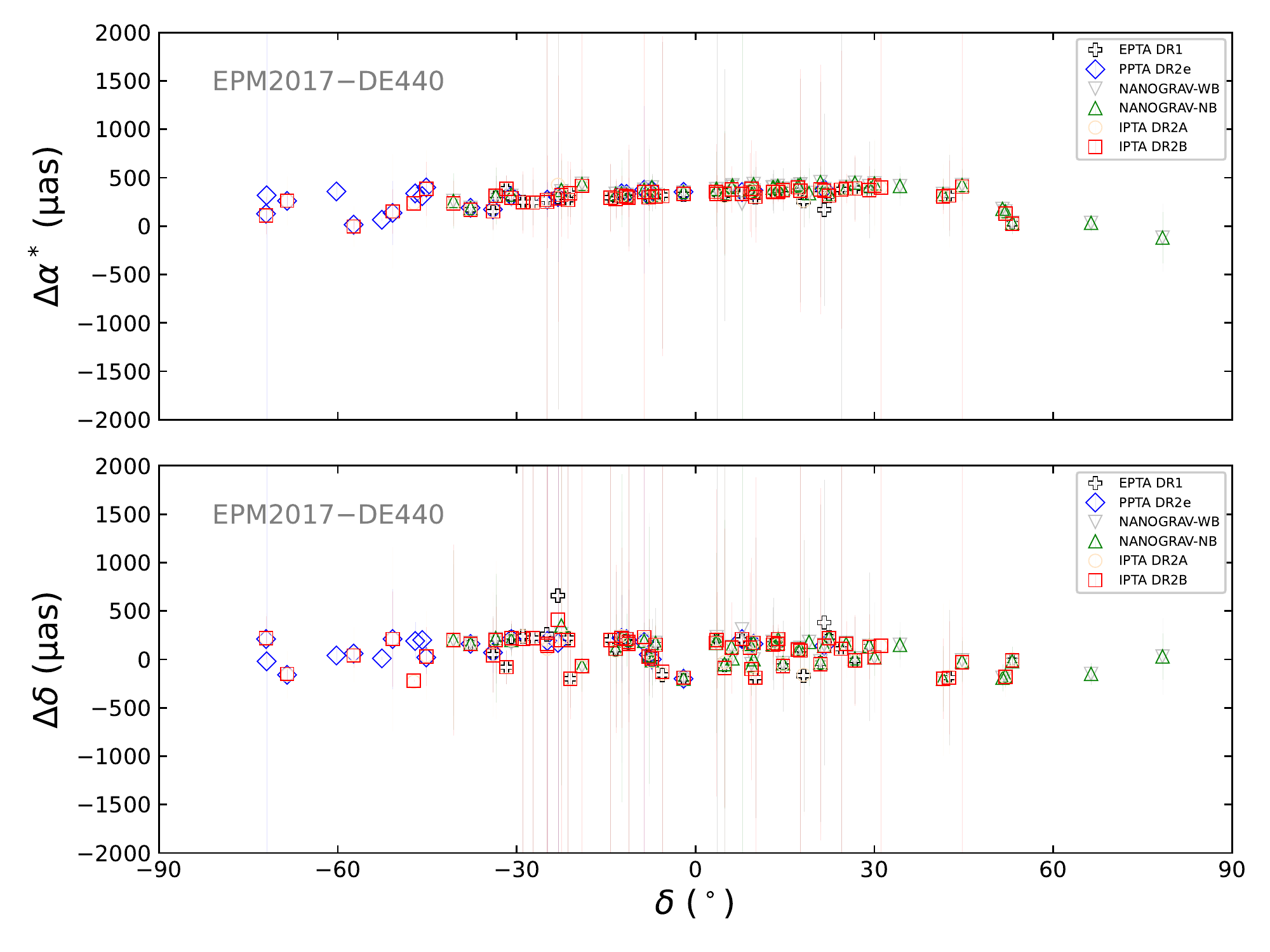}
  \caption[]{\label{fig:equ-pos-epm2017-vs-de440} %
  Offsets of the pulsar timing positions in the EPM2017 frame with referred to those in the DE440 frame as a function of the right ascension (left) and declination (right).
}
\end{figure*}

\begin{figure*} 
  \includegraphics[width=\columnwidth]{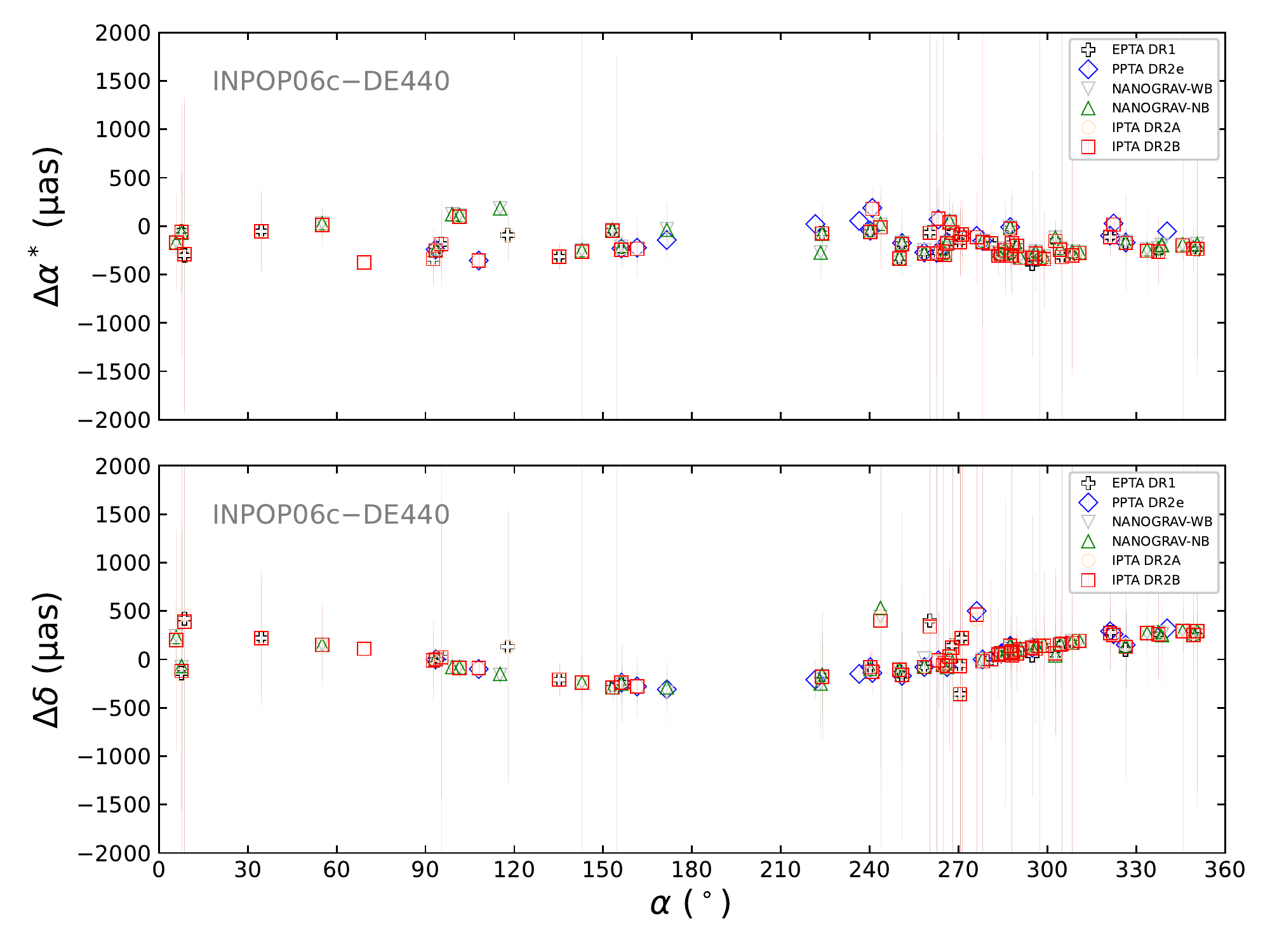}
  \includegraphics[width=\columnwidth]{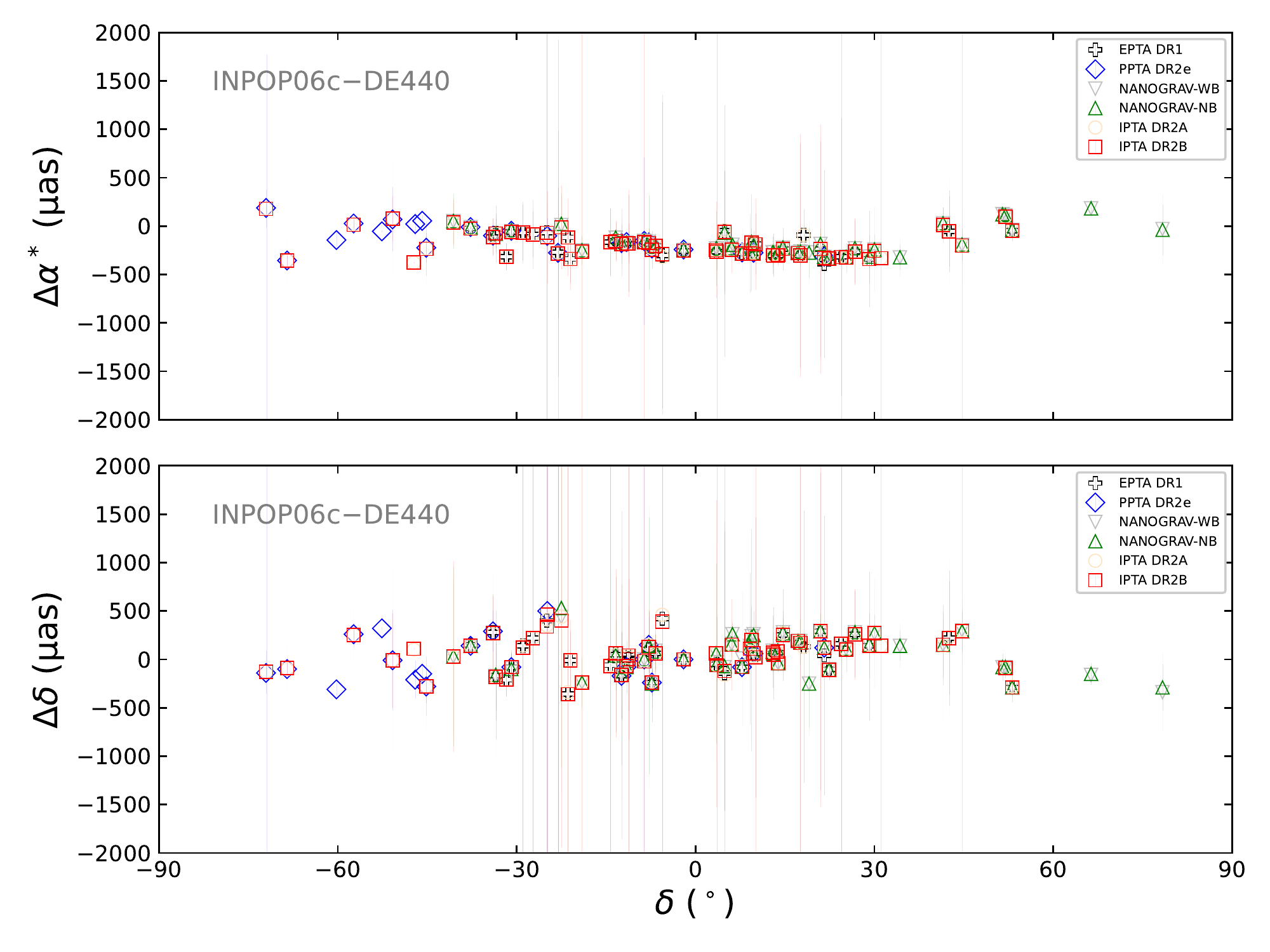}
  \caption[]{\label{fig:equ-pos-inpop06c-vs-de440} %
  Offsets of the pulsar timing positions in the INPOP06c frame with referred to those in the DE440 frame as a function of the right ascension (left) and declination (right).
}
\end{figure*}

\FloatBarrier

\begin{figure*} 
  \includegraphics[width=\columnwidth]{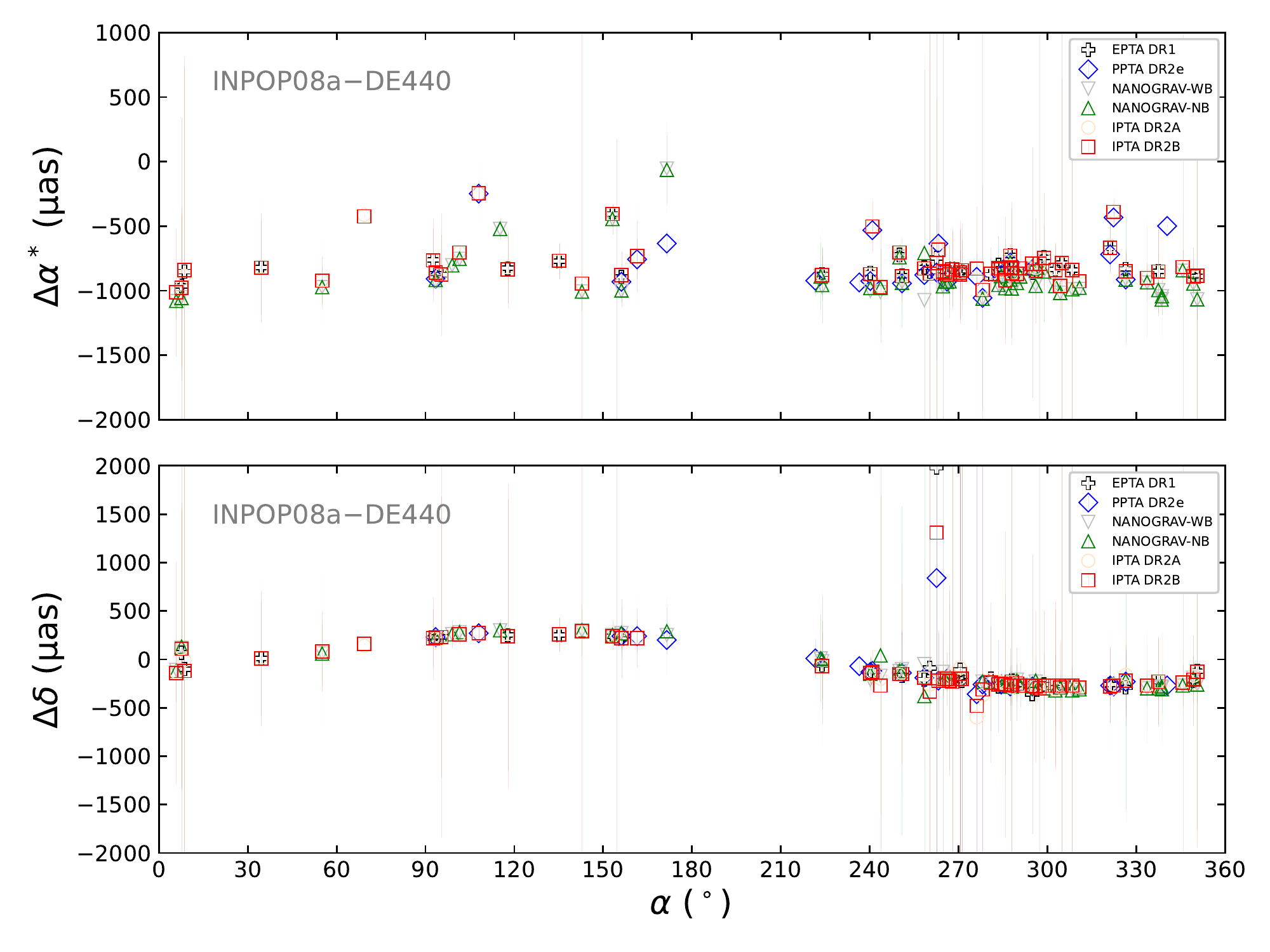}
  \includegraphics[width=\columnwidth]{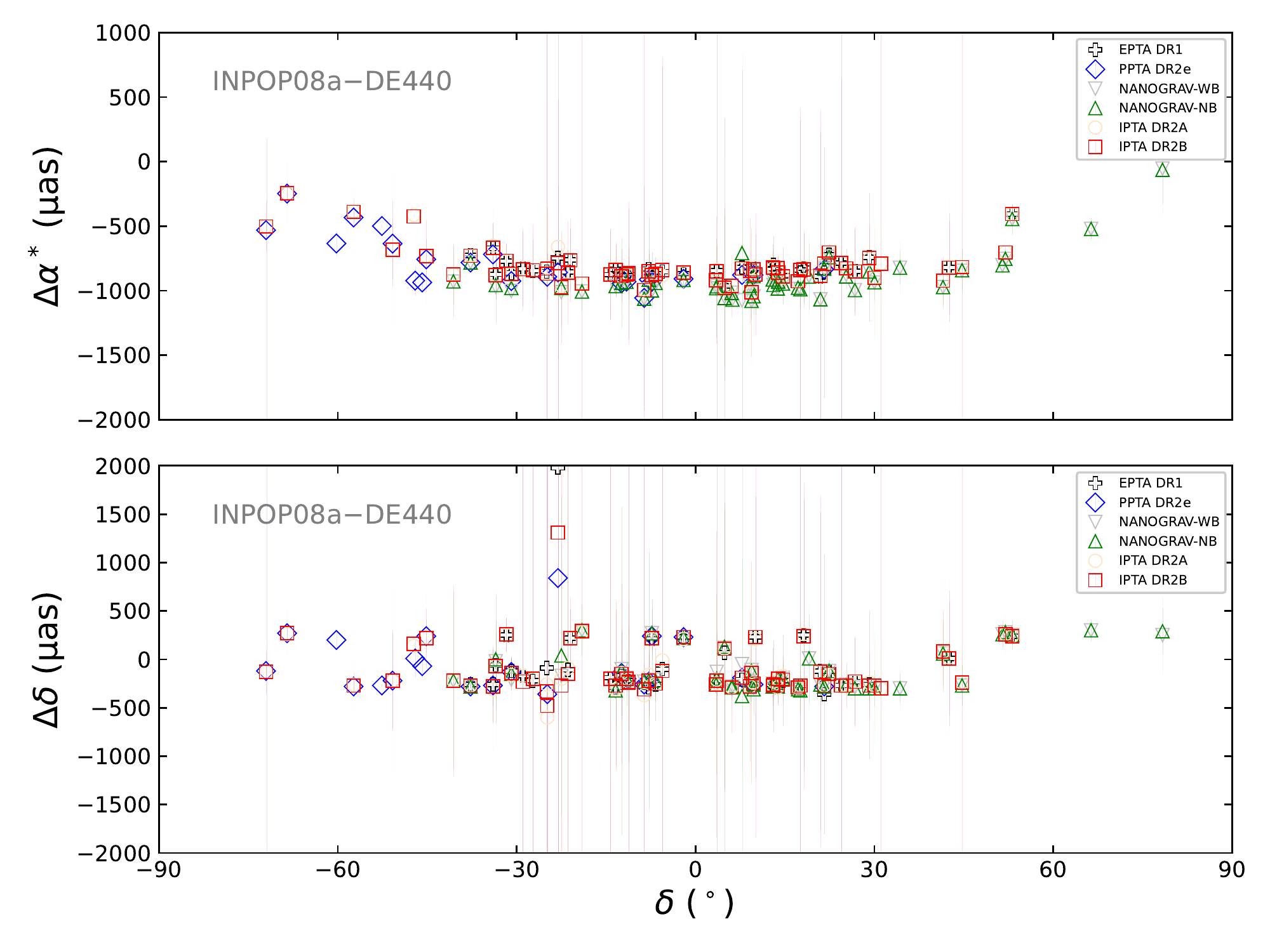}
  \caption[]{\label{fig:equ-pos-inpop08a-vs-de440} %
  Offsets of the pulsar timing positions in the INPOP08a frame with referred to those in the DE440 frame as a function of the right ascension (left) and declination (right).
}
\end{figure*}

\begin{figure*} 
  \includegraphics[width=\columnwidth]{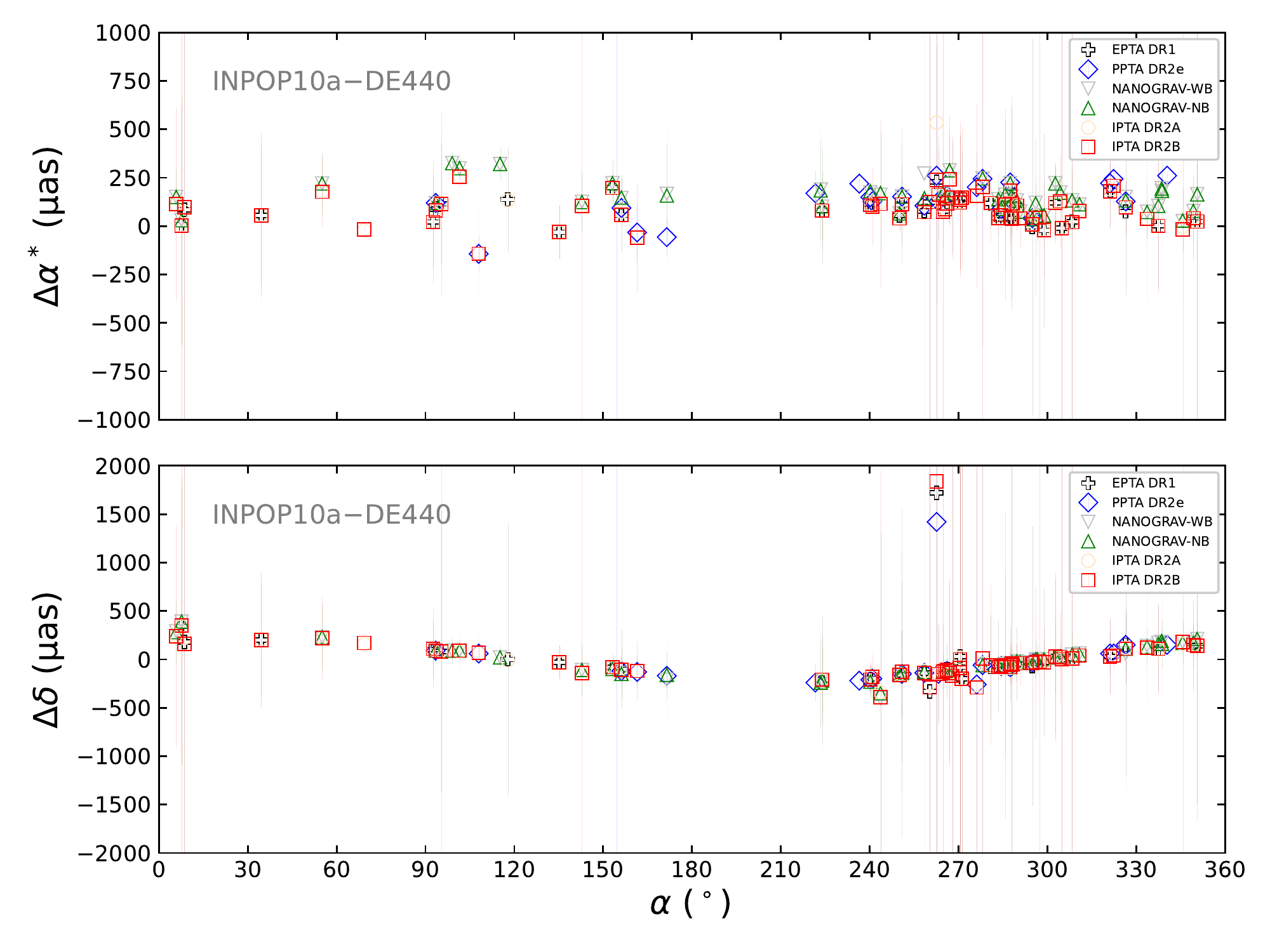}
  \includegraphics[width=\columnwidth]{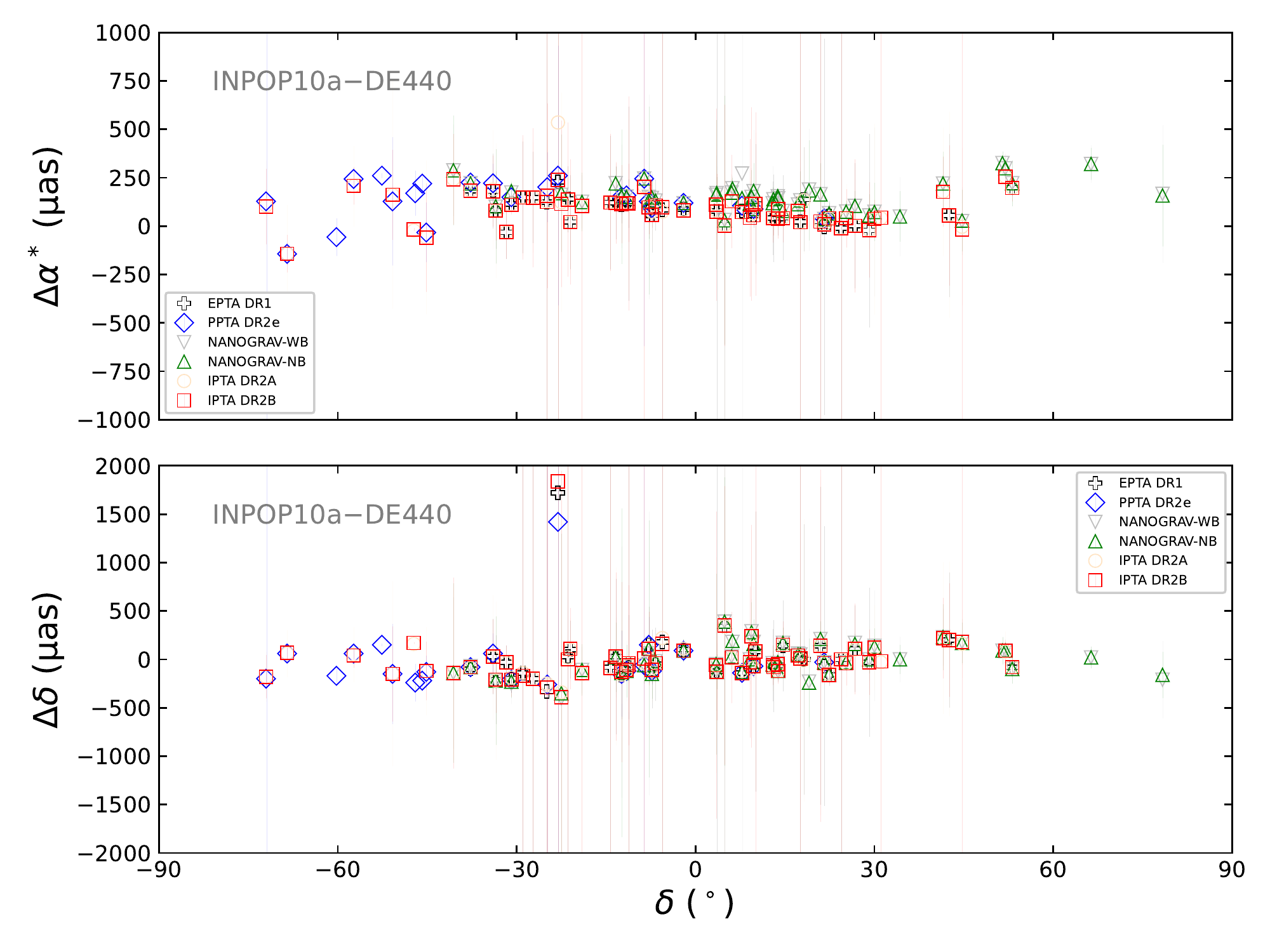}
  \caption[]{\label{fig:equ-pos-inpop10a-vs-de440} %
  Offsets of the pulsar timing positions in the INPOP10a frame with referred to those in the DE440 frame as a function of the right ascension (left) and declination (right).
}
\end{figure*}

\begin{figure*} 
  \includegraphics[width=\columnwidth]{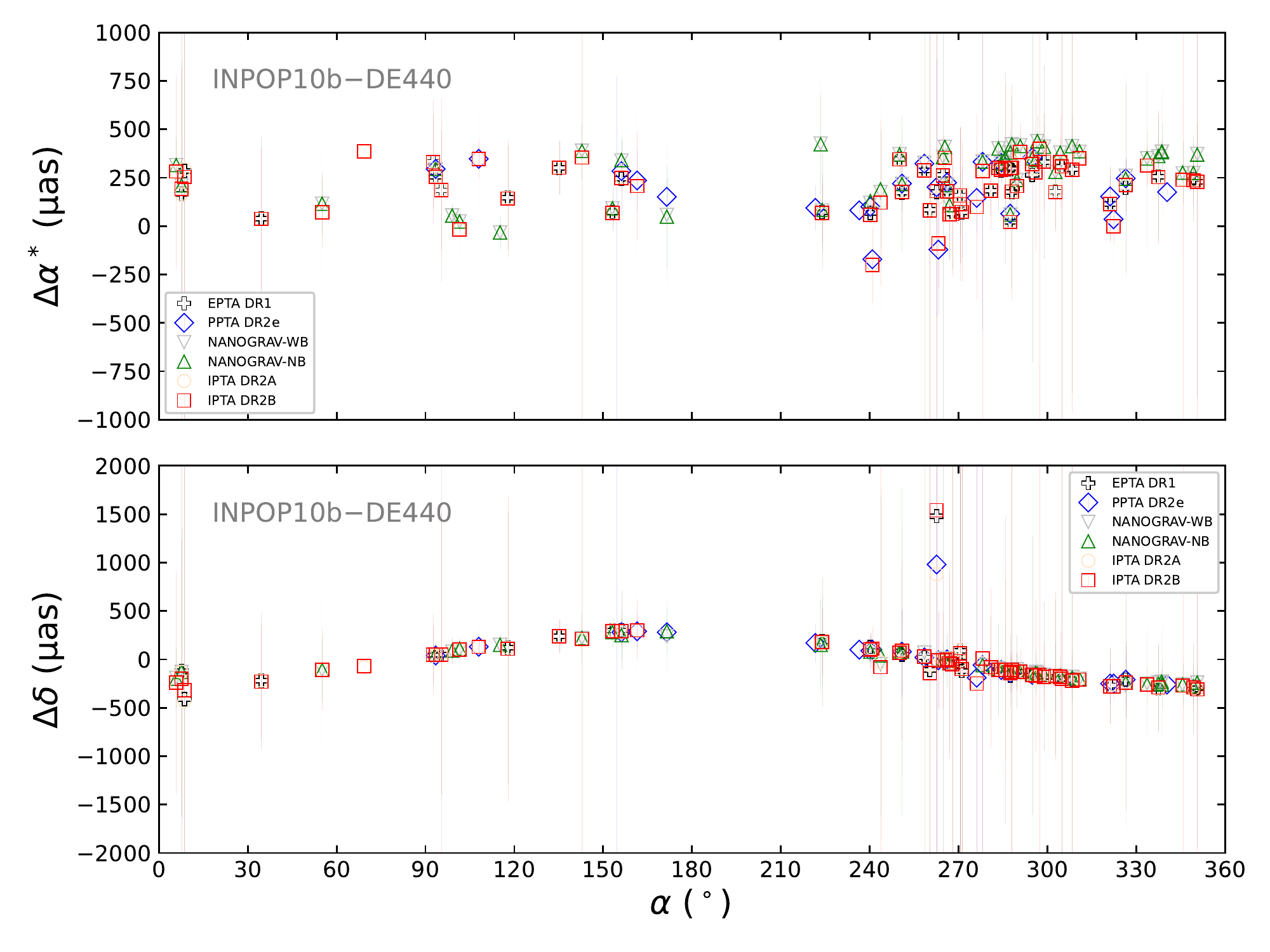}
  \includegraphics[width=\columnwidth]{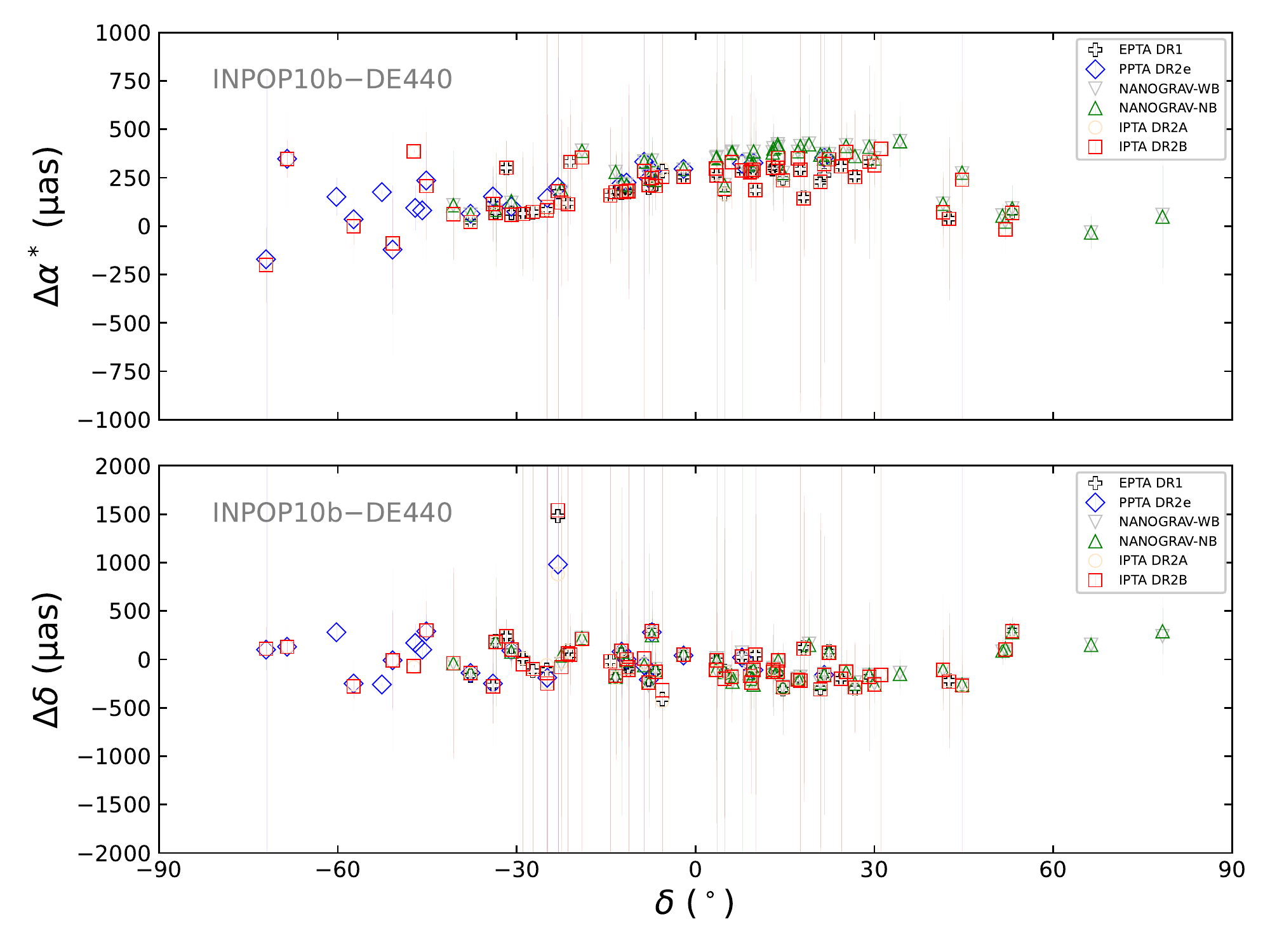}
  \caption[]{\label{fig:equ-pos-inpop10b-vs-de440} %
  Offsets of the pulsar timing positions in the INPOP10b frame with referred to those in the DE440 frame as a function of the right ascension (left) and declination (right).
}
\end{figure*}

\FloatBarrier

\begin{figure*} 
  \includegraphics[width=\columnwidth]{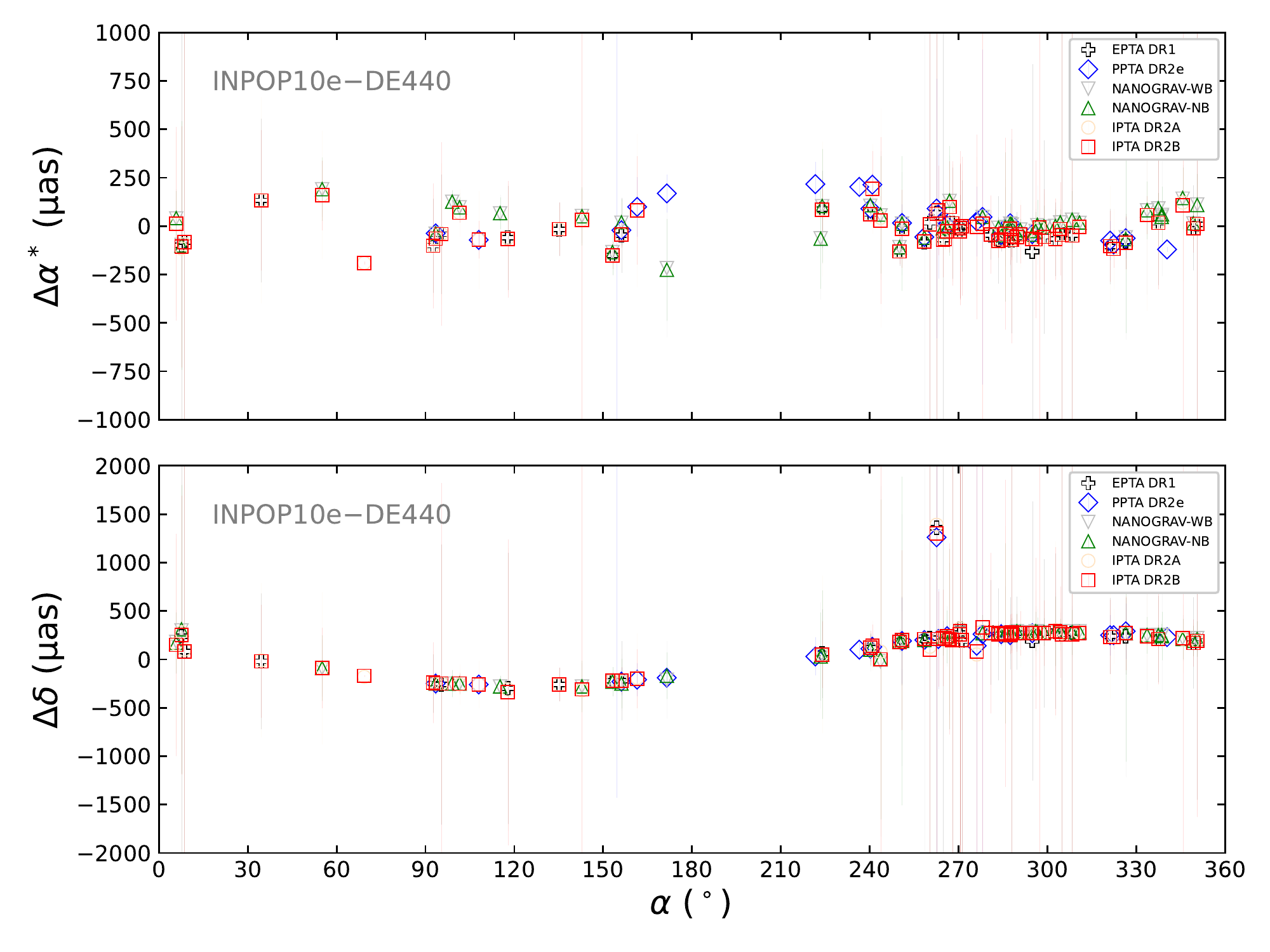}
  \includegraphics[width=\columnwidth]{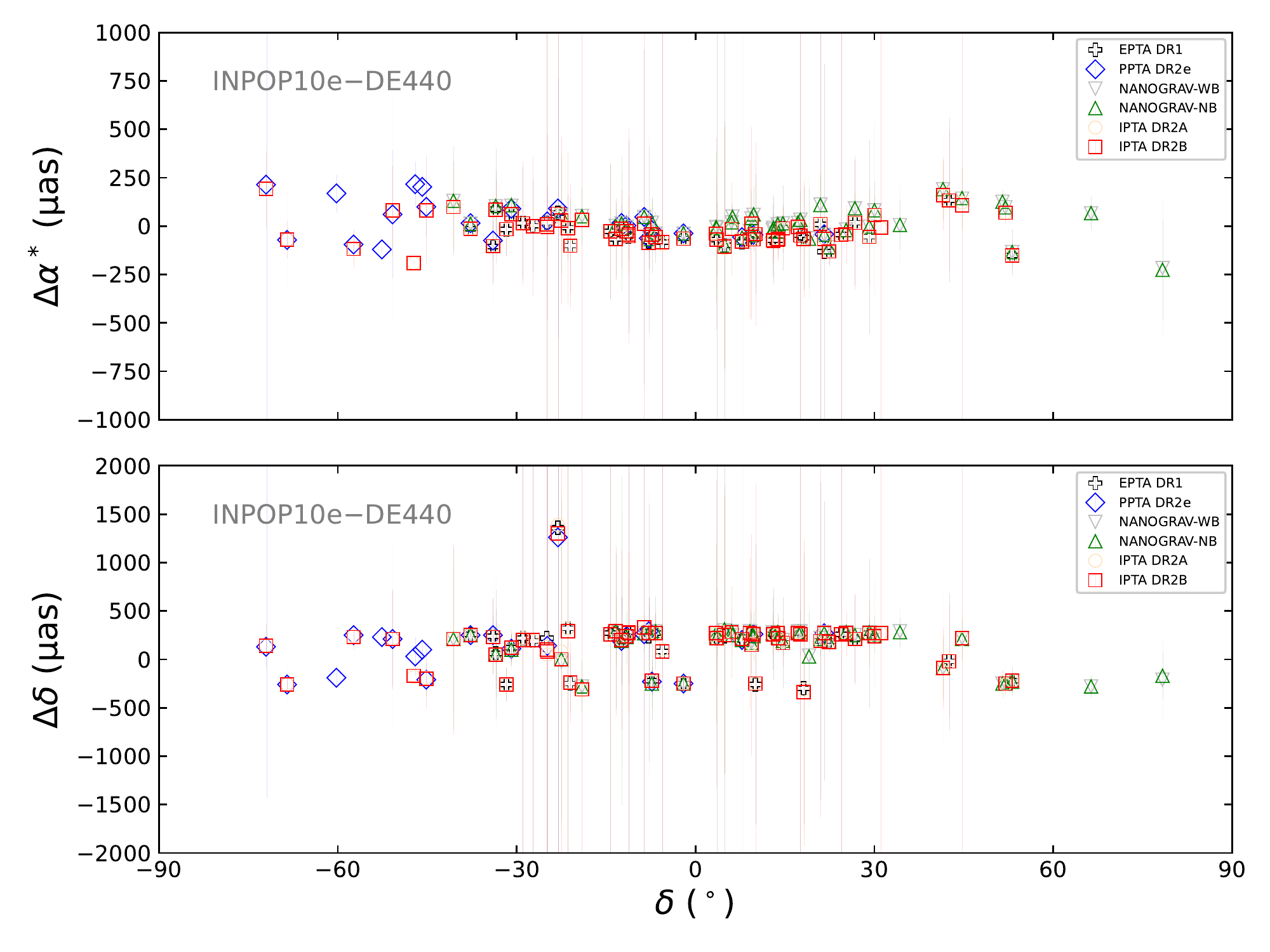}
  \caption[]{\label{fig:equ-pos-inpop10e-vs-de440} %
  Offsets of the pulsar timing positions in the INPOP10e frame with referred to those in the DE440 frame as a function of the right ascension (left) and declination (right).
}
\end{figure*}

\begin{figure*} 
  \includegraphics[width=\columnwidth]{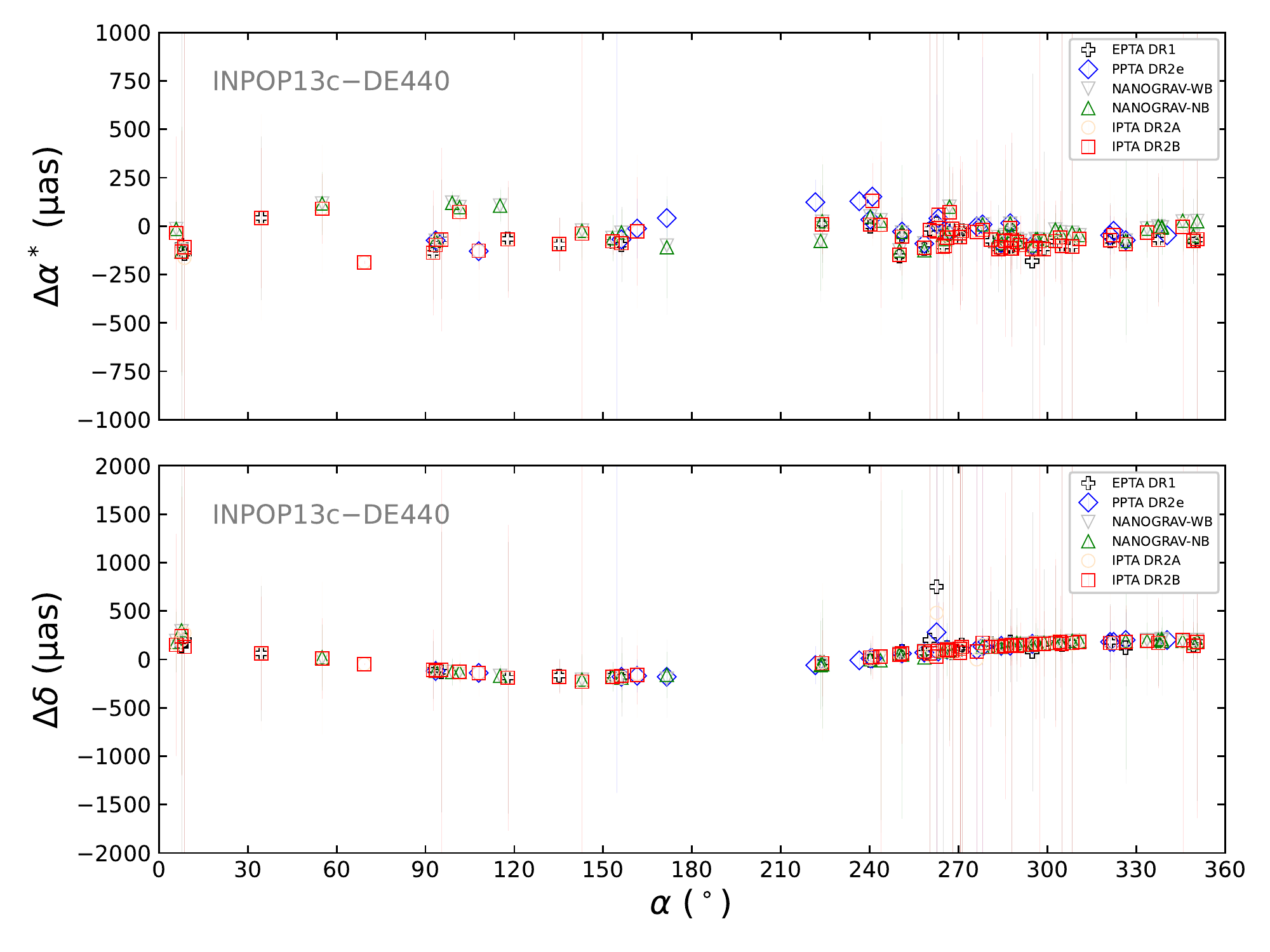}
  \includegraphics[width=\columnwidth]{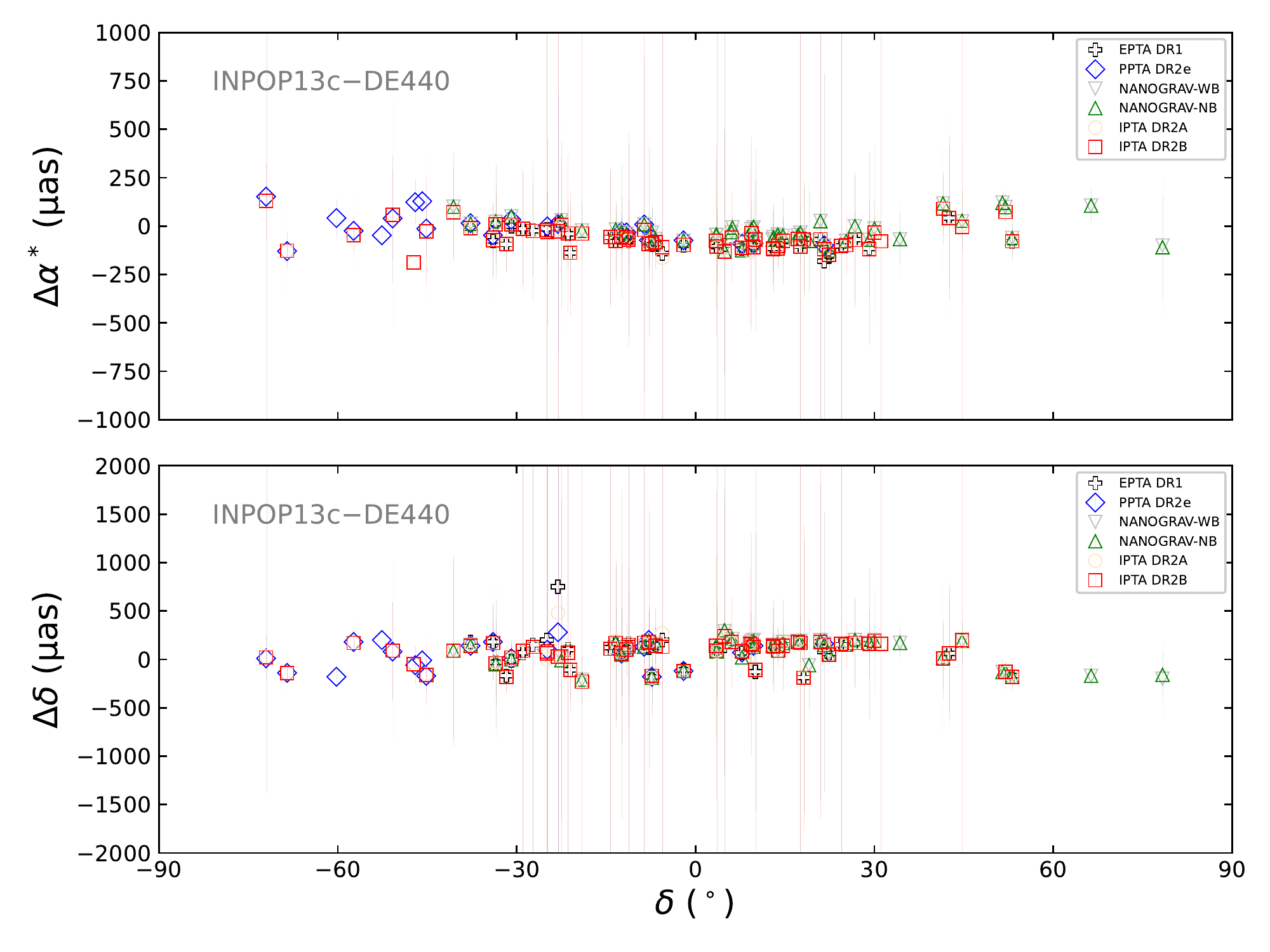}
  \caption[]{\label{fig:equ-pos-inpop13c-vs-de440} %
  Offsets of the pulsar timing positions in the INPOP13c frame with referred to those in the DE440 frame as a function of the right ascension (left) and declination (right).
}
\end{figure*}

\begin{figure*} 
  \includegraphics[width=\columnwidth]{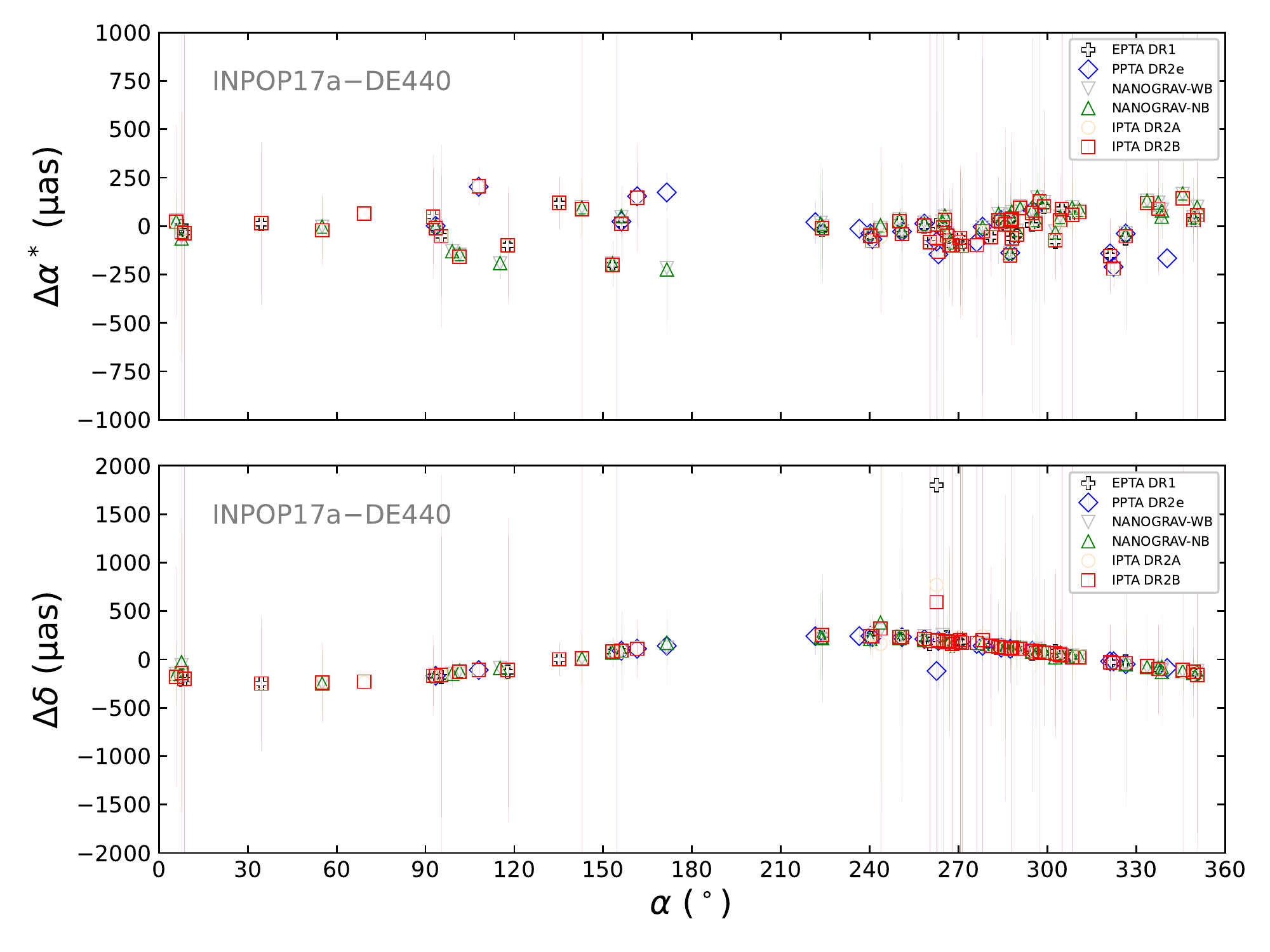}
  \includegraphics[width=\columnwidth]{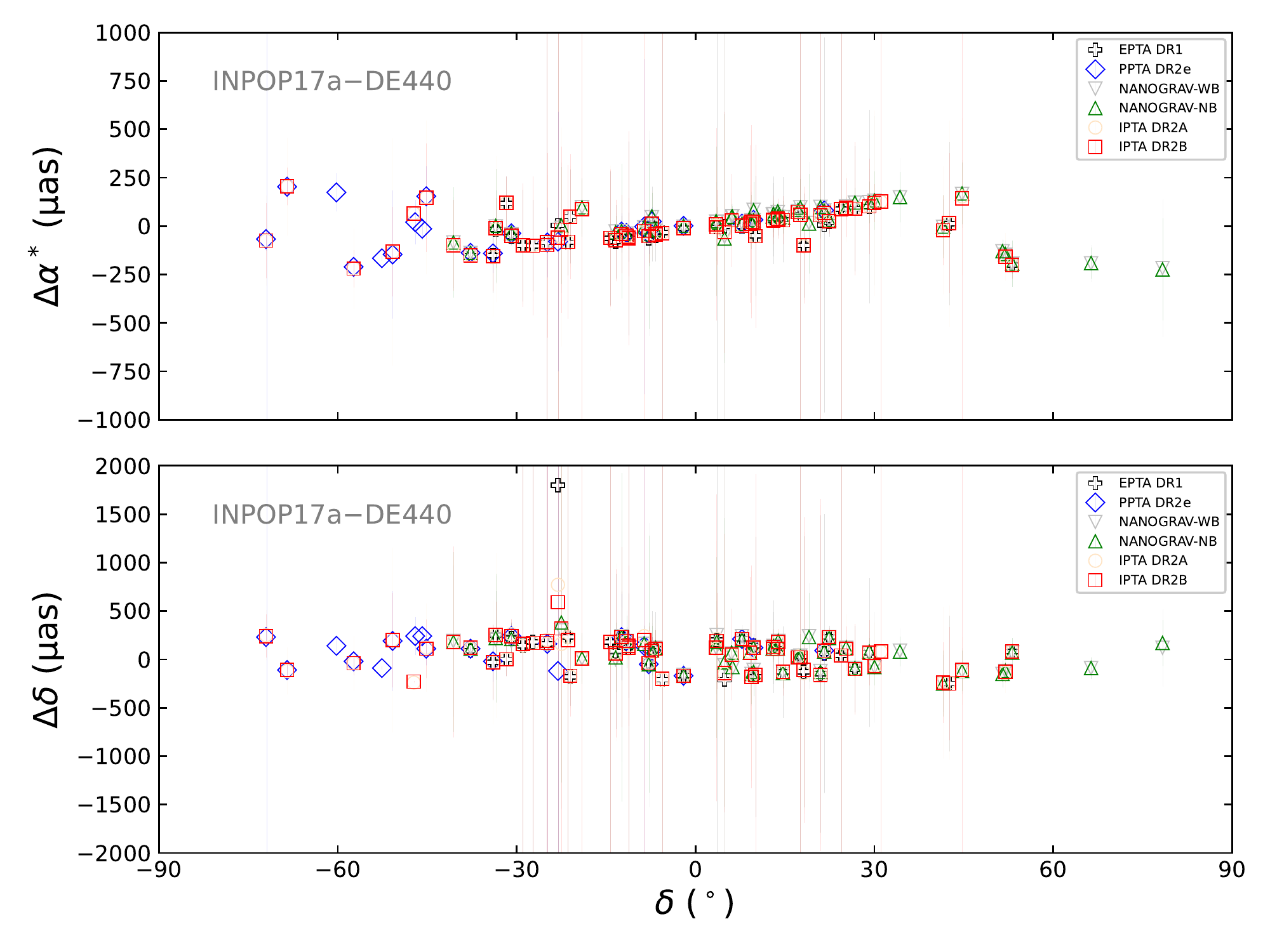}
  \caption[]{\label{fig:equ-pos-inpop17a-vs-de440} %
  Offsets of the pulsar timing positions in the INPOP17a frame with referred to those in the DE440 frame as a function of the right ascension (left) and declination (right).
}
\end{figure*}

\FloatBarrier

\begin{figure*} 
  \includegraphics[width=\columnwidth]{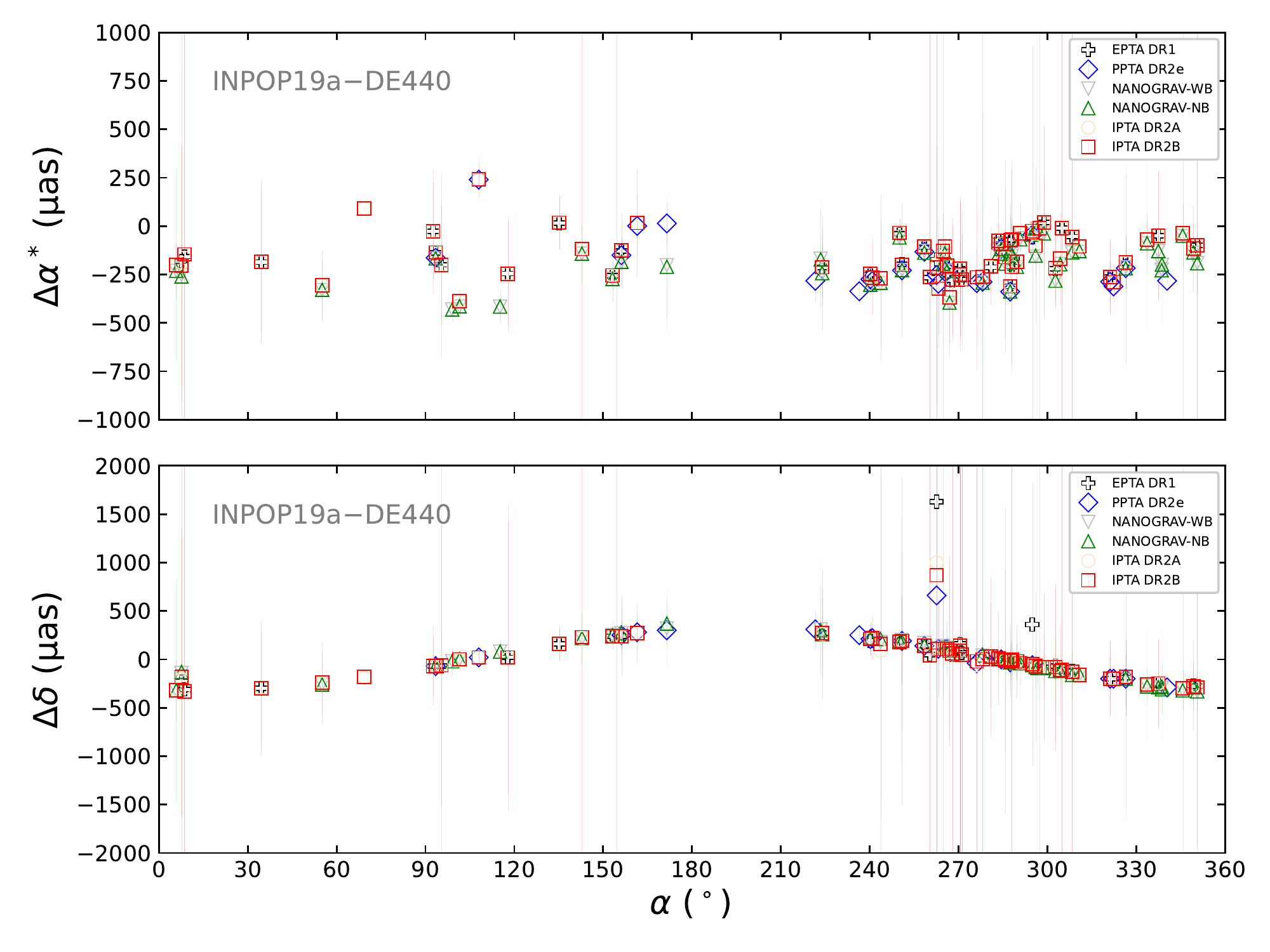}
  \includegraphics[width=\columnwidth]{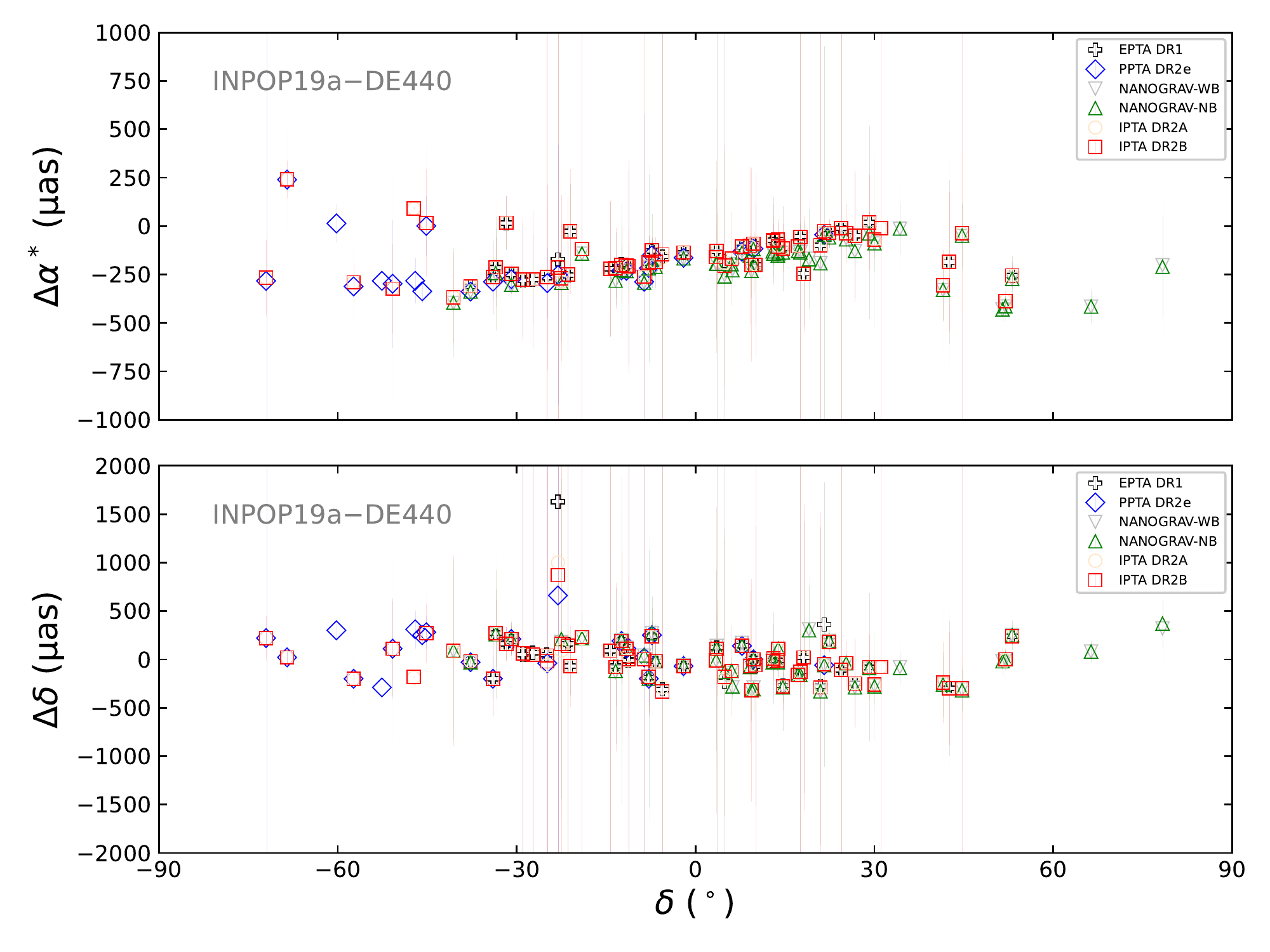}
  \caption[]{\label{fig:equ-pos-inpop19a-vs-de440} %
  Offsets of the pulsar timing positions in the INPOP19a frame with referred to those in the DE440 frame as a function of the right ascension (left) and declination (right).
}
\end{figure*}

\FloatBarrier

%% file: pm-diff-plot.tex
    \begin{figure*} [htbp]
      \includegraphics[width=\columnwidth]{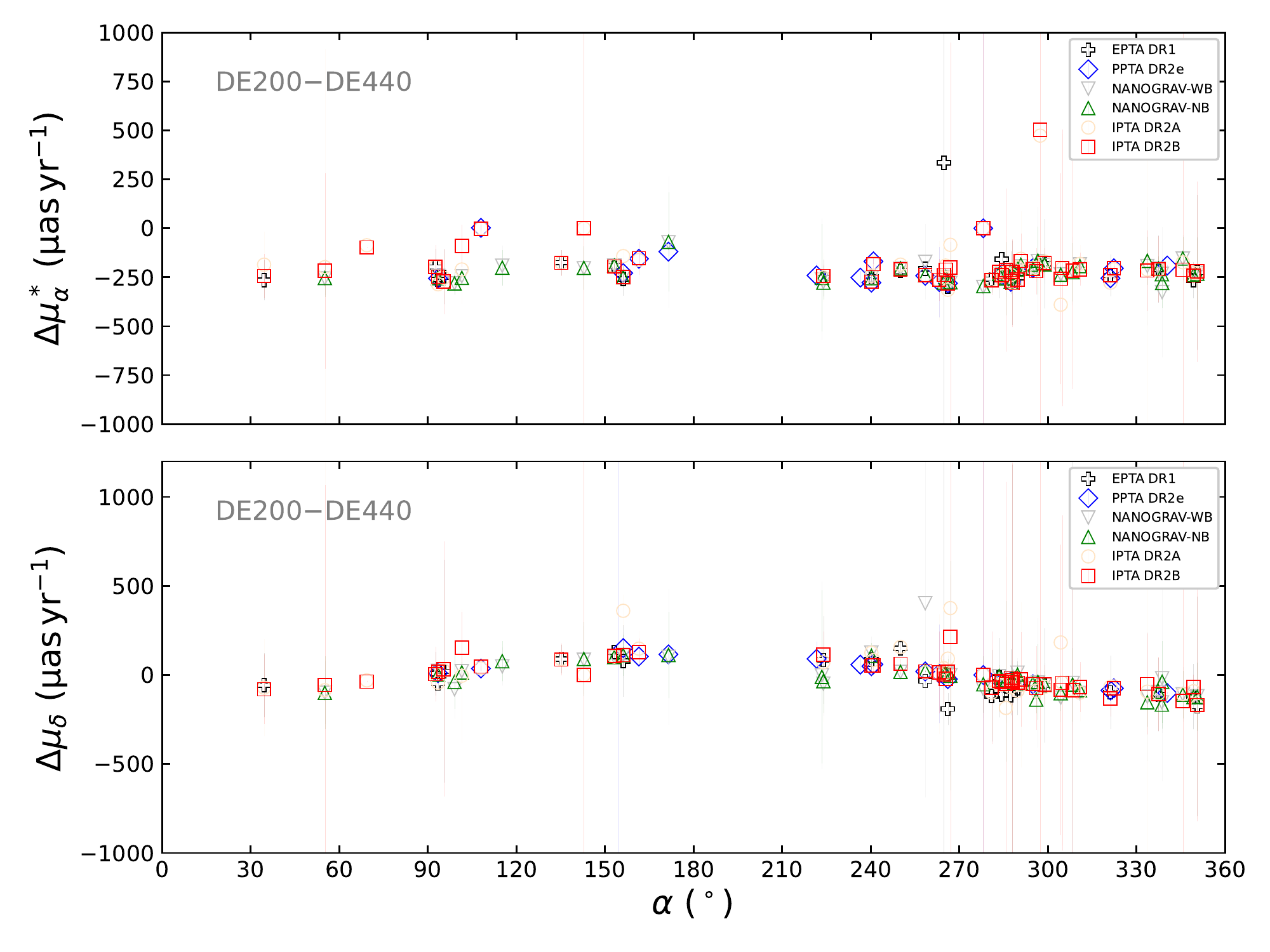}
      \includegraphics[width=\columnwidth]{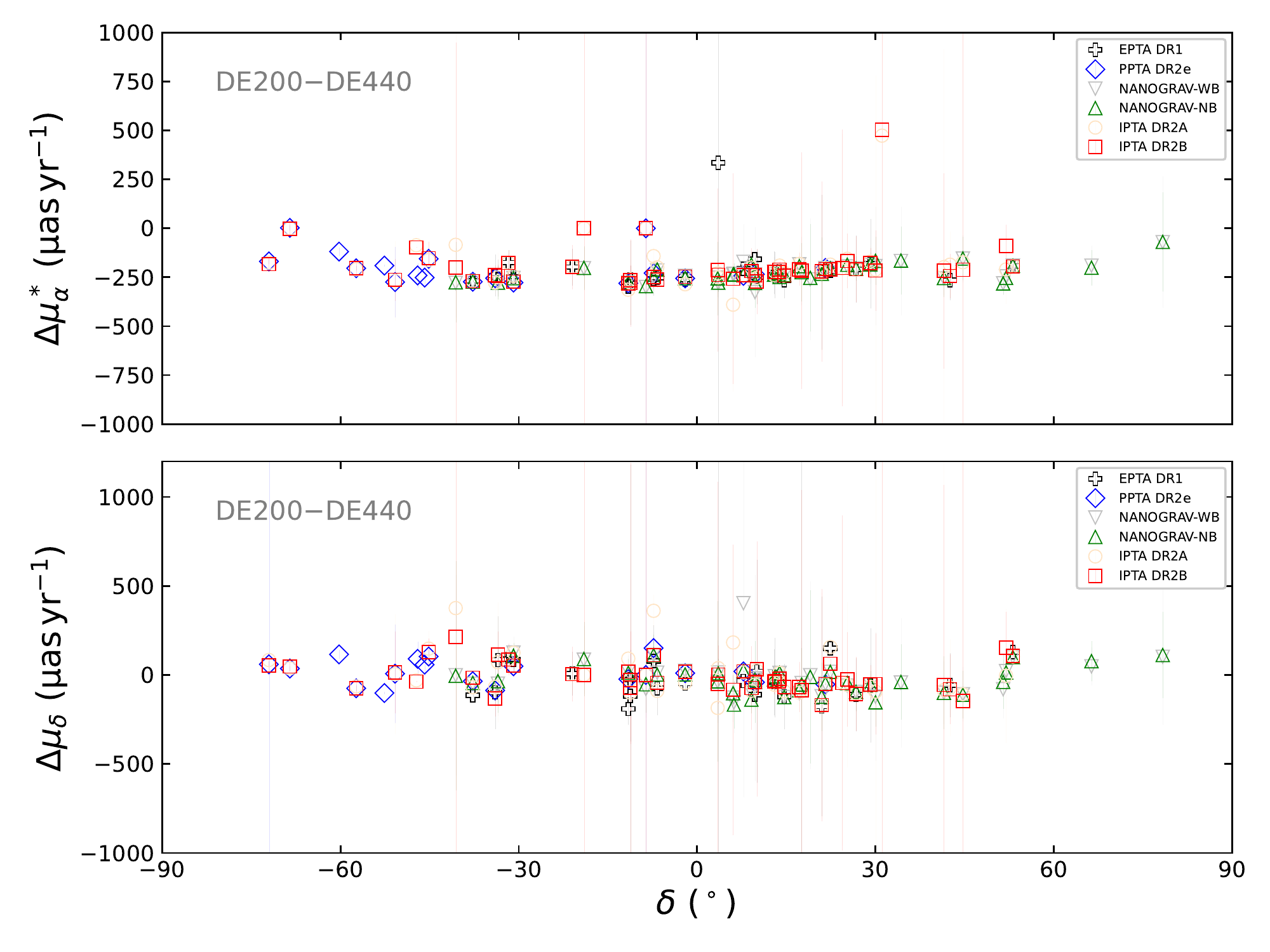}
      \caption[]{\label{fig:equ-pm-de200-vs-de440} %
      Offsets of the pulsar timing proper motions in the DE200 frame with referred to those in the DE440 frame as a function of the right ascension (left) and declination (right).
    }
    \end{figure*}

    \begin{figure*} 
      \includegraphics[width=\columnwidth]{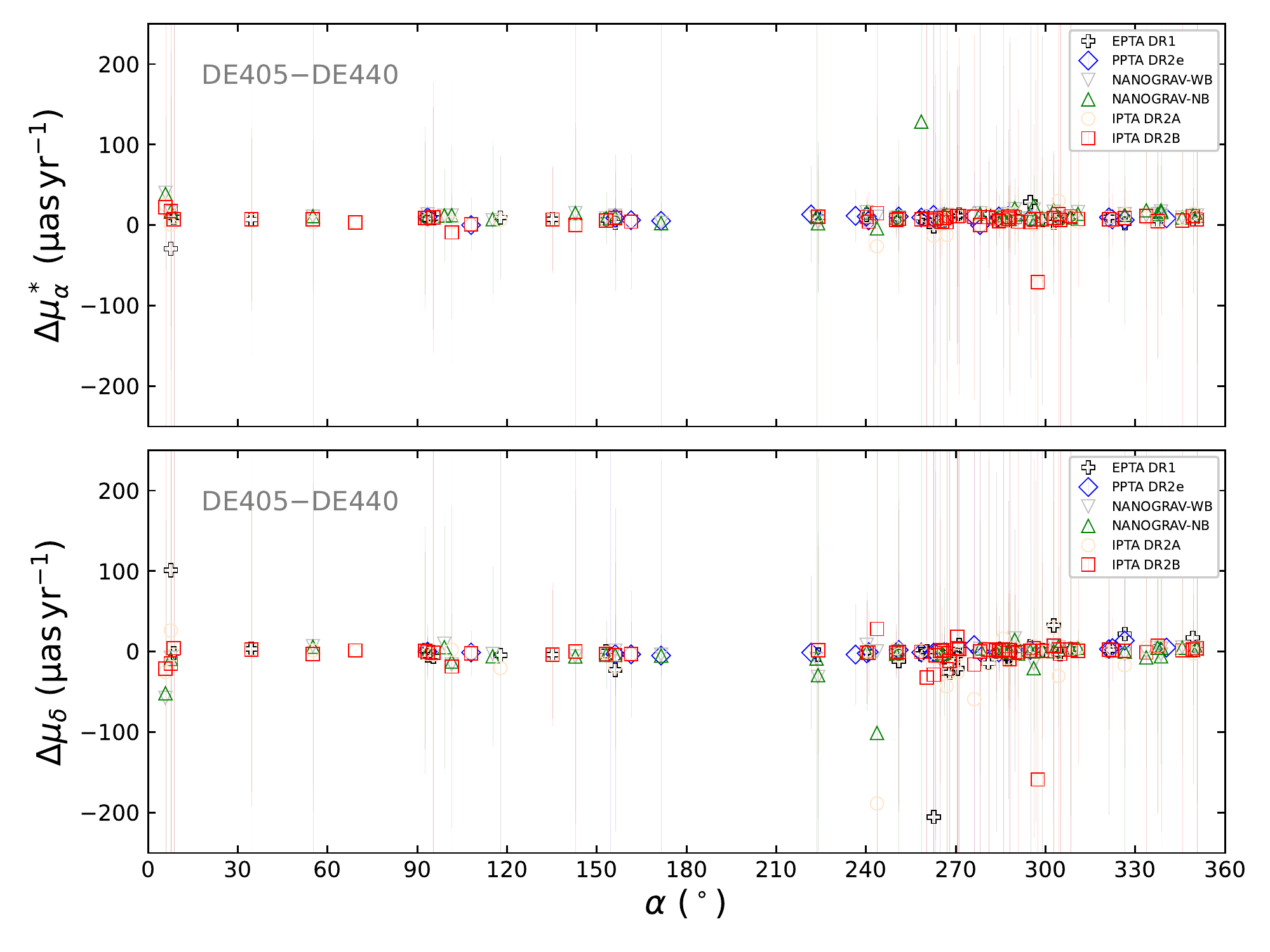}
      \includegraphics[width=\columnwidth]{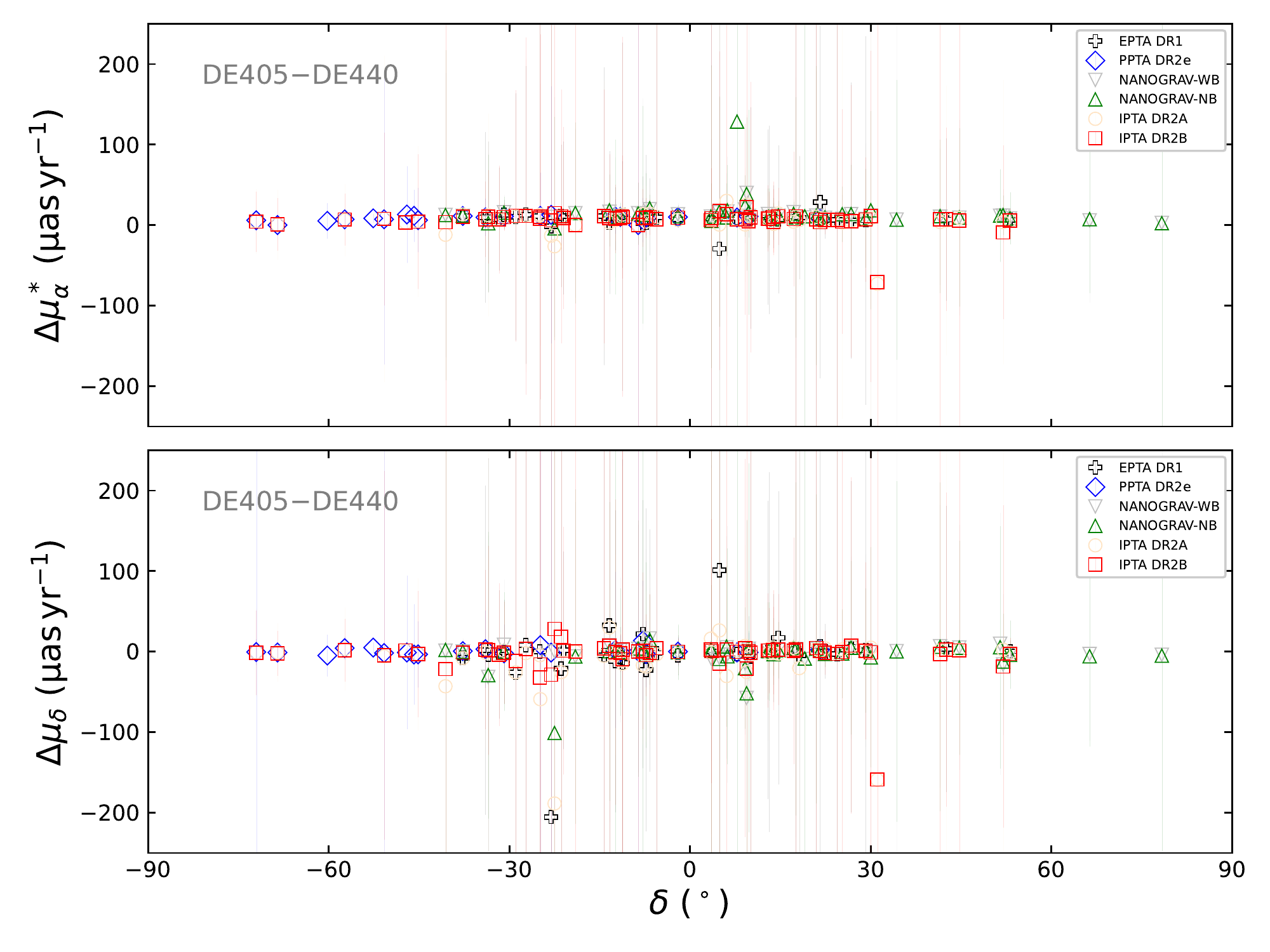}
      \caption[]{\label{fig:equ-pm-de405-vs-de440} %
      Offsets of the pulsar timing proper motions in the DE405 frame with referred to those in the DE440 frame as a function of the right ascension (left) and declination (right).
    }
    \end{figure*}

    \begin{figure*} 
      \includegraphics[width=\columnwidth]{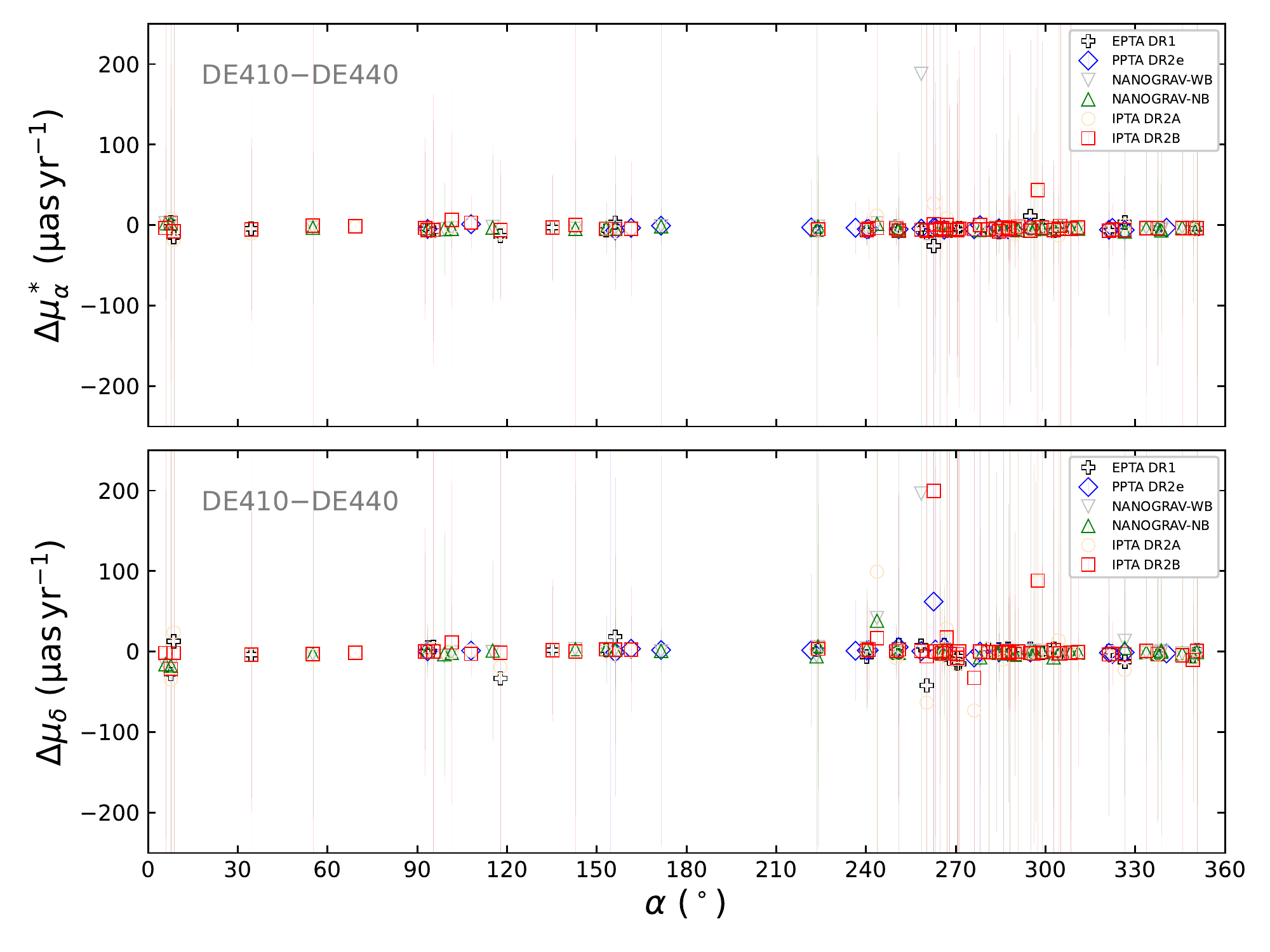}
      \includegraphics[width=\columnwidth]{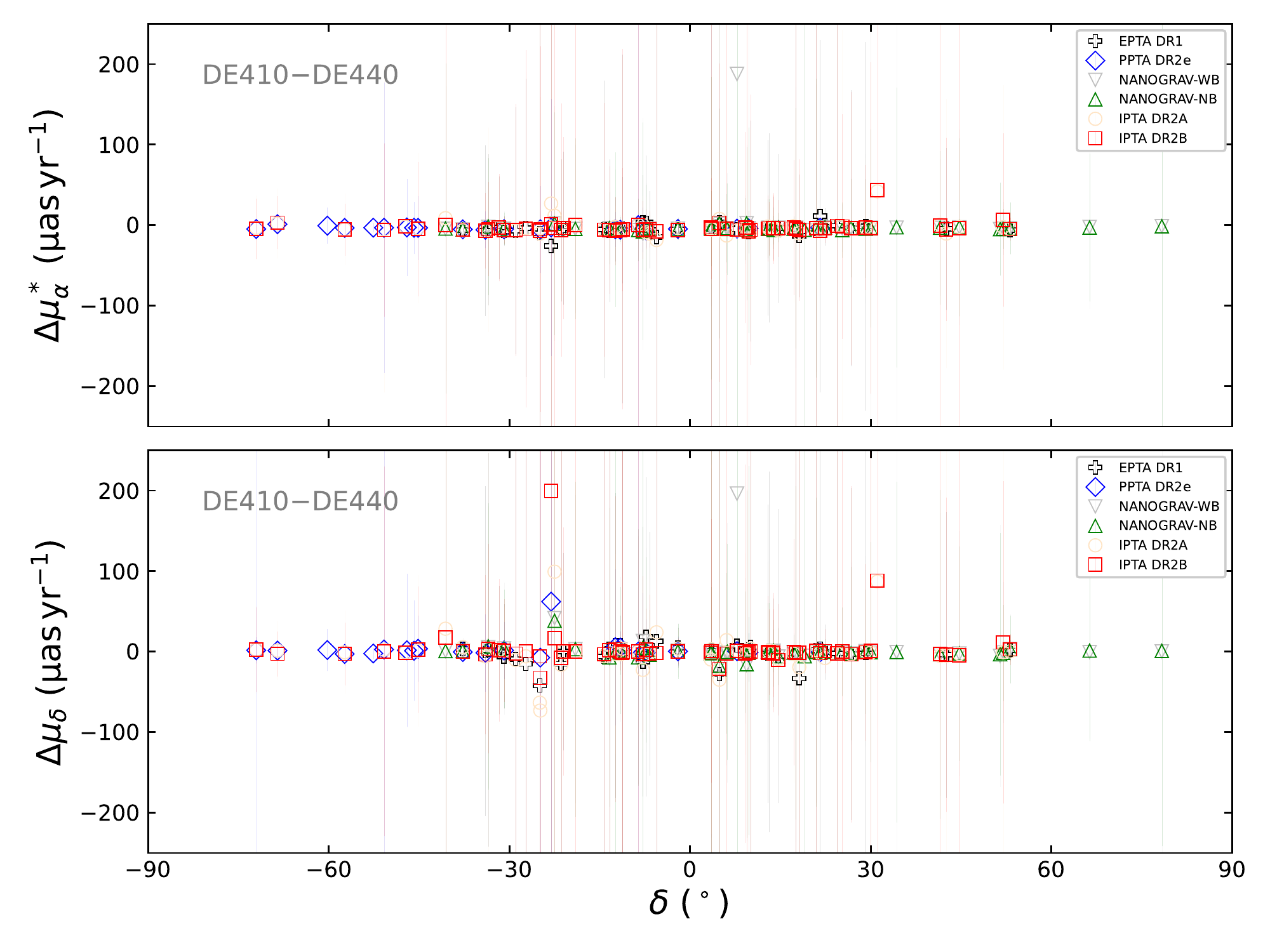}
      \caption[]{\label{fig:equ-pm-de410-vs-de440} %
      Offsets of the pulsar timing proper motions in the DE410 frame with referred to those in the DE440 frame as a function of the right ascension (left) and declination (right).
    }
    \end{figure*}

\FloatBarrier

    \begin{figure*} 
      \includegraphics[width=\columnwidth]{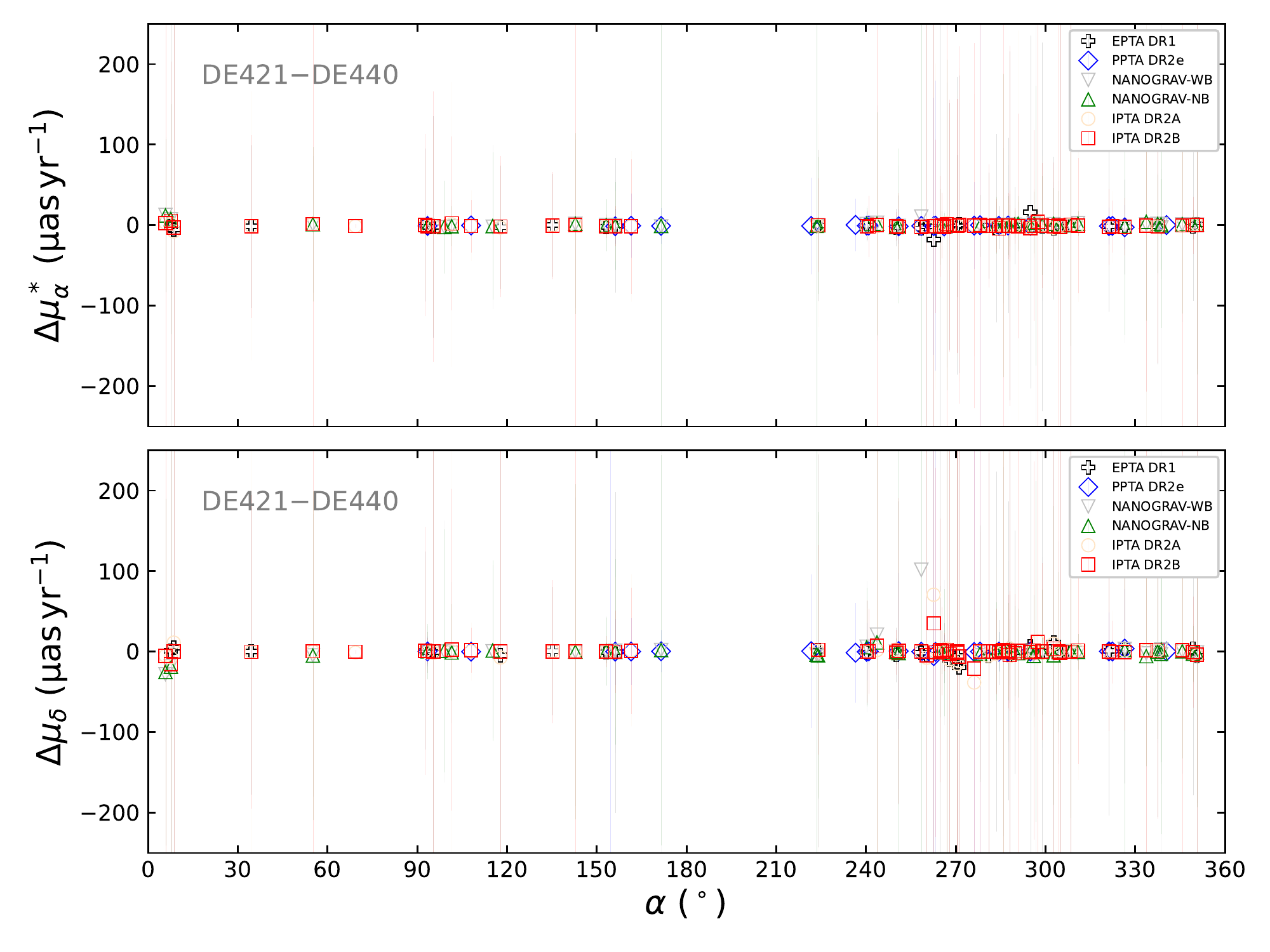}
      \includegraphics[width=\columnwidth]{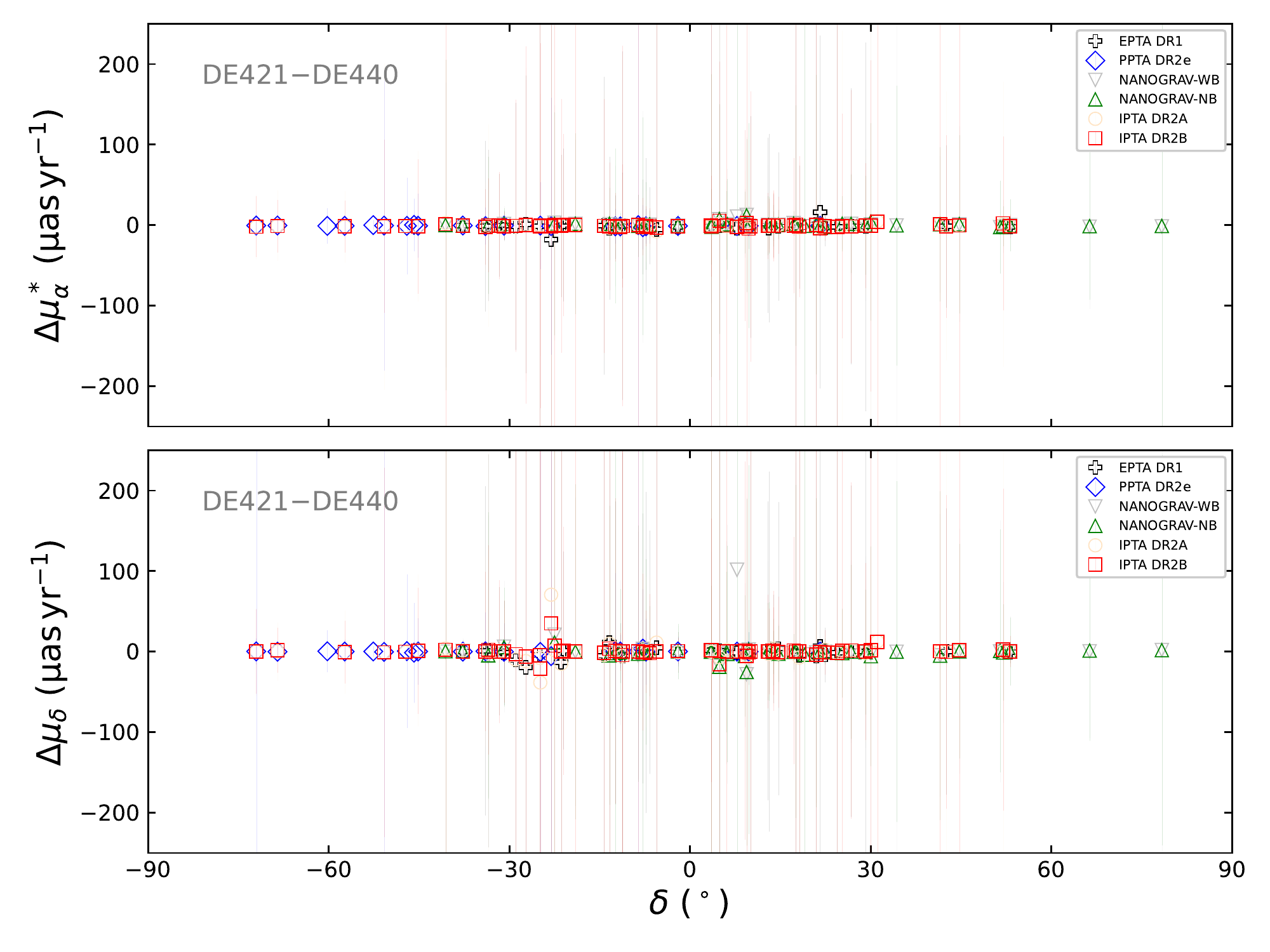}
      \caption[]{\label{fig:equ-pm-de421-vs-de440} %
      Offsets of the pulsar timing proper motions in the DE421 frame with referred to those in the DE440 frame as a function of the right ascension (left) and declination (right).
    }
    \end{figure*}

    \begin{figure*} 
      \includegraphics[width=\columnwidth]{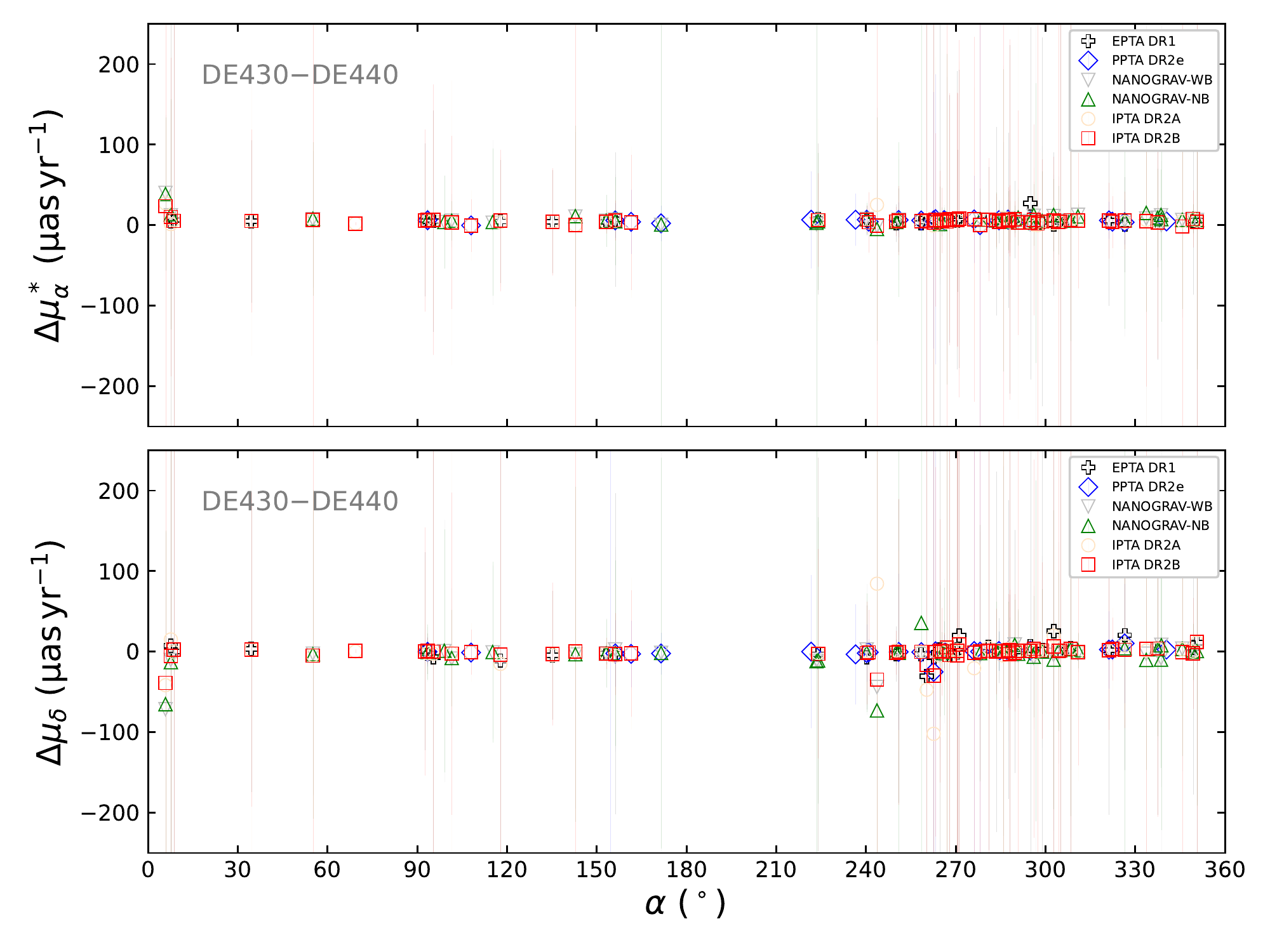}
      \includegraphics[width=\columnwidth]{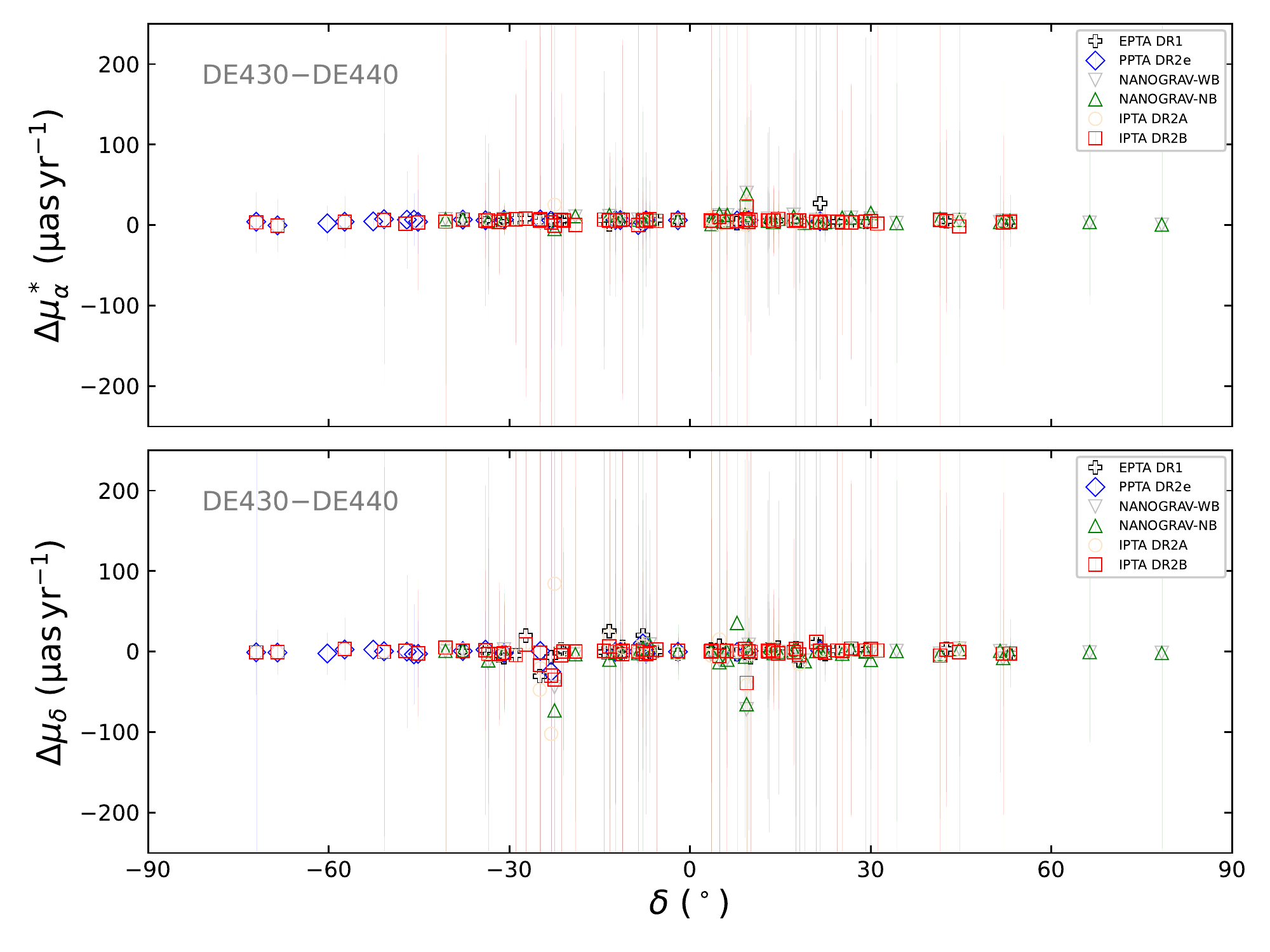}
      \caption[]{\label{fig:equ-pm-de430-vs-de440} %
      Offsets of the pulsar timing proper motions in the DE430 frame with referred to those in the DE440 frame as a function of the right ascension (left) and declination (right).
    }
    \end{figure*}

    \begin{figure*} 
      \includegraphics[width=\columnwidth]{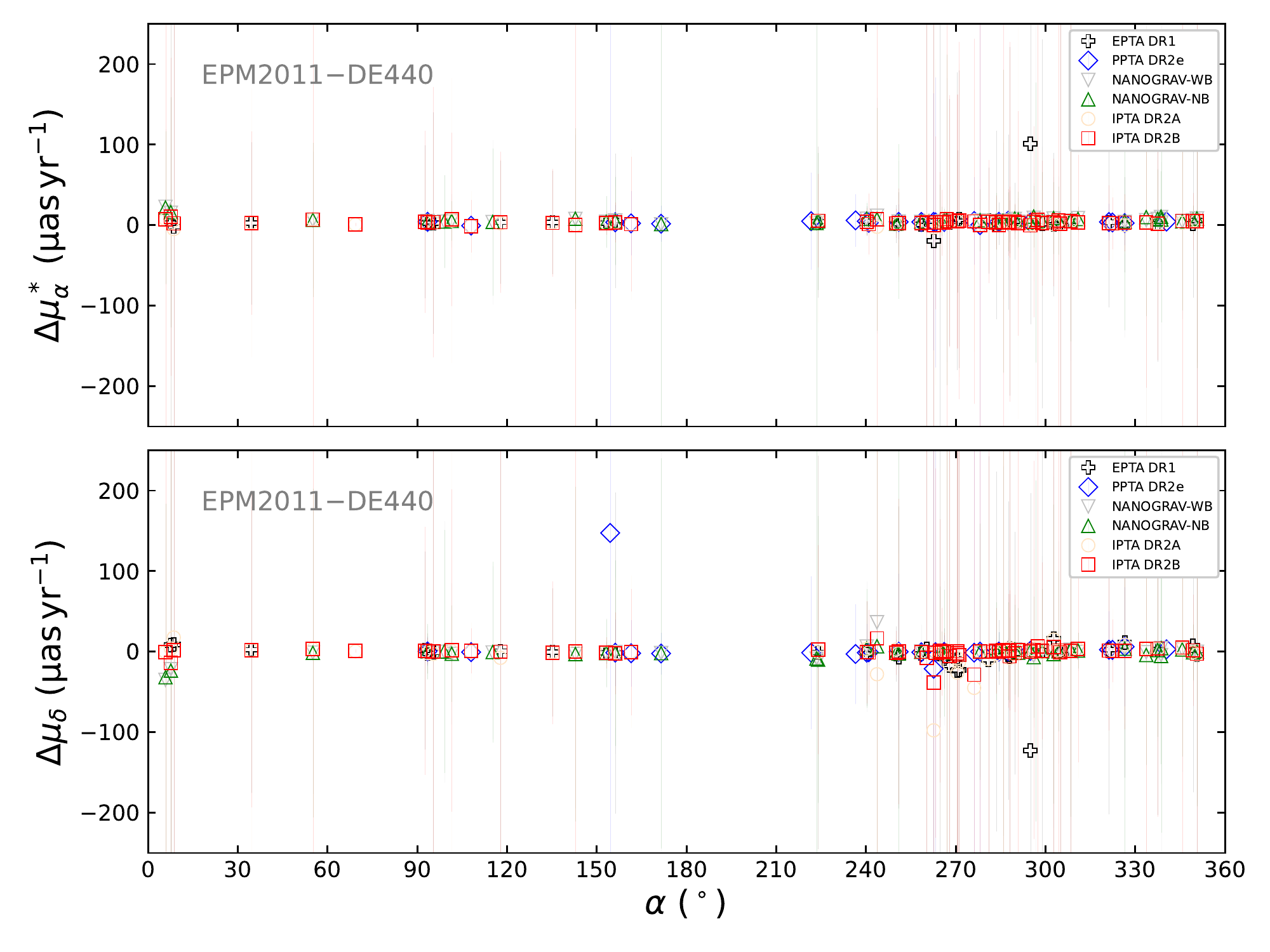}
      \includegraphics[width=\columnwidth]{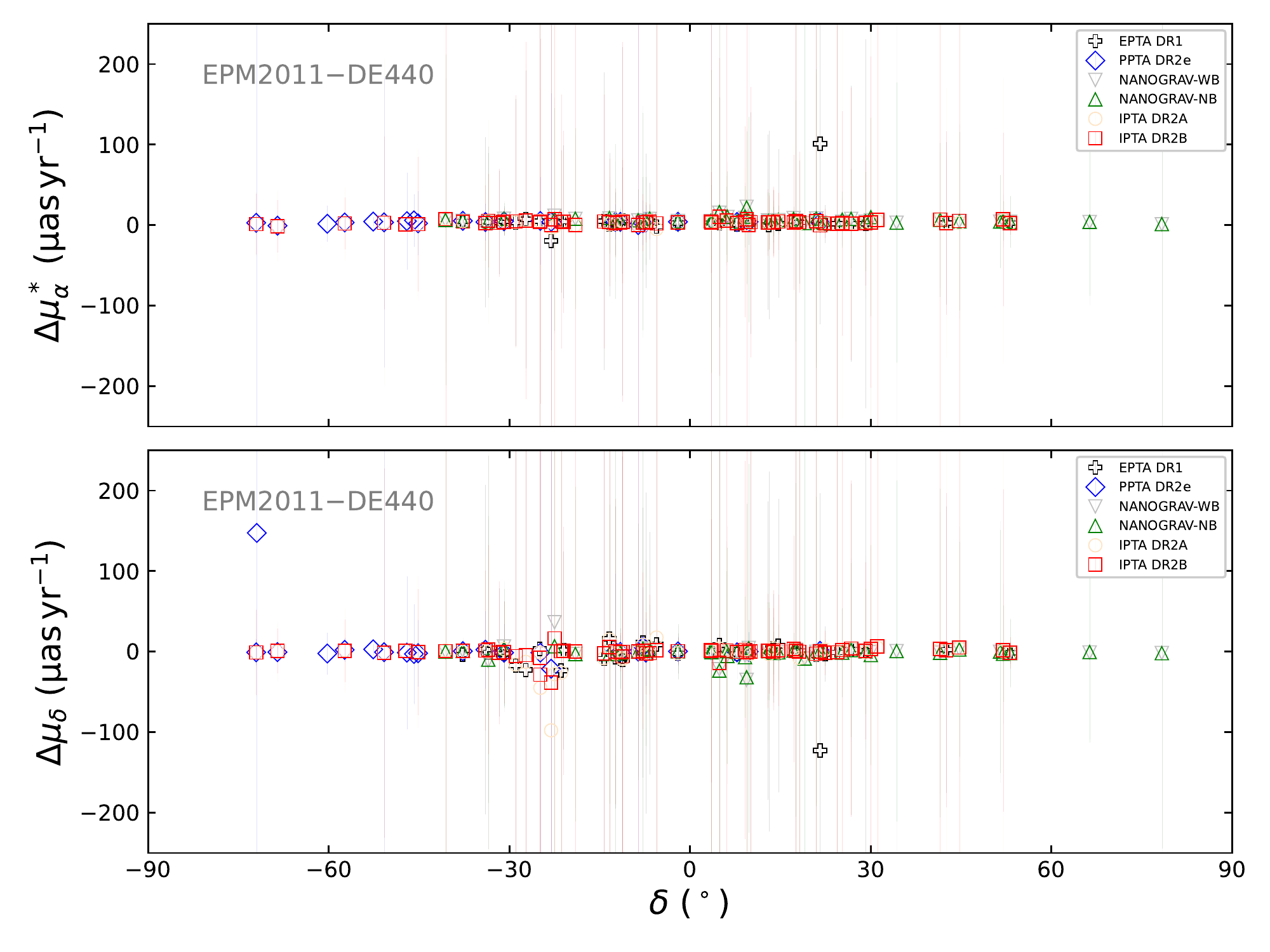}
      \caption[]{\label{fig:equ-pm-epm2011-vs-de440} %
      Offsets of the pulsar timing proper motions in the EPM2011 frame with referred to those in the DE440 frame as a function of the right ascension (left) and declination (right).
    }
    \end{figure*}

\FloatBarrier

    \begin{figure*} 
      \includegraphics[width=\columnwidth]{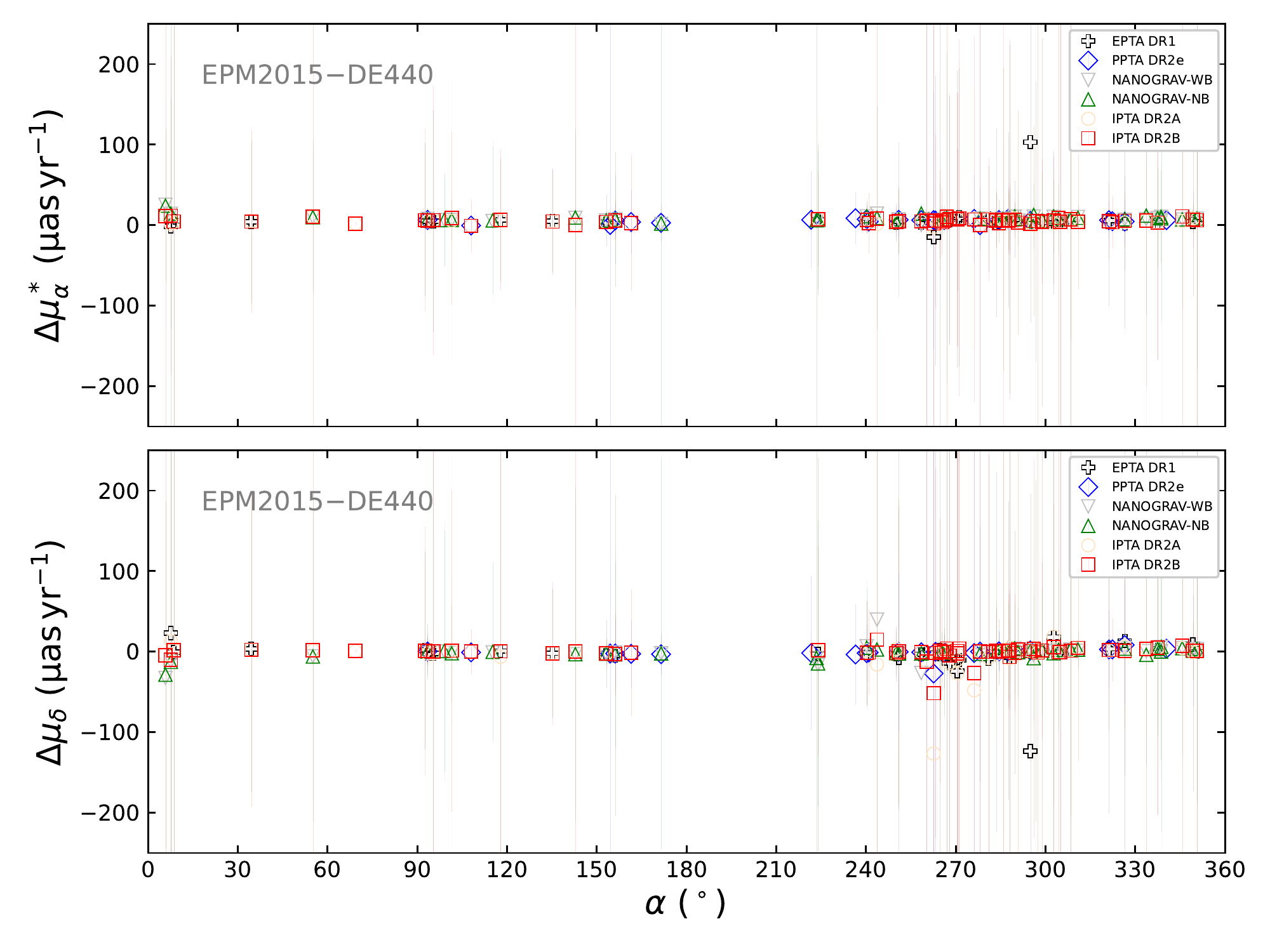}
      \includegraphics[width=\columnwidth]{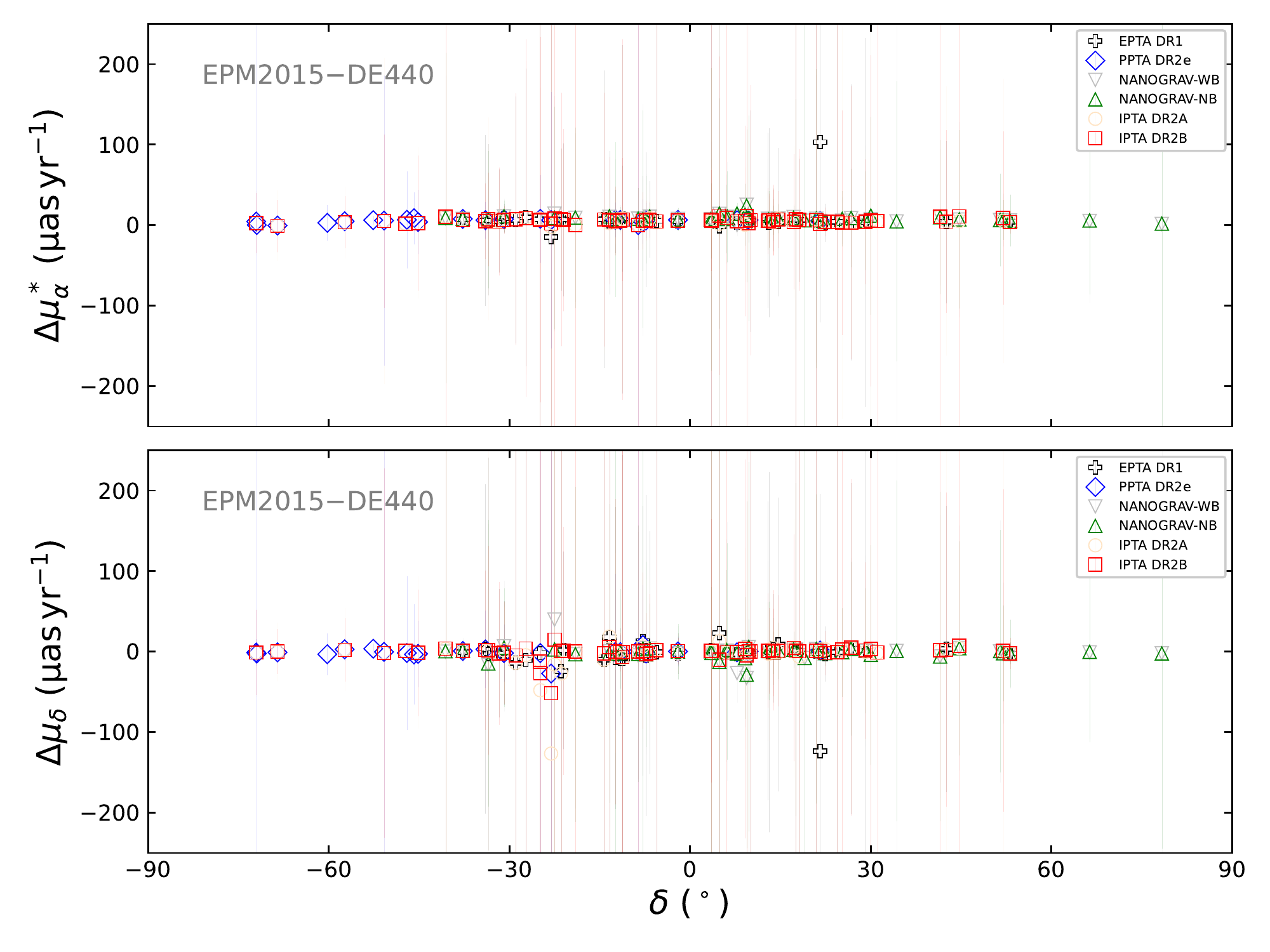}
      \caption[]{\label{fig:equ-pm-epm2015-vs-de440} %
      Offsets of the pulsar timing proper motions in the EPM2015 frame with referred to those in the DE440 frame as a function of the right ascension (left) and declination (right).
    }
    \end{figure*}

    \begin{figure*} 
      \includegraphics[width=\columnwidth]{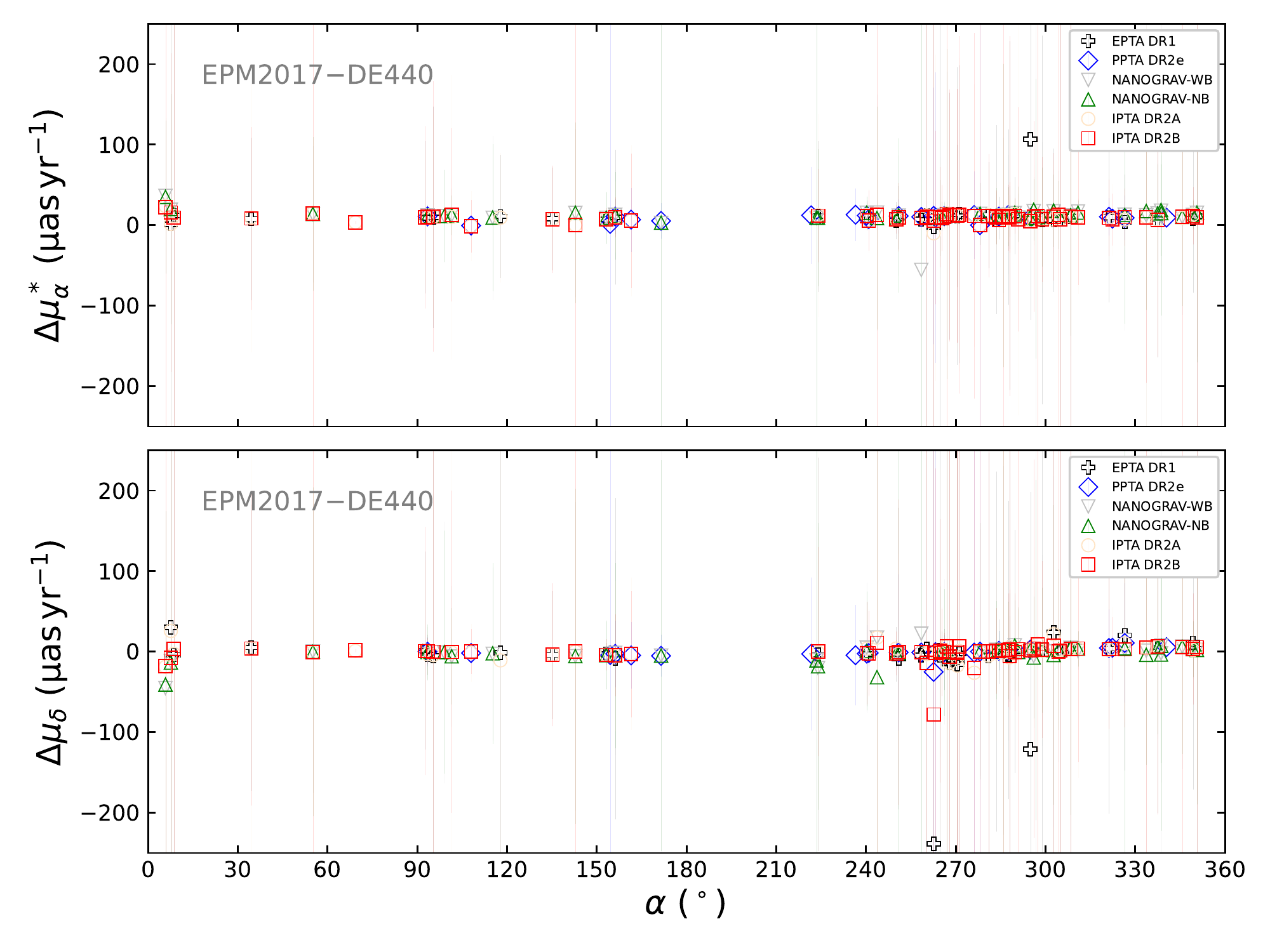}
      \includegraphics[width=\columnwidth]{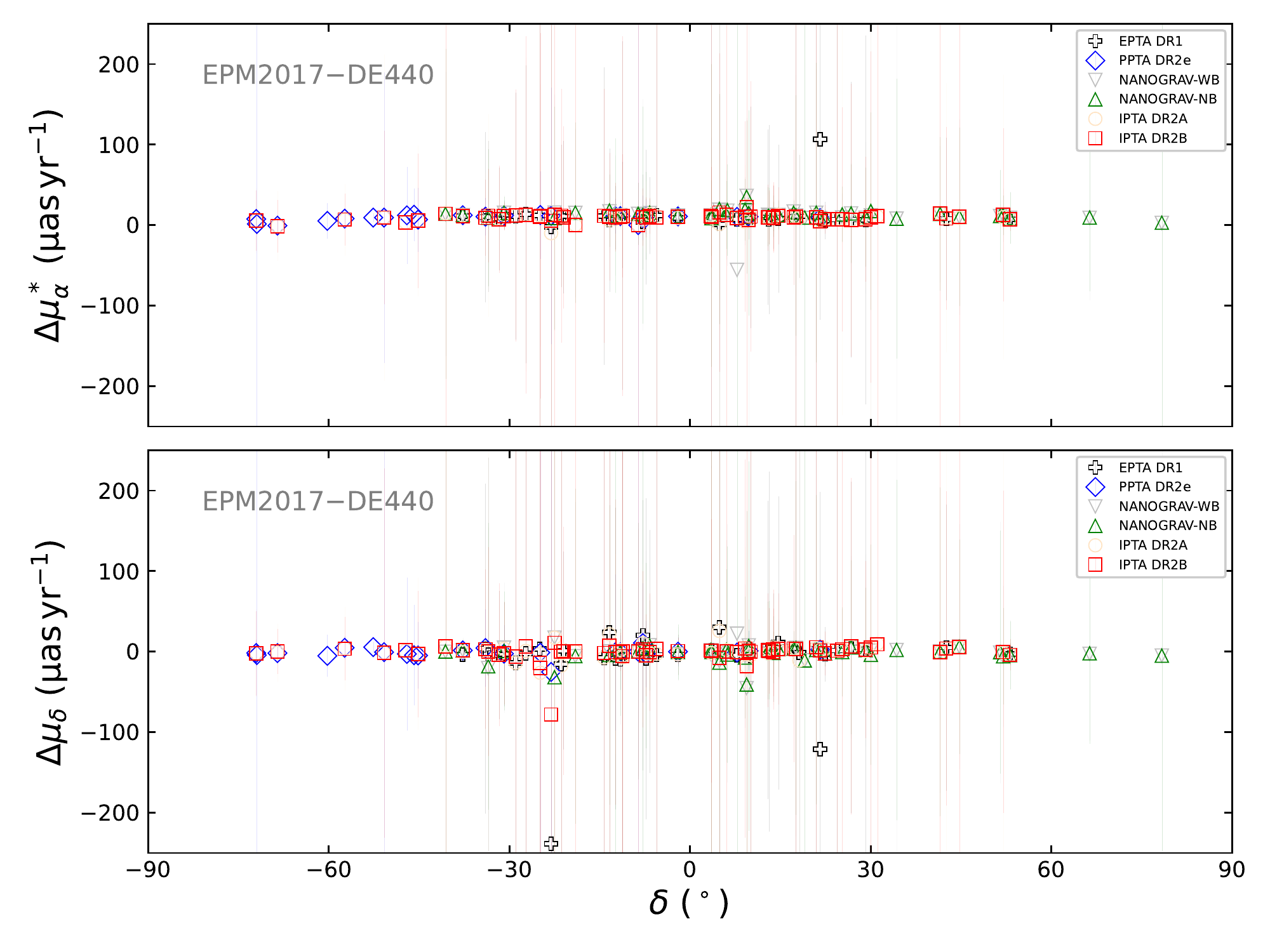}
      \caption[]{\label{fig:equ-pm-epm2017-vs-de440} %
      Offsets of the pulsar timing proper motions in the EPM2017 frame with referred to those in the DE440 frame as a function of the right ascension (left) and declination (right).
    }
    \end{figure*}

    \begin{figure*} 
      \includegraphics[width=\columnwidth]{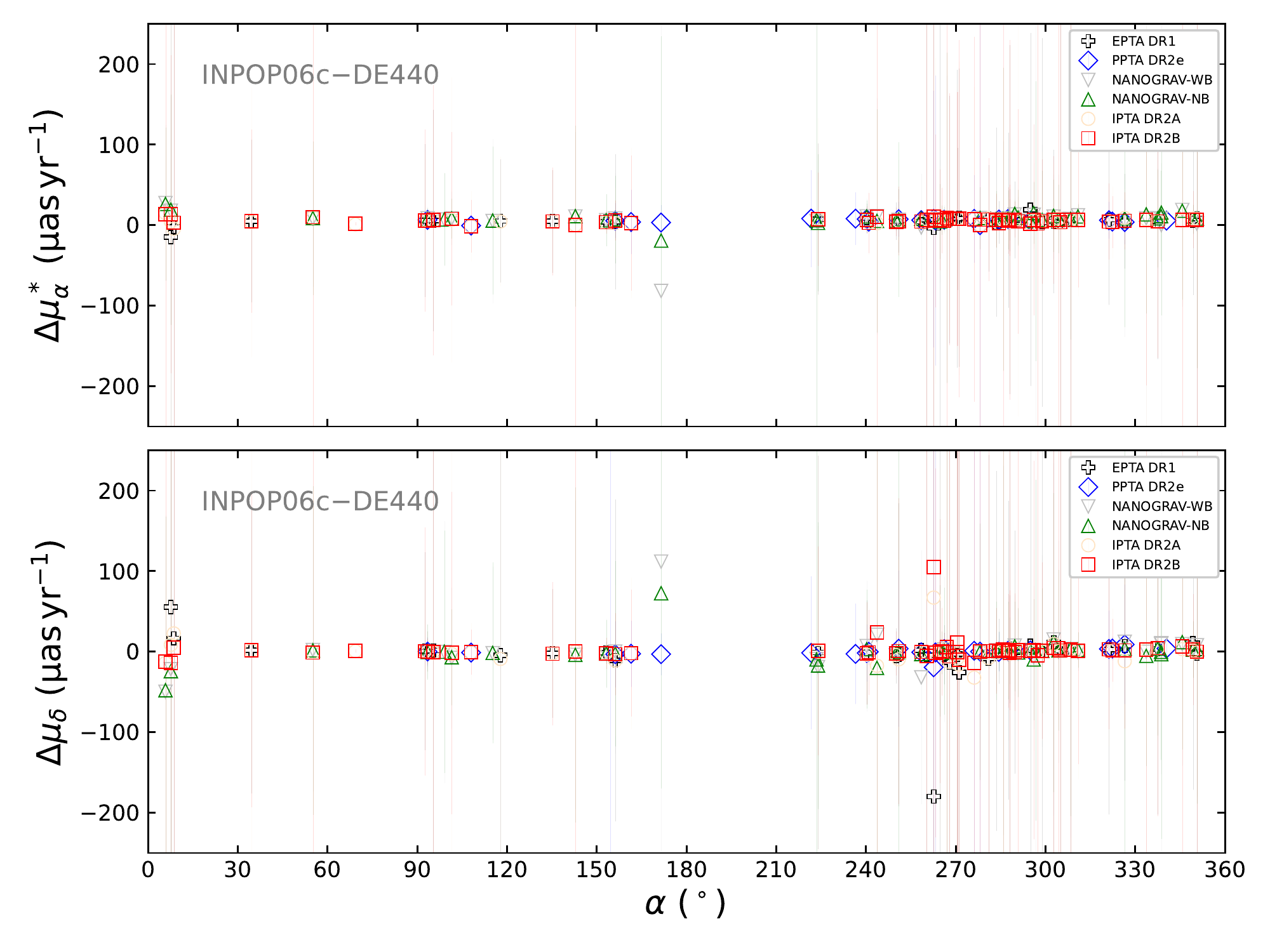}
      \includegraphics[width=\columnwidth]{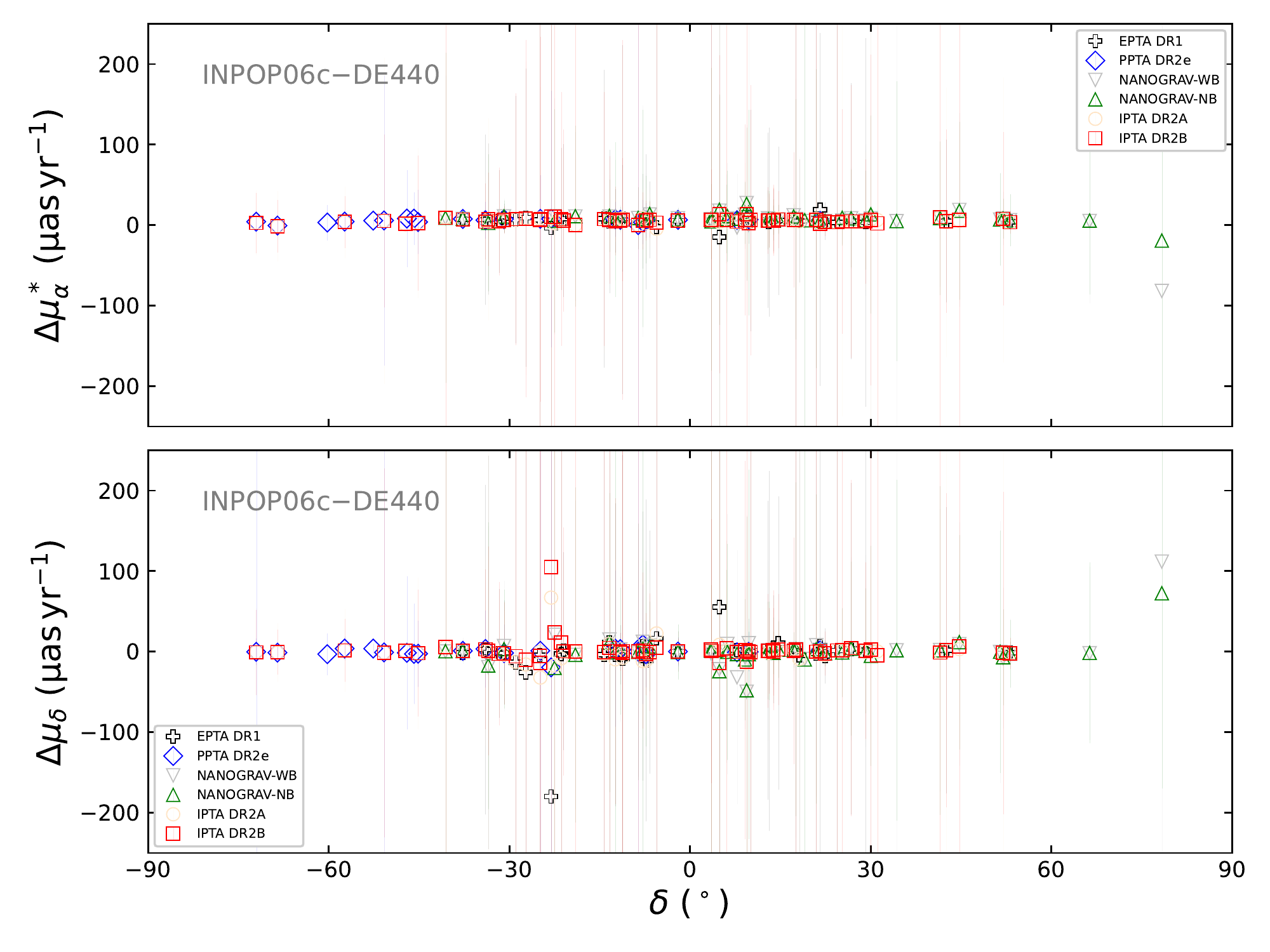}
      \caption[]{\label{fig:equ-pm-inpop06c-vs-de440} %
      Offsets of the pulsar timing proper motions in the INPOP06c frame with referred to those in the DE440 frame as a function of the right ascension (left) and declination (right).
    }
    \end{figure*}

\FloatBarrier

    \begin{figure*} 
      \includegraphics[width=\columnwidth]{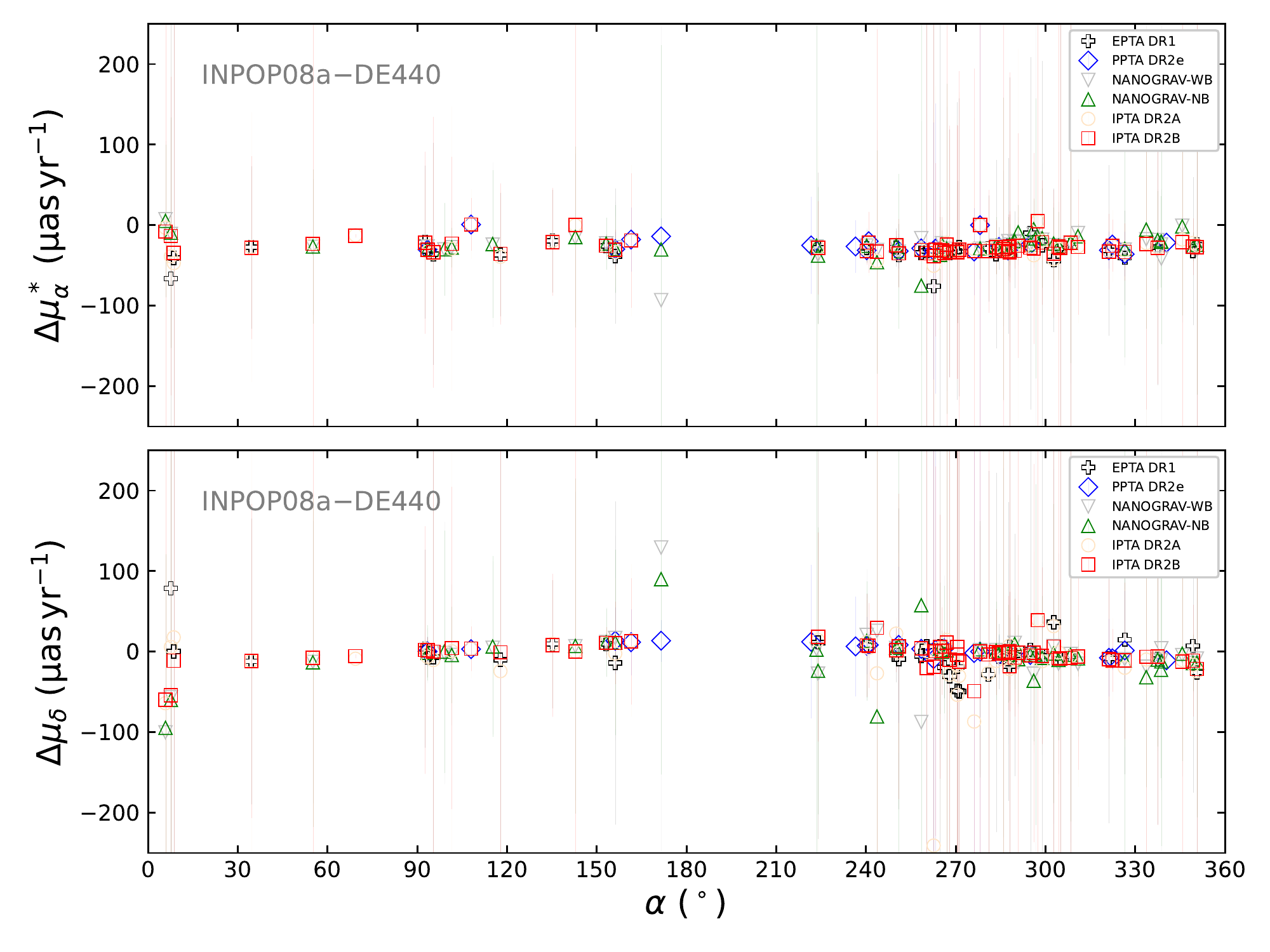}
      \includegraphics[width=\columnwidth]{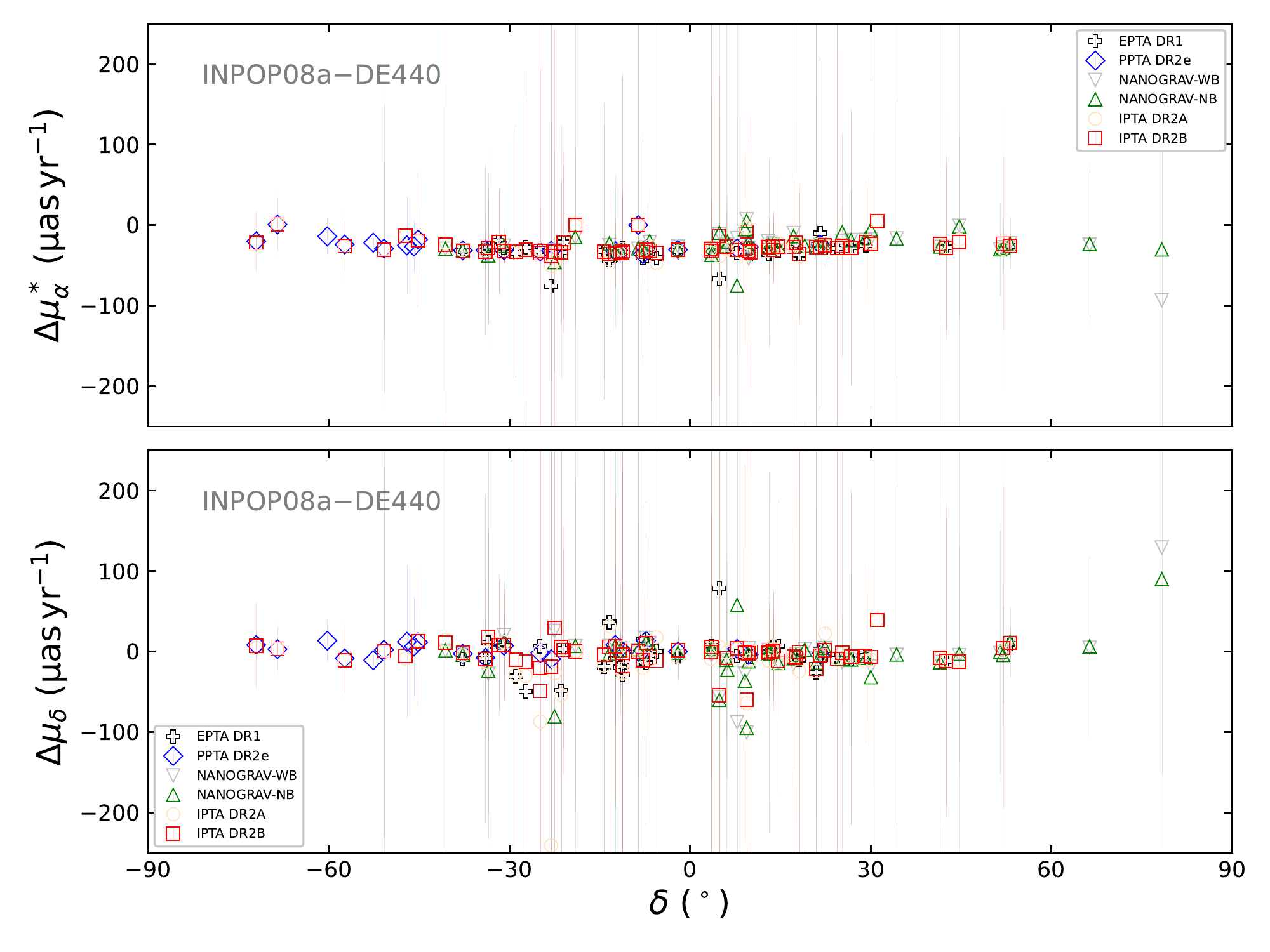}
      \caption[]{\label{fig:equ-pm-inpop08a-vs-de440} %
      Offsets of the pulsar timing proper motions in the INPOP08a frame with referred to those in the DE440 frame as a function of the right ascension (left) and declination (right).
    }
    \end{figure*}

    \begin{figure*} 
      \includegraphics[width=\columnwidth]{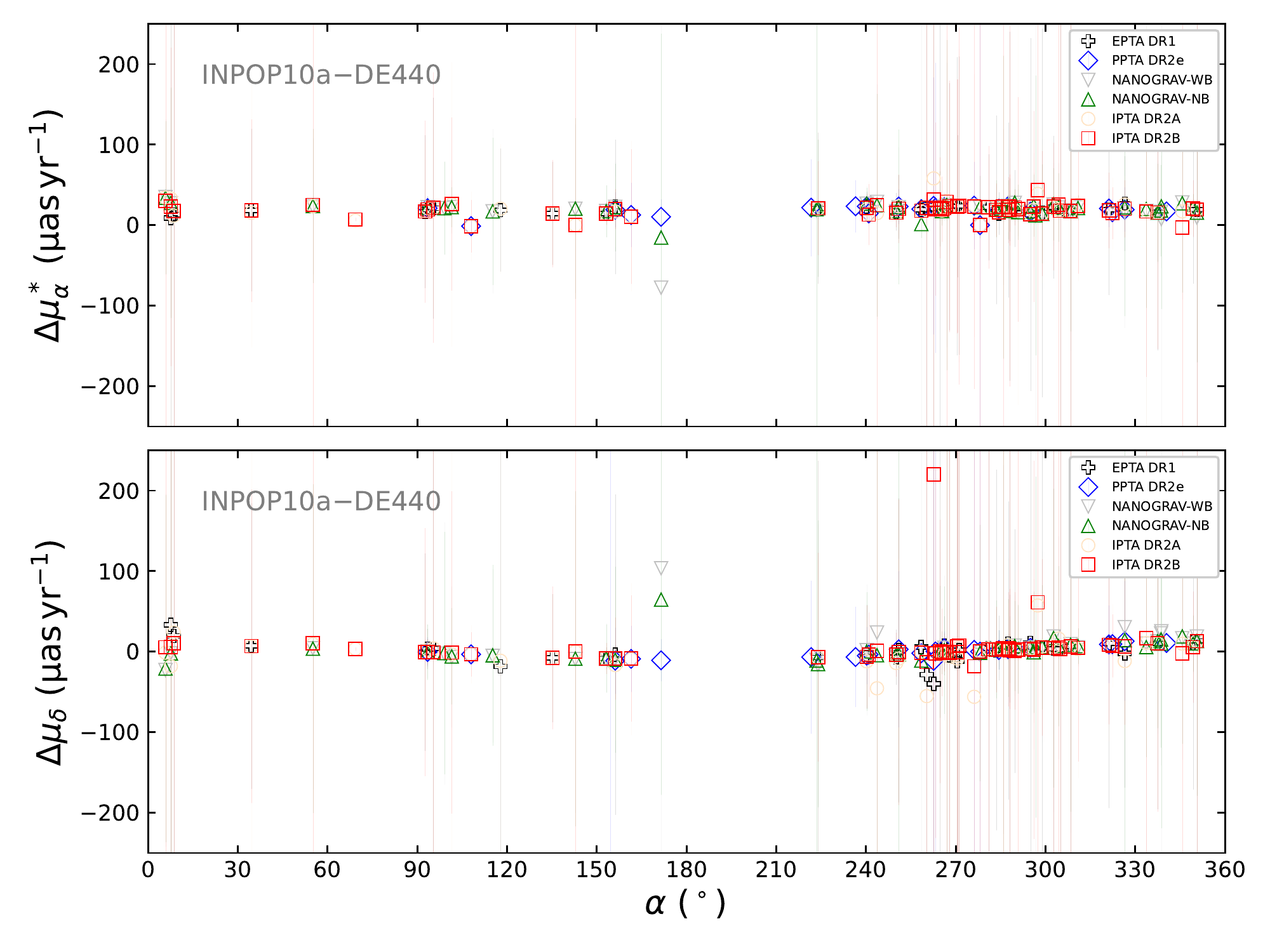}
      \includegraphics[width=\columnwidth]{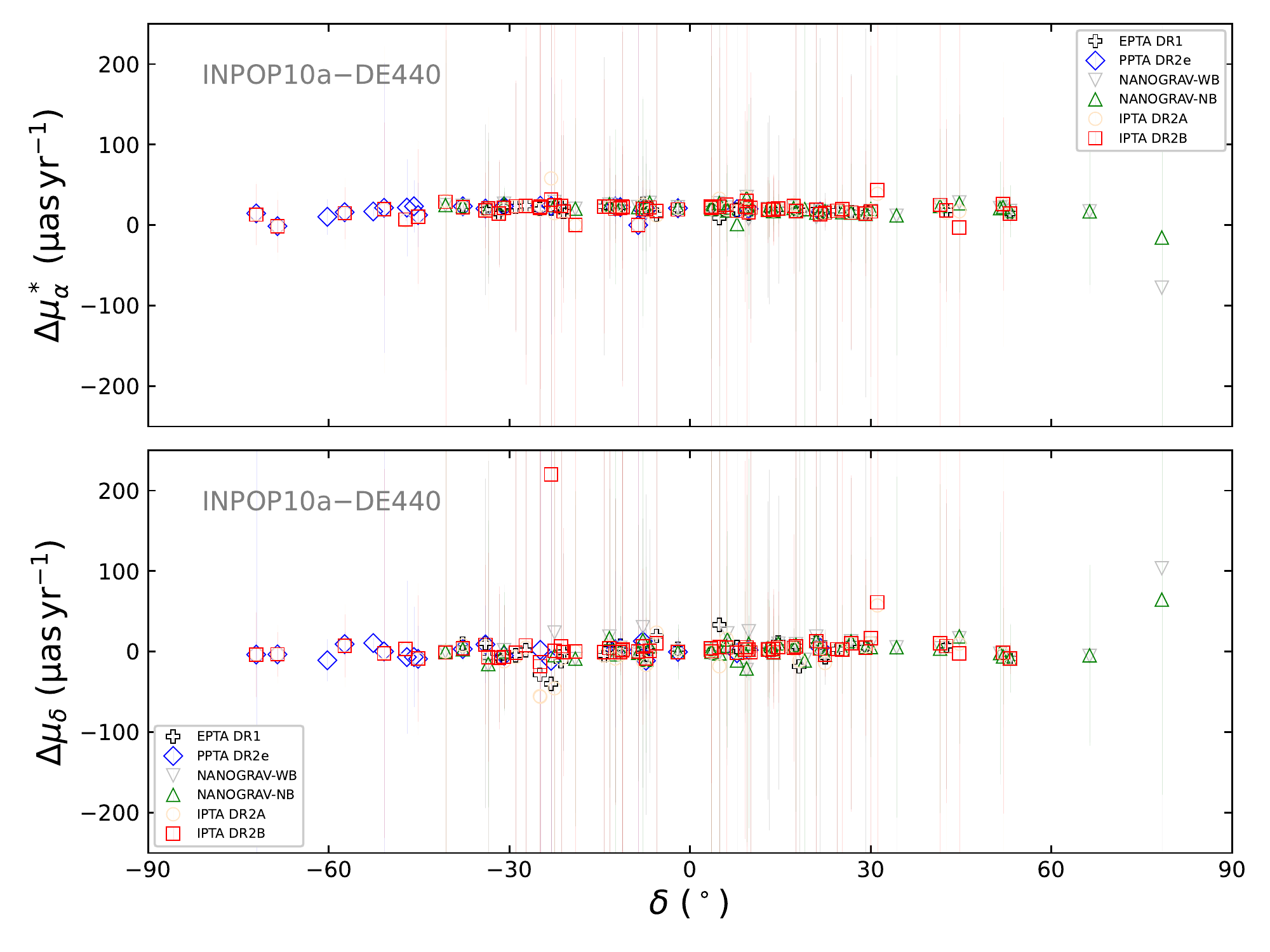}
      \caption[]{\label{fig:equ-pm-inpop10a-vs-de440} %
      Offsets of the pulsar timing proper motions in the INPOP10a frame with referred to those in the DE440 frame as a function of the right ascension (left) and declination (right).
    }
    \end{figure*}

    \begin{figure*} 
      \includegraphics[width=\columnwidth]{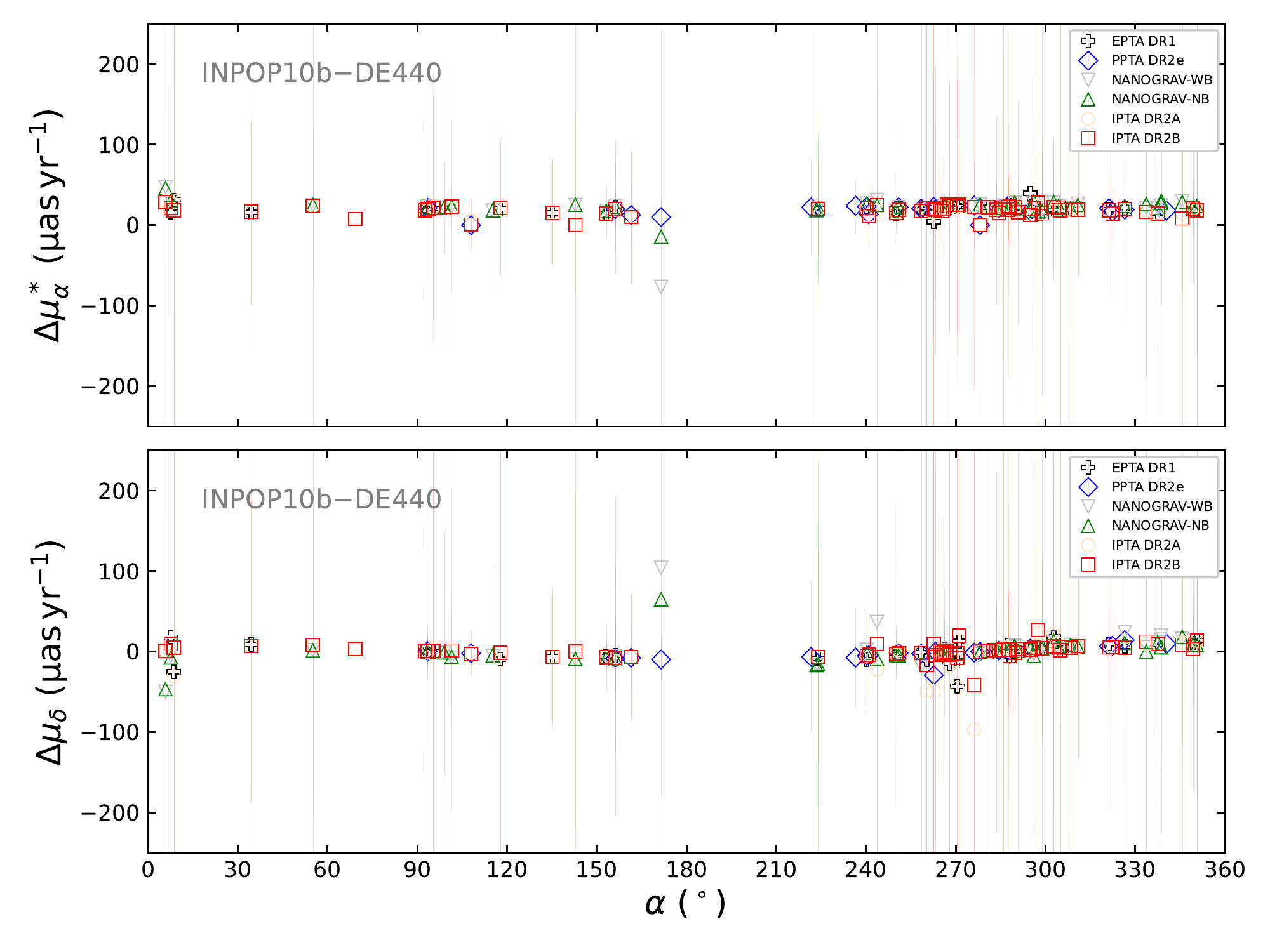}
      \includegraphics[width=\columnwidth]{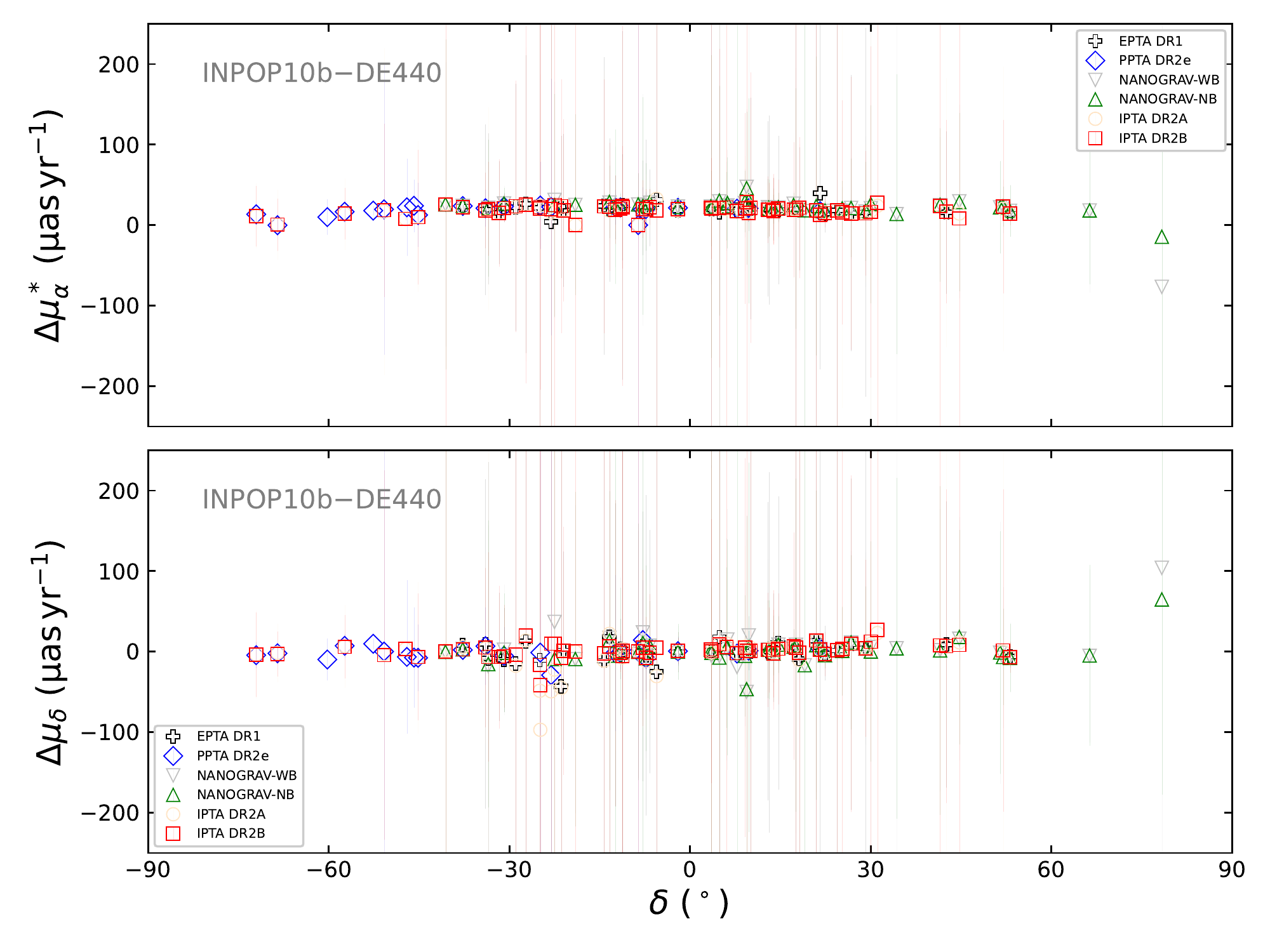}
      \caption[]{\label{fig:equ-pm-inpop10b-vs-de440} %
      Offsets of the pulsar timing proper motions in the INPOP10b frame with referred to those in the DE440 frame as a function of the right ascension (left) and declination (right).
    }
    \end{figure*}

\FloatBarrier

    \begin{figure*} 
      \includegraphics[width=\columnwidth]{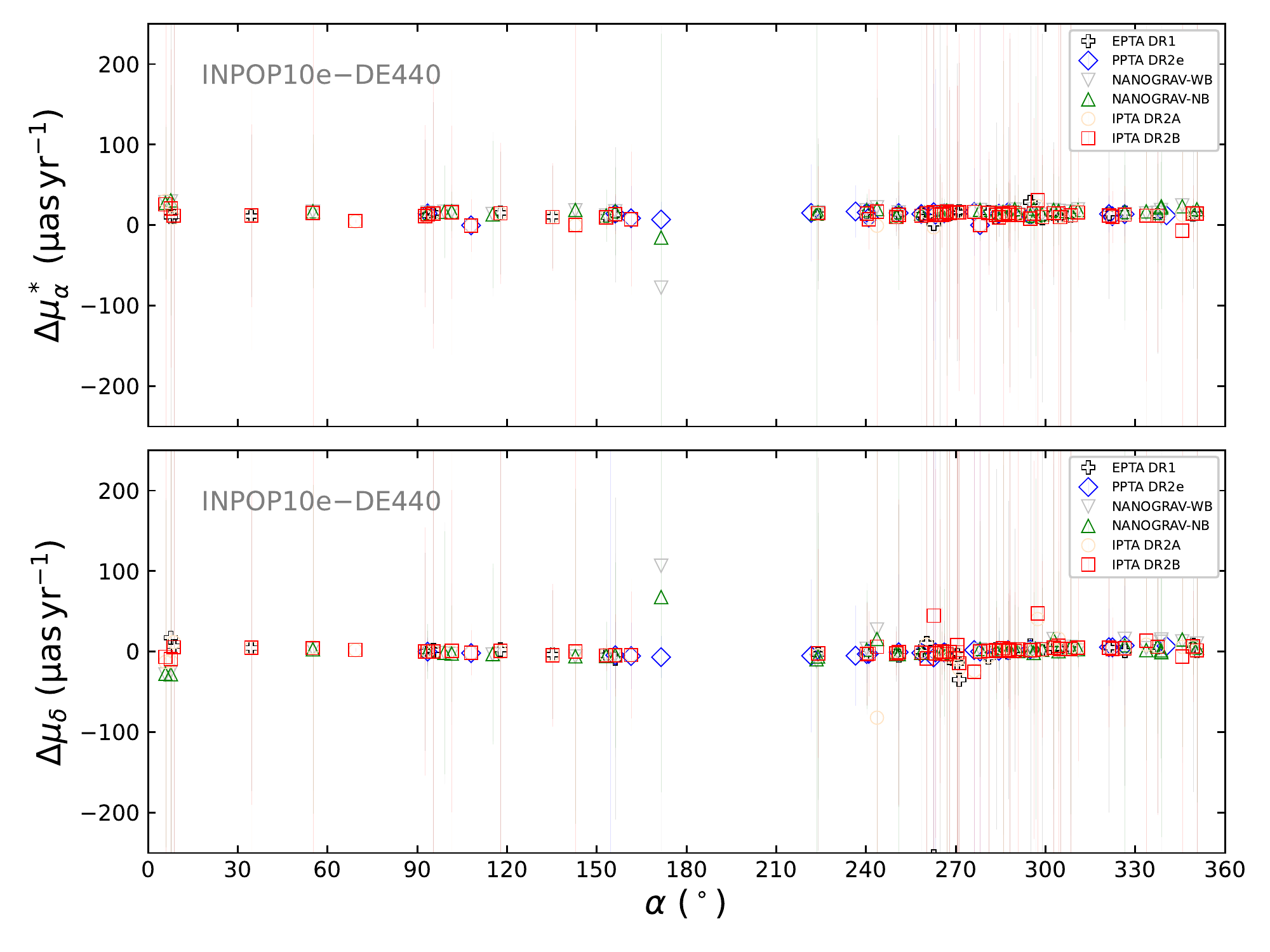}
      \includegraphics[width=\columnwidth]{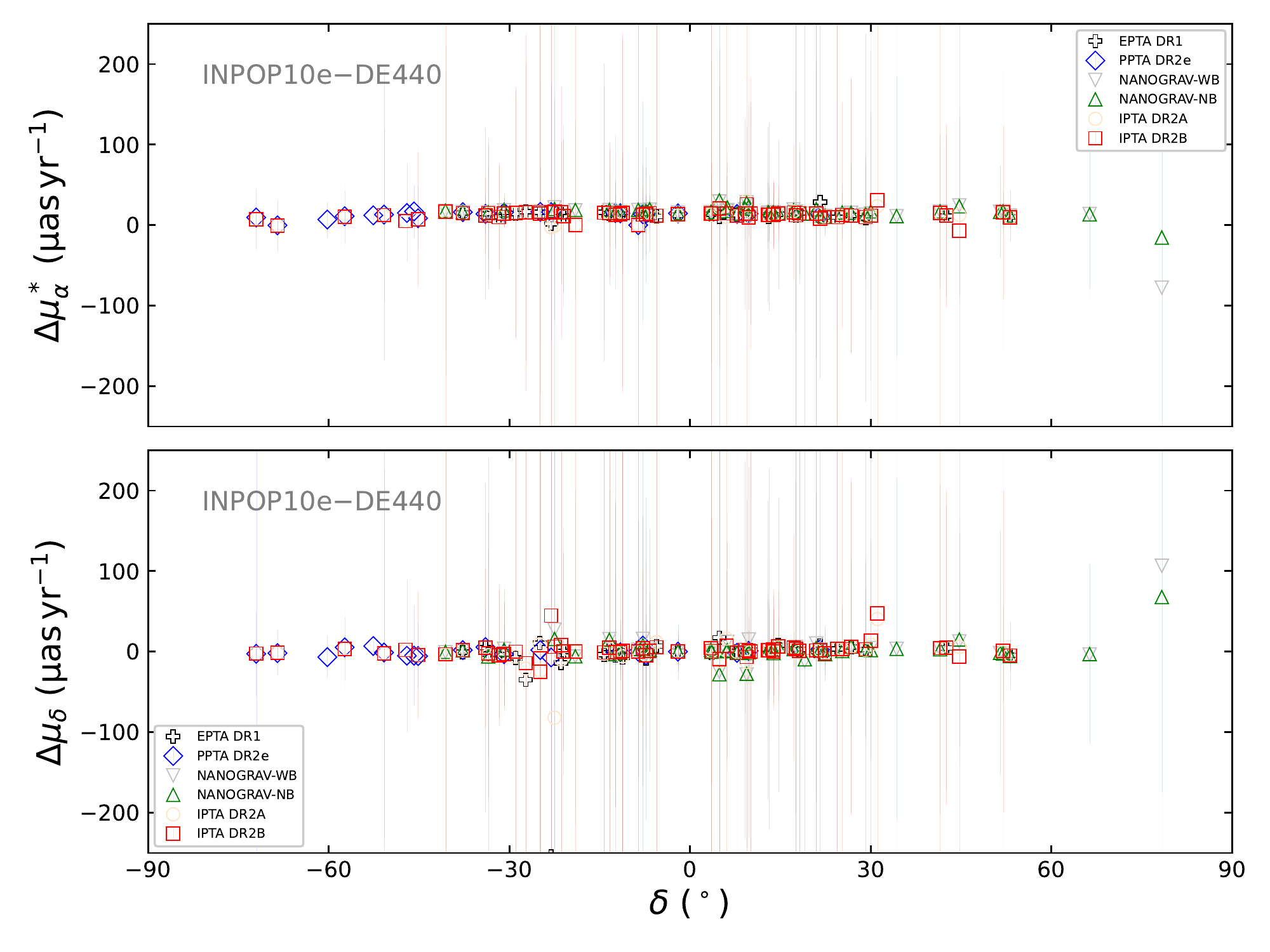}
      \caption[]{\label{fig:equ-pm-inpop10e-vs-de440} %
      Offsets of the pulsar timing proper motions in the INPOP10e frame with referred to those in the DE440 frame as a function of the right ascension (left) and declination (right).
    }
    \end{figure*}

    \begin{figure*} 
      \includegraphics[width=\columnwidth]{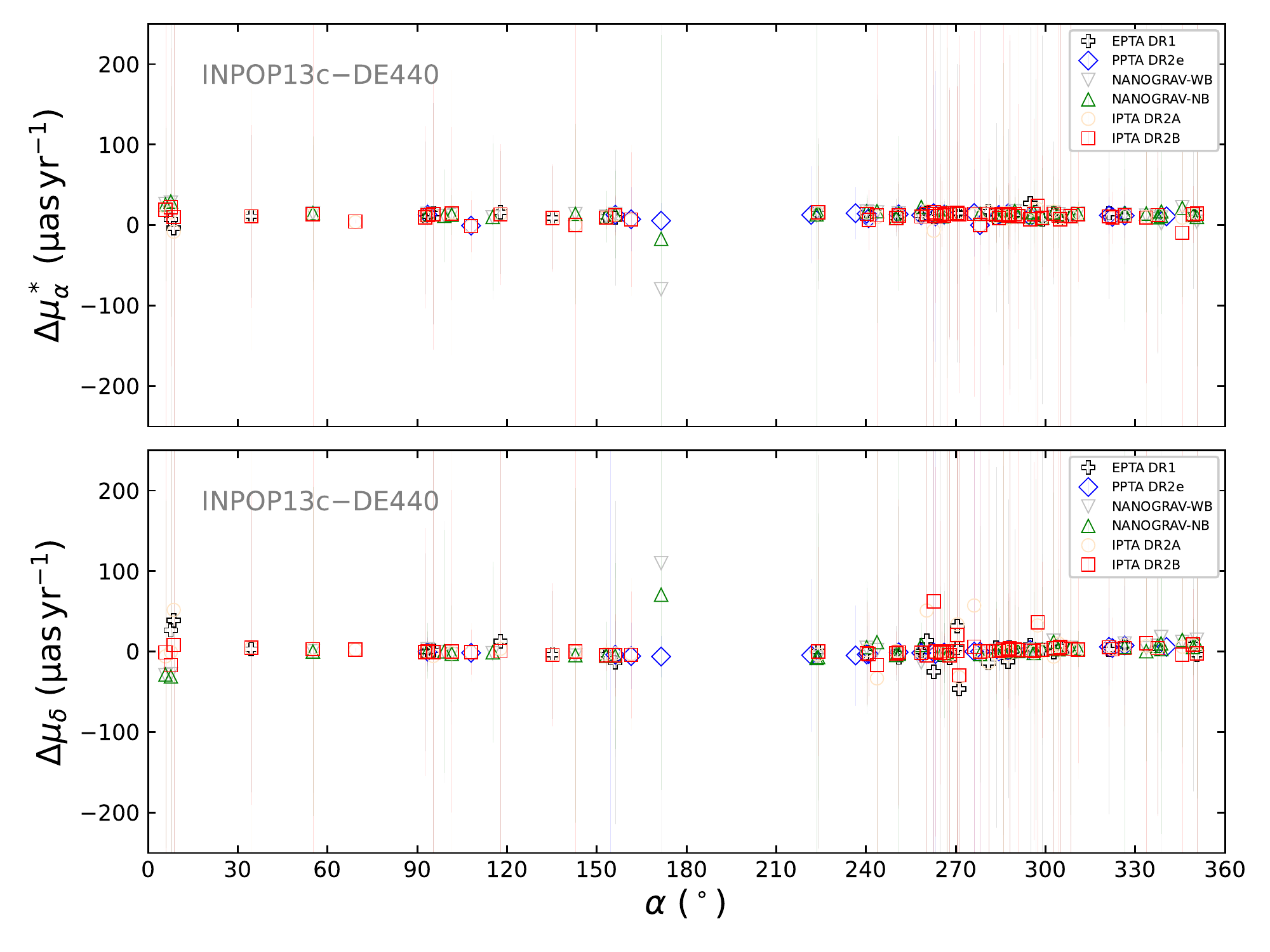}
      \includegraphics[width=\columnwidth]{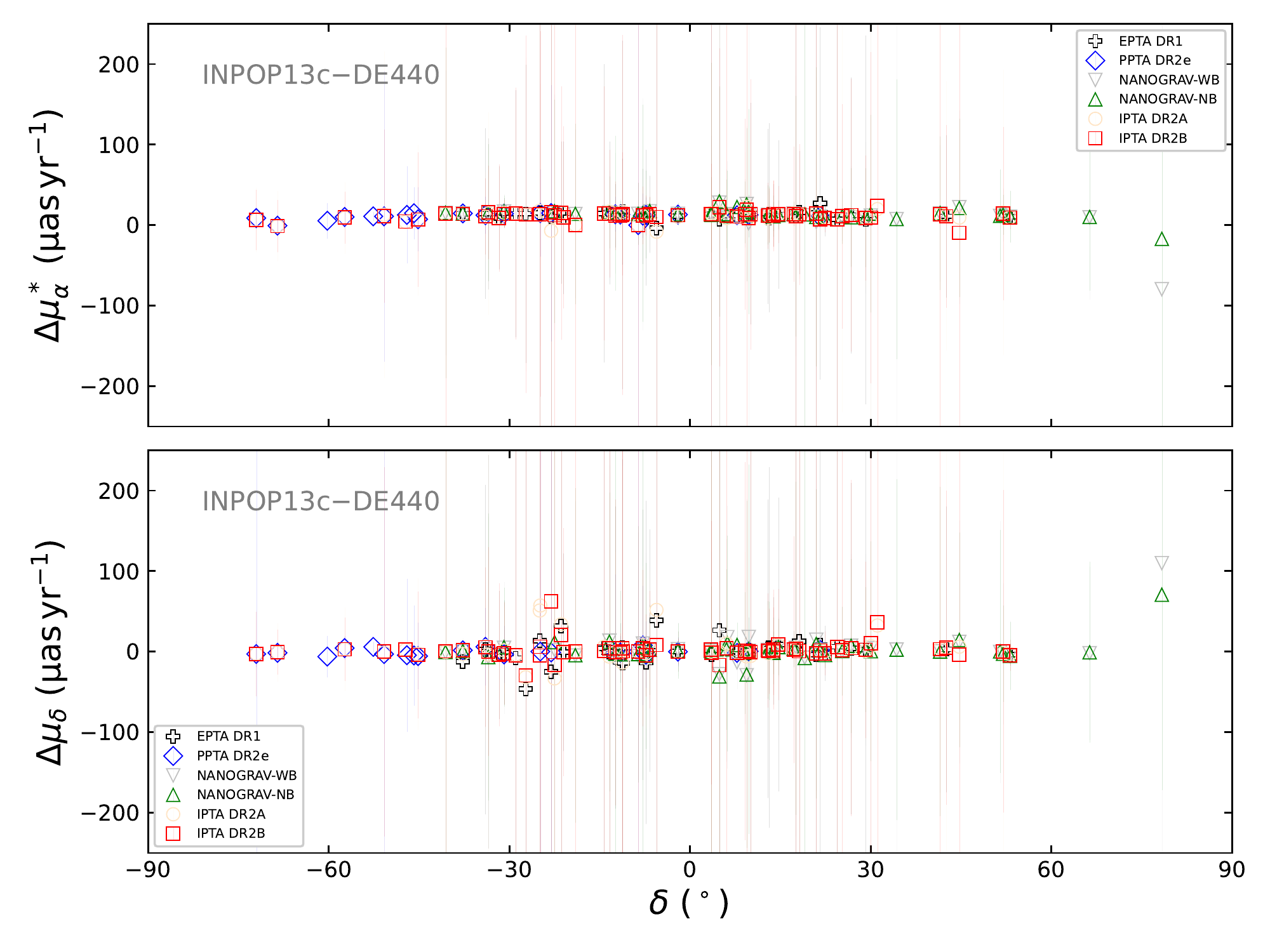}
      \caption[]{\label{fig:equ-pm-inpop13c-vs-de440} %
      Offsets of the pulsar timing proper motions in the INPOP13c frame with referred to those in the DE440 frame as a function of the right ascension (left) and declination (right).
    }
    \end{figure*}

    \begin{figure*} 
      \includegraphics[width=\columnwidth]{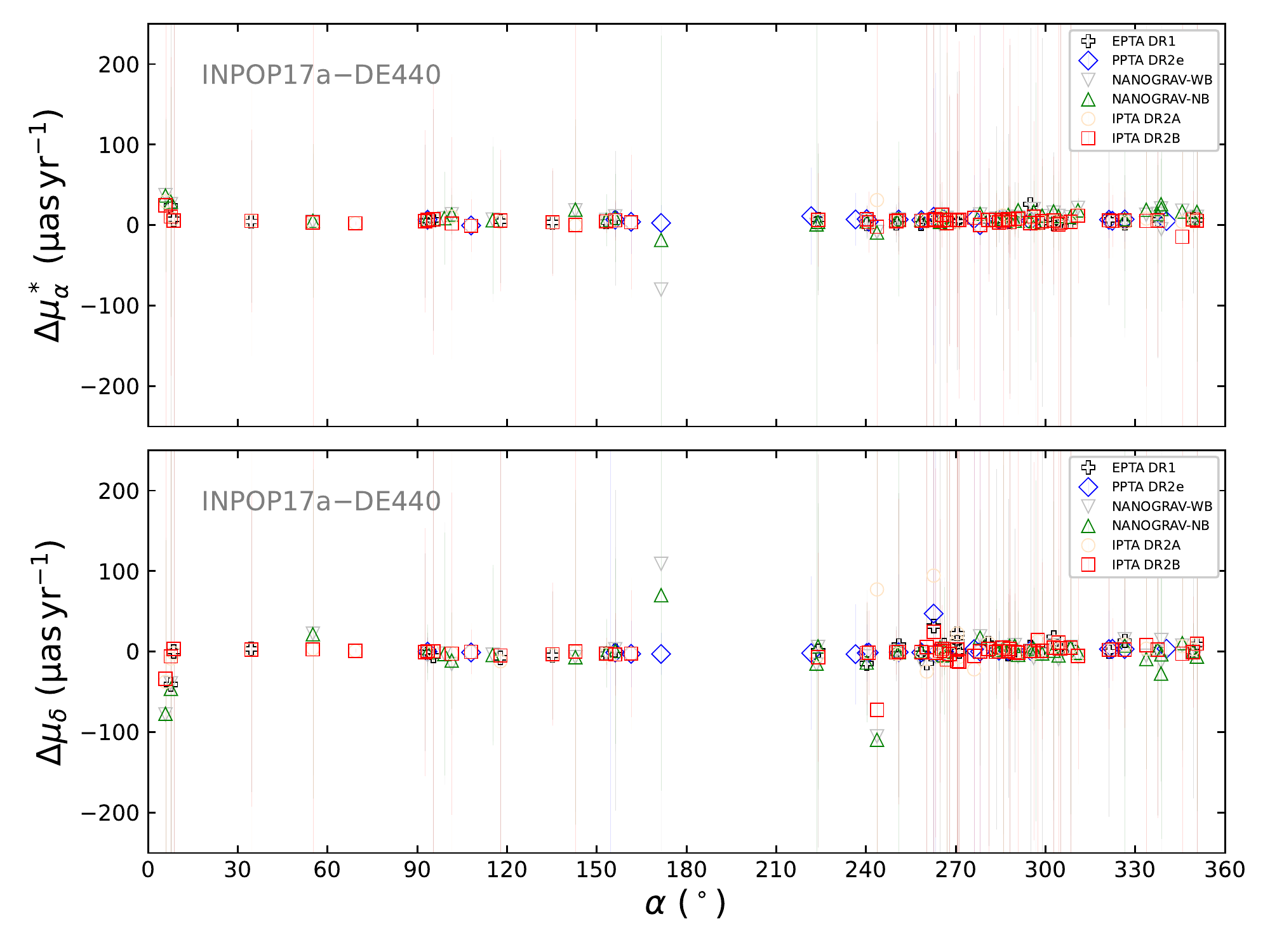}
      \includegraphics[width=\columnwidth]{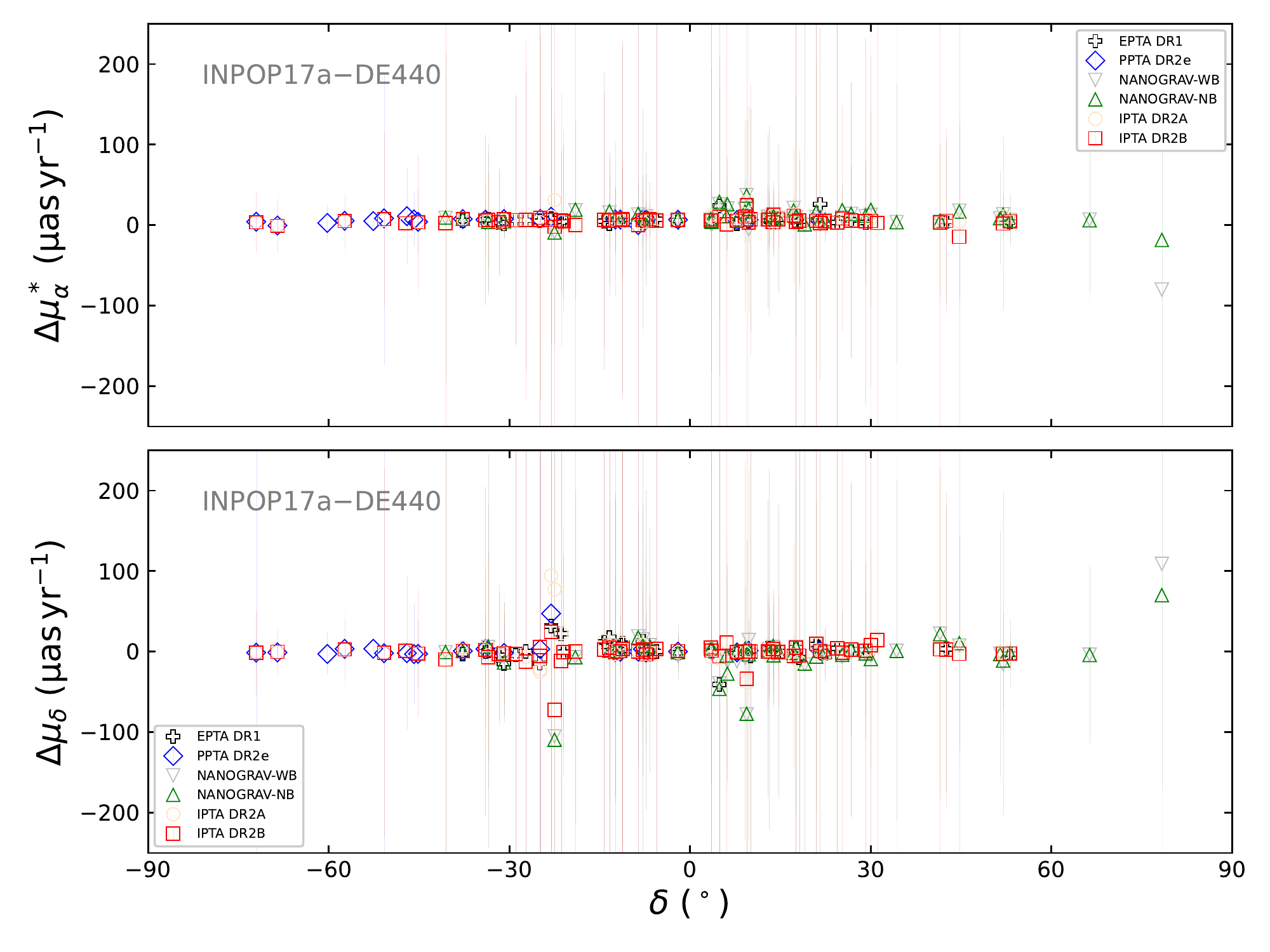}
      \caption[]{\label{fig:equ-pm-inpop17a-vs-de440} %
      Offsets of the pulsar timing proper motions in the INPOP17a frame with referred to those in the DE440 frame as a function of the right ascension (left) and declination (right).
    }
    \end{figure*}

\FloatBarrier

    \begin{figure*} 
      \includegraphics[width=\columnwidth]{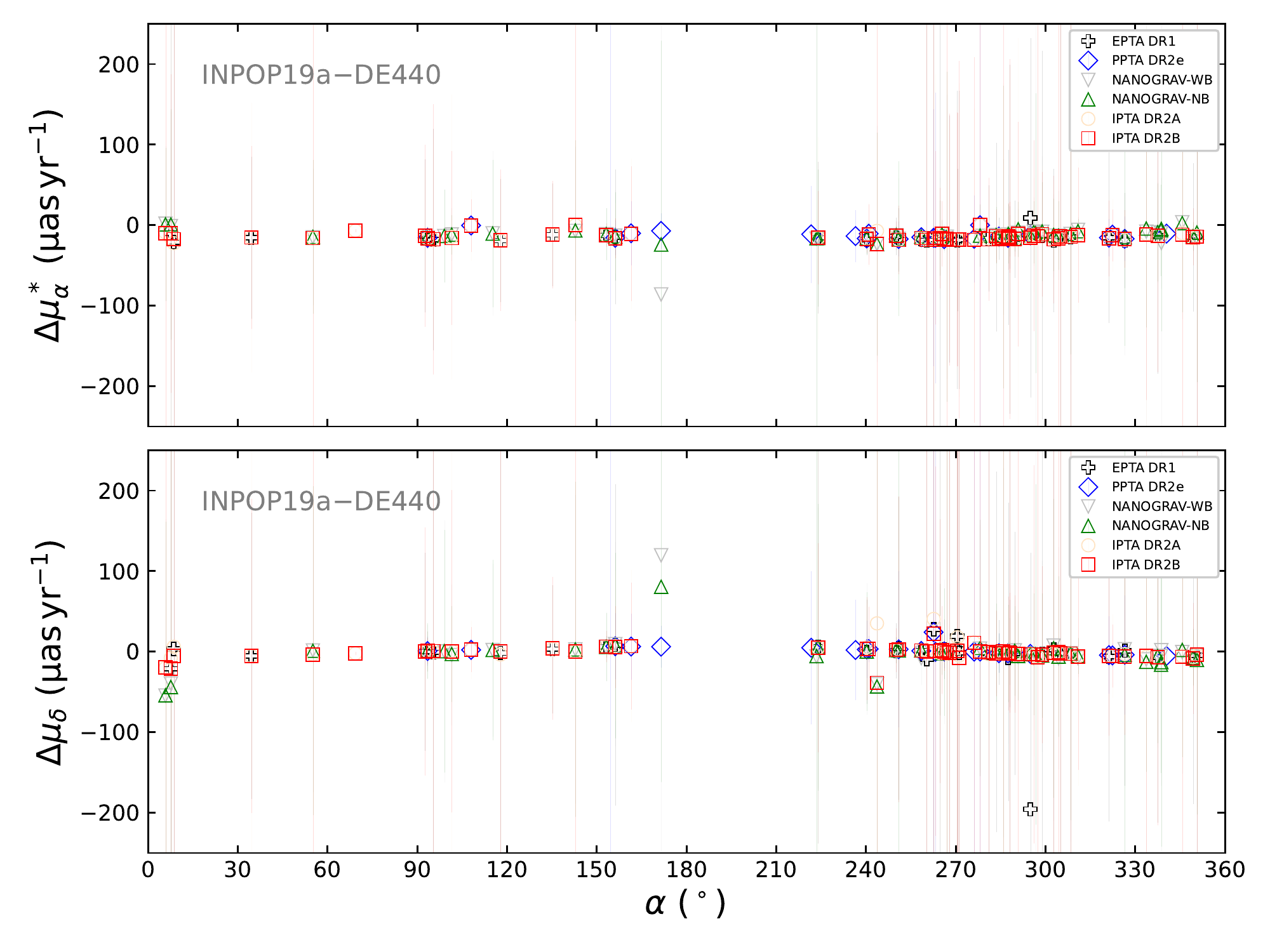}
      \includegraphics[width=\columnwidth]{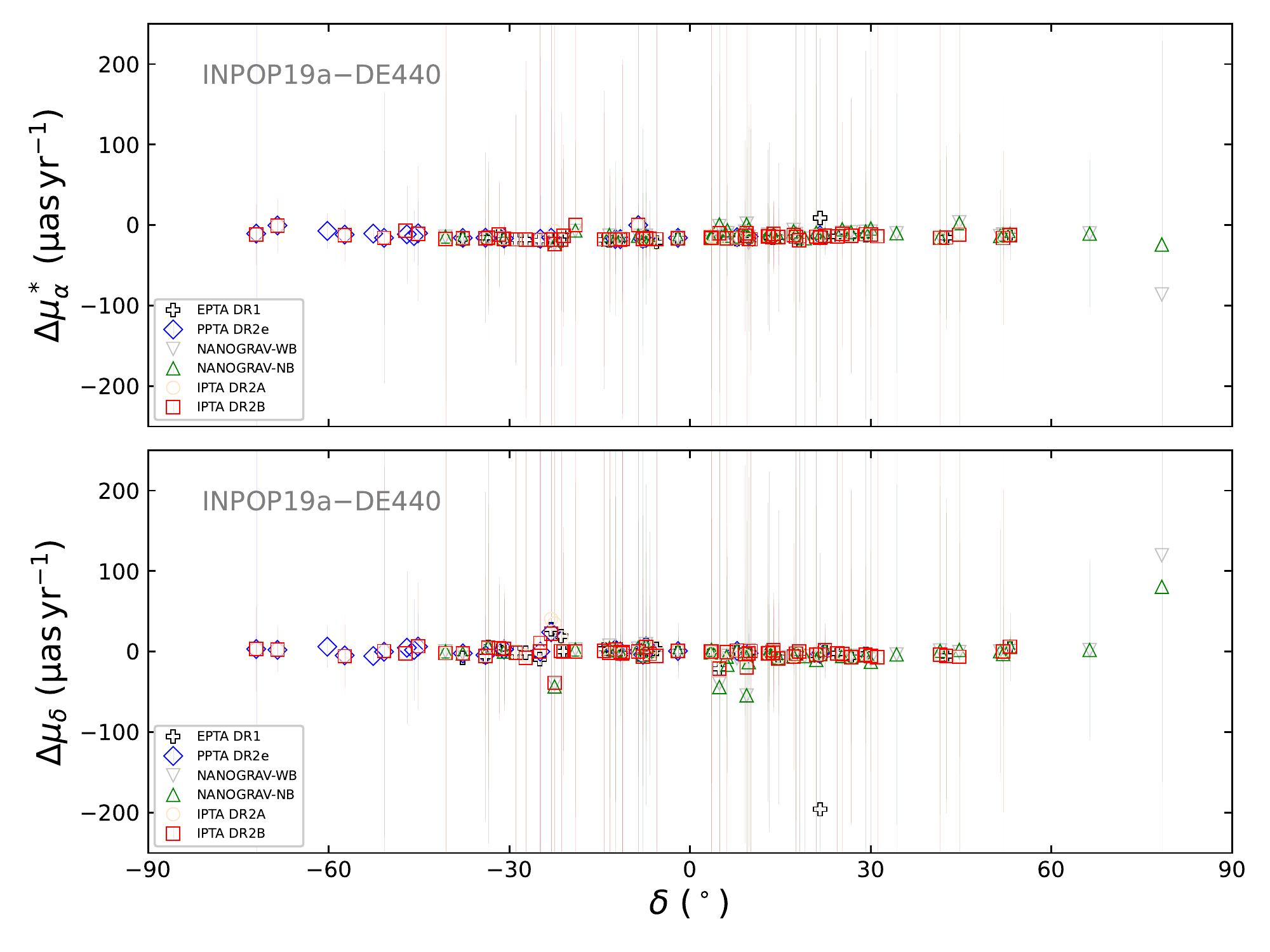}
      \caption[]{\label{fig:equ-pm-inpop19a-vs-de440} %
      Offsets of the pulsar timing proper motions in the INPOP19a frame with referred to those in the DE440 frame as a function of the right ascension (left) and declination (right).
    }
    \end{figure*}